\documentclass[fleqn,usenatbib]{mnras}
\usepackage{newtxtext,newtxmath}
\usepackage{comment}
\usepackage{multirow}
\usepackage{mathtools}
\usepackage[T1]{fontenc}
\usepackage{ae,aecompl}
\usepackage{gensymb}

\usepackage{graphicx}	
\usepackage{amsmath}	
\usepackage[ruled,vlined]{algorithm2e}
\usepackage{algpseudocode}
\usepackage{enumitem} 
\usepackage{xfrac} 

\setlist[itemize]{leftmargin=*}

\newcommand{\norm}[1]{\|#1\|}

\newcommand{\abs}[1]{\left|#1\right|}
\newcommand{\ud}{\mathrm{d}}

\newcommand{\vect}[1]{\vec{#1}}

\newcommand{\derfrac}[2]{\frac{\ud #1}{\ud #2}}

\renewcommand{\d}[1]{\ensuremath{\operatorname{d}\!{#1}}}

\newcommand{\bifrost}{\texttt{BIFROST}}

\newcommand{\sevn}{\texttt{SEVN}}
\newcommand{\parsec}{\texttt{PARSEC}}

\title[Asymmetric eccentric disks around SMBHs]{{Evolution of eccentric stellar disks around supermassive black holes: the complex disk disruption dynamics and the milliparsec stars}}
\author[A. Rantala \& T. Naab]{Antti Rantala$^{1}$\thanks{E-mail: anttiran@mpa-garching.mpg.de} and Thorsten Naab$^{1}$\\
$^{1}$Max-Planck-Institut f\"ur Astrophysik, Karl-Schwarzschild-Str. 1, 
D-85748, Garching, Germany\\
}
\date{Accepted XXX. Received YYY; in original form ZZZ}
\pubyear{2023}

\begin{document}
\label{firstpage}
\pagerange{\pageref{firstpage}--\pageref{lastpage}}
\maketitle

\begin{abstract}
We study the 10 Myr evolution of parsec-scale stellar disks with initial masses of $M_{\mathrm{disk}} = 1.0$ -- $7.5 \times 10^4 M_\odot$ and eccentricities $e_\mathrm{init}=0.1$--$0.9$ around supermassive black holes (SMBHs). Our disk models are embedded in a spherical background potential and have top-heavy single and binary star initial mass functions (IMF) with slopes of $0.25$--$1.7$. The systems are evolved with the N-body code \bifrost{} including post-Newtonian (PN) equations of motion and simplified stellar evolution. All disks are unstable and evolve on Myr timescales towards similar eccentricity distributions peaking at $e_\star \sim 0.3$--$0.4$. Models with high $e_\mathrm{init}$ also develop a very eccentric $(e_\star\gtrsim0.9)$ stellar population. For higher disk masses $M_\mathrm{disk} \gtrsim3 \times10^4\;\mathrm{M_\odot}$, the disk disruption dynamics is more complex than the standard secular eccentric disk instability with opposite precession directions at different disk radii - a precession direction instability. We present an analytical model describing this behavior. A milliparsec population of $N\sim10$--$100$ stars forms around the SMBH in all models. For low $e_\mathrm{init}$ stars migrate inward while for $e_\mathrm{init}\gtrsim0.6$ stars are captured by the Hills mechanism. Without PN, after $6$ Myr the captured stars have a sub-thermal eccentricity distribution. We show that including PN effects prevents this thermalization by suppressing resonant relaxation effects and cannot be ignored. The number of tidally disrupted stars is similar or larger than the number of milliparsec stars. None of the simulated models can simultaneously reproduce the kinematic and stellar population properties of the Milky Way center clockwise disk and the S-cluster.
\end{abstract}


\section{Introduction}\label{section: finalnumber-1}

\begin{figure*}
\includegraphics[width=0.8\textwidth]{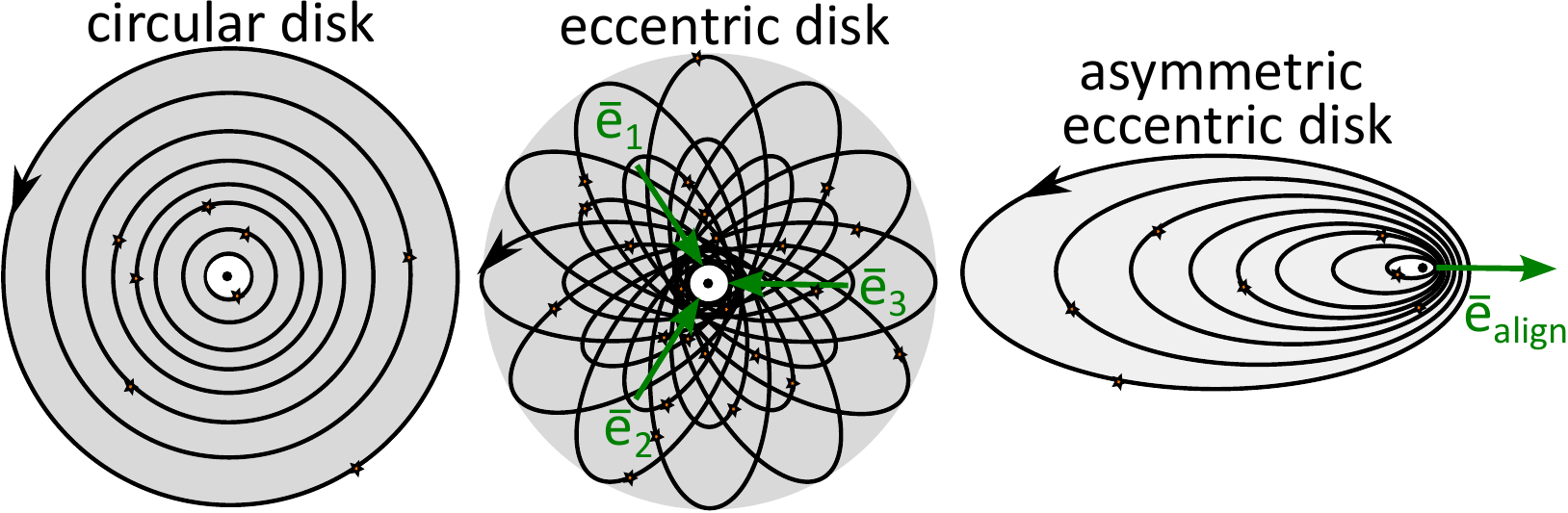}
\caption{Three types of disks around a SMBH. The most simple type of a disk is a Solar system like circular disk (left panel) in which all the disk stars have almost circular orbits. The orbits of the stars can also be eccentric in an axisymmetric disk (middle panel) with non-aligned the pericenter directions (eccentricity vectors) $\vect{e}_\mathrm{i}$. If the pericenter directions are apsidially aligned (right panel), i.e. $\vect{e}_\mathrm{i}=\vect{e}_\mathrm{aligned}$, the eccentric disk becomes asymmetric.}
\label{fig: disk-illustration}
\end{figure*}

Stellar disks with non-zero eccentricities around supermassive black holes (SMBHs) appear to be a common feature of galactic nuclei in the local Universe. In the Local Group, both the nuclei of the Milky Way and M31 galaxies host stellar disks within the gravitational influence radii of their central SMBHs. For M31, the observed double nucleus \citep{Light1974, Lauer1993, Kormendy1999,Statler1999,Bacon2001,Sambhus2002} is well explained by a apsidially aligned (i.e. lopsided) eccentric disk model \citep{Tremaine1995, Salow2001,Salow2004}. A schematic illustration of such a disk is provided in Fig. \ref{fig: disk-illustration}. The central SMBH of M31 appears to be surrounded by blue, bright stars \citep{Lauer1998,Brown1998,Lauer2012} with an estimated age of $200$ Myr \citep{Bender2005}. In addition, a number of complex observed features in the nuclei of local non-obscured, both dwarf and massive, early-type galaxies, such as offset or double nuclei and nuclei with central minima \citep{Binggeli2000,Lauer2002,Lauer2005,Houghton2006}, can be explained by aligned eccentric disks and viewing angle effects (e.g. \citealt{Sridhar1999,Madigan2018}).

At the distance of the Milky Way center, $\sim 8.2$ kpc \citep{GRAVITYCollaboration2019}, stellar kinematics can be studied in unprecedented level of detail compared to other galaxies. The central parsec of the Milky Way around the SMBH Sgr A* (e.g. \citealt{Morris1996,Tal2005,Genzel2010,Tal2011,Mapelli2016}) contains $\sim 100$--$200$ young, bright and massive stars (e.g. \citealt{Krabbe1991,Genzel1994,Blum1996,Lu2006}) of mainly types O and WR. The ages of these stars must be very young, $\lesssim10$ Myr from life-times O-type stars. \cite{Paumard2006} gives an age estimate of $6\pm2$ Myr while \cite{Lu2013} prefers a somewhat younger age of $2.5$--$6$ Myr. The total mass of the young stars can be as high as $1.4\times10^4\;\mathrm{M_\odot}$--$3.7\times10^4\;\mathrm{M_\odot}$ above $1\;\mathrm{M_\odot}$ \citep{Lu2013}. The stellar initial mass function (IMF) in the Milky Way center is very top-heavy, the IMF ($ \xi(m) \propto m^{-\alpha}$) slope estimates ranging from $\alpha=0.45\pm0.3$ \citep{Bartko2010} to $\alpha=1.7\pm0.2$ \citep{Lu2013} in various structures in the central $0.5$ pc. Resolved stellar kinematics (\citealt{Genzel2000,Ghez2000,Levin2003,Beloborodov2006,Lu2006,Gillessen2009,Bartko2010}) has revealed multiple disky and filamentary structures around the central SMBH Sgr A*, and most of the bright stars (up to $\sim75\%$, \citealt{vonFellenberg2022}) reside in these disk structures. The main structures are the two disks at very large angle with respect to each other, the clock-wise (CW) and the counter-clockwise disk (CCW) (e.g. \citealt{Genzel2003,Tanner2006,Paumard2006,vonFellenberg2022}). The CW disk is significantly warped (\citealt{Bartko2009}, see also \citealt{Yelda2014})  and is moderately eccentric with a median eccentricity of $0.4$-$0.5$. The CCW disk has similar median eccentricity as the CW disk but also contains high-eccentricity stars, though not beyond $e>0.9$ \citep{vonFellenberg2022}. Additionally, two outer disky or filamentary structures have been recently reported \citep{vonFellenberg2022}. The median eccentricity of these structures is higher than in the CW and CCW disks, $e\sim 0.7$. 

For the formation and evolution of such nuclear disks, hydrodynamical simulations have been used to study fragmentation and star formation both in circular \citep{Nayakshin2007} and eccentric \citep{Alexander2008} gas disks and rings around SMBHs. The molecular cloud infall and disruption simulations \citep{Sanders1998, Bonnell2008,Mapelli2008,Mapelli2012,Lucas2013,Trani2018} have shown to result in clumpy, filamentary, eccentric and rapidly star-forming disks with a top-heavy IMF. Various other dissipative scenarios have been investigated, including gas clouds engulfing SMBHs and forming accretion disks \citep{Wardle2008,Alig2011}, collisions of clouds with circumnuclear gas rings \citep{Alig2013} or clouds colliding with other clouds \citep{Hobbs2009}.

In this work we specifically focus on the dynamics of low-mass asymmetric eccentric disk systems for which $M_\mathrm{disk} \ll M_\mathrm{\bullet}$. The stability of such disks around SMBHs depends on how massive the disk is compared to the mass of the background stellar cusp at the spatial scale of the disk. Low-mass disks with $M_\mathrm{disk} \ll M_\mathrm{cusp}$ are unstable due to the so-called secular eccentric disk instability (\citealt{Madigan2009,Madigan2011}, see also \citealt{Haas2016,Subr2016}). The present-day individual Milky Way center disks around the $M_\bullet=4\times10^6\;\mathrm{M_\odot}$ SMBH \citep{GRAVITYCollaboration2019,GRAVITYCollaboration2020,EHT2022_IV} are in this range as $M_\mathrm{disk}\sim10^4\;\mathrm{M_\odot}$ for the CW disk and $M_\mathrm{disk} \sim 5\times 10^3\;\mathrm{M_\odot}$ for the CCW disk \citep{Bartko2010} and the background cusp mass within $r_\mathrm{0}=1$ pc is $M_\mathrm{cusp}(r_\mathrm{0})\sim 5 \times 10^5\;\mathrm{M_\odot}$ (e.g. \citealt{Genzel2003,Schodel2007}). These Milky Way values are collected into Table \ref{table: masses_timescales}. On the other hand, massive eccentric disks ($M_\mathrm{disk} \gg M_\mathrm{cusp}$) such as the asymmetric M31 center structure can be stable over gigayear timescales \citep{Tremaine1995,Jacobs2001,Madigan2018}.

A long-time motivation for studying the stellar dynamics in the central parsec of the Milky Way has been the goal of understanding its very central arc-second ($\sim 0.04$ pc). Going towards the central SMBH from the inner edge of the CW disk, the properties of the orbiting stars dramatically change. The innermost O/WR stars have a semi-major axis close to $a_\star\sim0.04$ pc inside which most of the observed stars are of type B. This compact cluster around the Milky Way SMBH is the famous S-cluster \citep{Eckart1996,Ghez1998}.  Rather than residing in a disky structure, the orbits of the S-stars are isotropic. The eccentricities of the S-stars are predominantly high compared to the outer disk stars, following a thermal or even superthermal distribution (e.g. \citealt{Schodel2002,Ghez2003,Eisenhauer2005,Ghez2005,Gillessen2008,Gillessen2009,Gillessen2017}). Just as the O/WR stars in the disks, the B-stars of the S-cluster must be young, having an age below $\lesssim400$ Myr from their main sequence lifetimes. Whether the S-cluster stars and the disk originate from a common formation scenario still remains an open question. The S-cluster lacks O/WR stars, which are prominently present in the disks, and being isotropic and very eccentric the orbits of S-cluster stars considerably differ from the orbits of the disk stars \citep{Gillessen2017,vonFellenberg2022}.

One of the scenarios for the formation of the S-cluster relies on the Hills mechanism. Given a small enough pericenter distance $r_\mathrm{p}$, SMBHs can disrupt binary stars \citep{Hills1988}. One binary component is captured by the SMBH on a very eccentric orbit whereas the other component becomes unbound is ejected as a high-velocity star \citep{Bromley2012,Kobayashi2012,Zhang2013,Chen2014,Madigan2014,Zhang2013,Generozov2020,Generozov2021}. An A-type high-velocity star with a velocity of $1800\; \mathrm{km/s}$ and trajectory consistent with ejection by the Milky Way SMBH $\sim 5$ Myr ago has recently been discovered \citep{Koposov2020}, which might indicate that the Hills mechanism indeed operates in the Milky Way center. However, a single high-velocity star still does not prove that the Hills mechanism is the dominant formation pathway for the S-stars. Various competing scenarios to the Hills picture involve a low-eccentricity origin for the S-stars \citep{Genzel2010}. The low-eccentricity scenarios include disk migration in the initial (gaseous) disk \citep{Levin2007}  and migration aided by a massive perturber \citep{Perets2009b}, such as an intermediate-mass black hole (e.g. \citealt{MastrobuonoBattisti2014}). It should be noted that both the high and the low-eccentricity origin channels require an efficient relaxation mechanism to change their initial eccentricity distribution to the observed one, either from above or below \citep{Genzel2010}.

Various N-body simulation studies have been devoted to studying the origin and dynamics of the S-cluster stars (e.g. \citealt{Berukoff2006,Kenyon2008,Perets2009b,Madigan2009,Fujii2010,Perets2010,Antonini2013,Hamers2014,Subr2016,Generozov2020}). The simulations can be classified in two broad categories: the S-star origin simulations and S-cluster relaxation simulations. There is substantial overlap between the two categories, but only few simulation studies have explored the full dynamical picture using a single simulation code in a self-consistent manner. For the disk binary origin scenario we highlight the simulations of \cite{Subr2016} as the one of the most self-consistent setups to date. Their simulations begin from an eccentric disk with an IMF and a binary star population, follow the disk disruption and Hills mechanism of binary disruption, and finally the relaxation of the captured stars using the \texttt{NBODY6} code version of \cite{Aarseth2003}. However, their full setup did not include a stellar background population, post-Newtonian effects or stellar evolution. Thus, the eccentric disk instability and Hills mechanism scenario for the S-cluster formation has not yet been fully explored in the literature in a self-consistent manner.

In this study we examine the evolution of initially asymmetric eccentric disks and the dynamical consequences of the eccentric disk instability using direct N-body simulations. Our simulations have a number of important differences compared to the previous studies regarding the simulation setup and the numerical code used. Our initial conditions include an asymmetric eccentric disk model (with single and binary stars) around a SMBH embedded in a smooth external background potential representing the background stellar cusp. The gravitational dynamics of the system from disk disruption to Hills mechanism and the evolution of the population of captured stars around the SMBH is integrated using a single simulation code \bifrost{} \citep{Rantala2021,Rantala2023}. The code combines a hierarchical version of the fourth-order forward symplectic integrator \citep{Chin1997,Chin2005,Chin2007a,Dehnen2017a} with specialized secular and regularized solvers for binaries, close fly-bys, multiple systems and small clusters around massive black holes \citep{Rantala2020}. As described above, most simulation studies have concentrated in separate aspects such as the eccentric disk instability, Hills mechanism or the relaxation of the S-cluster, or included several but have switched from one simulation code to another during the study. We also employ single stellar evolution, tidal disruption event (TDE) prescription and post-Newtonian equations of motion for the stars in our simulation setups.

Next, we study the dynamics of the asymmetric eccentric disk instability in the intermediate-mass regime, i.e. the disk is still less massive than the background cusp, but its mass is not negligible (with still assuming $M_\mathrm{disk} \ll M_\bullet$). Thus, the inequality $M_\mathrm{disk} < M_\mathrm{cusp}$ still holds, but the relation $M_\mathrm{disk} \ll M_\mathrm{cusp}$ does not. Because of this, the disk disruption process is more complex than the standard secular eccentric disk instability scenario of \cite{Madigan2009}, as the assumption of the negligible contribution of the disk mass itself to the precession of the disk star orbits does not hold anymore. We argue that the Milky Way center CW and CCW disks might have been in this mass regime at their formation time $\sim6$ Myr ago, and being top-heavy they have since lost mass due to stellar evolution, reaching $M_\mathrm{disk} \sim10^4\;\mathrm{M_\odot}$ at present day. Simulation studies in the literature have focused on studying the disruption of asymmetric eccentric disks with a mass of the present-day CW disk (e.g. \citealt{Madigan2009,Subr2016}) and not considered the fact that the disks were more massive at their formation time in the past.

The article is structured as follows. After the introduction we briefly review the relevant stellar-dynamical processes around SMBHs for this study in Section \ref{section: finalnumber-2}. The updated simulation code and the initial conditions for numerical simulations are described in Section \ref{section: finalnumber-3} and Section \ref{section: finalnumber-4}. Next, we discuss the simulation results. The evolution of the intermediate-mass asymmetric eccentric disks, their disruption and later evolution are examined in Section \ref{section: finalnumber-5} while Section \ref{section: finalnumber-6} presents the properties of high-velocity stars and stars accreted by the SMBH in the simulations. The origin and the properties of milliparsec-scale stellar population are investigated in Section \ref{section: finalnumber-7}. We discuss our results in the context of Milky Way center observations and previous simulation studies in \ref{section: finalnumber-8}. Finally, we summarize and conclude in Section \ref{section: finalnumber-9}.

\section{Stellar dynamics around supermassive black holes}\label{section: finalnumber-2}

\begin{table}
\begin{tabular}{l c l }
\hline
component & present-day mass & notes\\
\hline
SMBH & $M_\bullet = 4\times10^6\;\mathrm{M_\odot}$ &\\
disk & $M_\mathrm{disk} \sim 1\times10^4\;\mathrm{M_\odot}$ & CW disk\\
disk & $M_\mathrm{disk} \sim 5\times10^3\;\mathrm{M_\odot}$ & CCW disk\\
cusp & $M_\mathrm{cusp} \sim 5\times10^5\;\mathrm{M_\odot}$ & old stars within $1$ pc\\
\hline
\end{tabular}
\caption{The observed present-day masses of the Milky Way SMBH \citep{GRAVITYCollaboration2019}, the clockwise and the counterclockwise disk \citep{Bartko2010} and the old spherical cusp population \citep{Genzel2003,Schodel2007}.}
\label{table: masses_timescales}
\end{table}

\begin{table}
\begin{tabular}{l c l l}
\hline
physical effect & timescale & at $0.01$ pc & at $0.1$ pc\\
\hline
relativistic PN1.0 precession & $t_\mathrm{GR}$ & $0.6$ Myr & $190$ Myr\\
Lense-Thirring precession & $t_\mathrm{LT}$ & $182$ Myr & $\gg t_\mathrm{Hubble}$\\
mass precession & $t_\mathrm{M}$ & $1.3$ Myr & $1.0$ Myr\\
\hline
two-body relaxation & $t_\mathrm{NR}$ & $420$ Myr & $1.9$ Gyr\\
scalar resonant relaxation & $t_\mathrm{SRR}$ & $40$ Myr & $1.3$ Gyr\\
vector resonant relaxation & $t_\mathrm{VRR}$ & $0.2$ Myr & $1.8$ Myr\\
\hline
\end{tabular}
\caption{The approximate estimates for a number of precession and relaxation timescales in the Milky Way center. The orbital eccentricity is assumed to be $e_\star = 0.5$. The Lense-Thirring timescale is calculated for a rapidly spinning black hole with $s=1$. The relaxation timescales $t_\mathrm{NR}$, $t_\mathrm{SRR}$ and $t_\mathrm{VRR}$ are adopted from the recent estimates in the literature \citep{Panamarev2022}. Lower estimates for the scalar resonant relaxation timescale have also been proposed, down to $10$ Myr near $0.01$ pc \citep{Antonini2013}.
}
\label{table: precession_timescales}
\end{table}

\subsection{Relativistic and Newtonian and precession effects}
\subsubsection{Relativistic precession}

We very briefly review the most important stellar dynamics processes around supermassive black holes relevant for this study. If the SMBH is considerably more massive than the background stellar cusp ($M_\bullet\gg M_\mathrm{cusp}$), the motion of stars around the SMBH is nearly Keplerian. The Keplerian orbital period of a star with a mass $m_\star$ and a semi-major axis $a_\star$ around a SMBH with mass $M_\bullet$ is
\begin{equation}
P_\mathrm{\star} = 2 \pi \left( \frac{a_\mathrm{\star}^3}{G M} \right)^{1/2}    
\end{equation}
in which $M=M_\bullet+m_\star$. If the stellar cusp is spherically symmetric, the reduced angular momentum vector $\vect{h}=\vect{r}\times\vect{v}$ of an example star is constant and the orbit of the star is confined into a plane. The eccentricity vector (also known as the Laplace-Runge-Lenz vector) $\vect{e}$ of the star pointing towards the pericenter defined as
\begin{equation}\label{eq: laplacerungelenz}
    \vect{e} = \frac{1}{G M} \vect{v}\times \vect{h} - \frac{\vect{r}}{r}
\end{equation}
is another constant of motion in the pure Keplerian case, and the direction of the periapsis of the orbit does not change over time. The eccentricity of the stellar orbit is the norm of the eccentricity vector, $e_\star = \norm{\vect{e}}$.

Both Newtonian (non-Keplerian) and relativistic effects can cause the orbit of the star to precess, i.e. the direction of the eccentricity vector will change with time. In the following we assume $M_\bullet \gg m_\star$. In the lowest-order post-Newtonian order PN1.0 the precession time-scale $t_\mathrm{GR}$ of the orbit is
\begin{equation}\label{eq: timescale-pnprecession}
t_\mathrm{GR} = \frac{ c^2 a_\mathrm{\star} (1-e_\mathrm{\star}^2) P_\mathrm{\star} }{3 G M_\mathrm{\bullet}} = \frac{ a_\mathrm{\star} (1-e_\mathrm{\star}^2) P_\mathrm{\star} }{3 r_\mathrm{g}}
\end{equation}
in which $r_\mathrm{g}$ is the gravitational radius of the SMBH. The PN1.0 precession is commonly referred to as the geodetic, mass, or de Sitter precession. The precession direction is famously prograde \citep{Einstein1915}. The spin $\vect{S}_\bullet$ of the SMBH is the source of the frame dragging or the Lense-Thirring nodal precession \citep{Lense1918}. The precession time-scale \citep{Poisson2014} is
\begin{equation}\label{eq: timescale-ltprecession}
t_\mathrm{LT} = \frac{\pi c^3 a_\star^3 \left(1-e_\star^2\right)^{3/2}}{G^2 M_\bullet^2 s} = \frac{c \left(1-e_\star^2\right)^{3/2} P_\star^2}{2 \pi s r_\mathrm{g}}
\end{equation}
in which $0 \leq s \leq 1$ is the dimensionless spin parameter $s=c\norm{\vect{S}_\bullet}/G M_\bullet^2$. On the milliparsec scales we are interested in this study the Lense-Thirring precession is weak compared to the PN1.0 precession for moderately eccentric stellar orbits and would become important only on spatial scales close to the gravitational radius $r_\mathrm{g}$ \citep{Merritt2013}.

\begin{figure}
\includegraphics[width=\columnwidth]{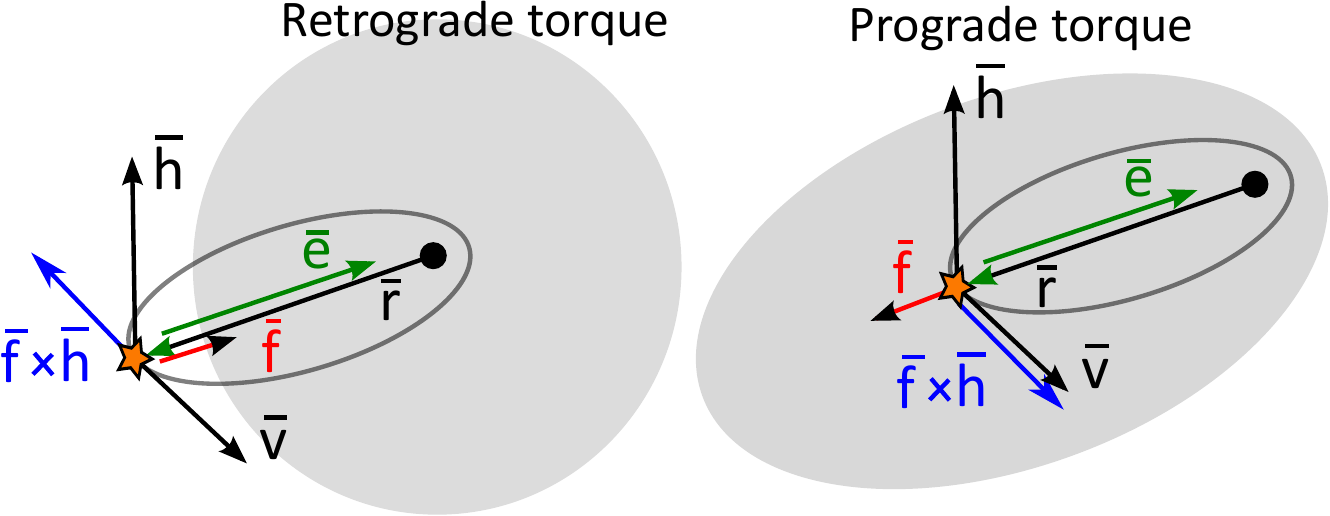}
\caption{A schematic illustration of Newtonian retrograde (left panel) and prograde (right panel) torques on a star (yellow symbol) orbiting the central SMBH (solid black circle). The eccentricity vector $\vect{e}$ pointing towards the pericenter is defined in Eq. \eqref{eq: laplacerungelenz} while the reduced angular momentum vector $\vect{h}$ is defined as $\vect{h}=\vect{v}\times\vect{r}$. At the apocenter the torque $\vect{\tau}$ due to a small non-Keplerian perturbation $\vect{f}$ is $\vect{\tau} = \vect{f} \times \vect{h}$. If the perturbing background is spherically symmetric (left panel), the torque and thus the precession direction is always retrograde. In the asymmetric case (right panel) both retrograde and prograde torques may occur at different points of the orbit, and the net torque integrated over the entire orbit determines the precession direction.}
\label{fig: disk-torque}
\end{figure}

\subsubsection{Mass precession -- the spherical case}\label{section: precession-theory-sphere}
Precession effects are present in the Newtonian case as well as long as the potential deviates from the Keplerian point mass potential. Following e.g. \cite{Madigan2018}, the rates of change of the reduced angular momentum vector $\vect{h}$ and the eccentricity vector $\vect{e}$ of a star orbiting the SMBH are
\begin{equation}\label{eq: torque-h-e}
\begin{split}
\derfrac{\vect{h}}{t} &= \vect{r} \times \vect{f}\\
\derfrac{\vect{e}}{t} &= \vect{e}' = \frac{1}{G M_\bullet} \left( \vect{f} \times \vect{h}  + \vect{v} \times (\vect{r} \times \vect{f})\right),
\end{split}
\end{equation}
in which $\vect{f}$ is a small non-Keplerian contribution to the total acceleration from the extended mass distribution around the SMBH. This is schematically illustrated in Fig. \ref{fig: disk-torque}. In the simple case of a spherically symmetric but extended mass distribution, the non-Keplerian perturbing acceleration will always point towards the SMBH due to Newton's shell theorem, so the change of the eccentricity vector will always be in the retrograde direction, $\vect{e}' \cdot \vect{v}_\mathrm{p}<0$. Here $\vect{v}_\mathrm{p}$ is the periapsis velocity. If the stellar orbit is embedded in a stellar cusp with mass $M_\mathrm{cusp}\ll M_\bullet$, the precession time-scale \citep{Merritt2011,Merritt2013} is
\begin{equation}\label{eq: timescale-massprecession}
t_\mathrm{M} = \frac{M_\mathrm{\bullet} P_\mathrm{\star}  }{M_\mathrm{cusp}(<a_\mathrm{\star})} \frac{1+\sqrt{1-e_\mathrm{\star} ^2}}{\sqrt{1-e_\mathrm{\star} ^2}} = \frac{M_\mathrm{\bullet} P_\mathrm{\star}  }{M_\mathrm{cusp}(<a_\mathrm{\star})} f(e_\star).
\end{equation}
The direction of this so-called mass precession is retrograde. For a spherical power-law cusp with a slope of $\gamma$ the mass precession time-scale scales as
\begin{equation}
t_\mathrm{M} \propto a_\star^{\gamma-3/2} f(e_\star),
\end{equation}
so for $\gamma=3/2$ the precession rate is independent of $a_\star$. It is very important to note that the mass precession timescale $t_\mathrm{M}$ monotonically increases with increasing orbital eccentricity as $df(e)/de>0$ in Eq. \eqref{eq: timescale-massprecession}. Thus, stars on more eccentric orbits  precess at a lower rate than stars on more less eccentric orbits, and almost radial stellar orbits in the limit $e_\star\rightarrow1$ do not precess at all. Estimates for both the relativistic and Newtonian precession timescales for the Milky Way center are collected in Table \ref{table: precession_timescales}.

\subsubsection{Mass precession -- the asymmetric disk case}\label{section: precession-theory-disk}
Another relevant setup for our study is an asymmetric eccentric disk around a SMBH. Such a setup is illustrated in the right panel of Fig. \ref{fig: disk-illustration}. The mass precession process of a stellar orbit in an asymmetric eccentric disk is more complicated than in the simple spherical case. Naively, one might expect that the contribution of the disk mass within the orbit 
of the star would cause retrograde precession just as in the spherical case. However, the eccentric disk models are asymmetric and thus intuition from Newton's shell theorem does not hold. The perturbing non-Keplerian acceleration $\vect{f}$ as in Fig. \ref{fig: disk-torque} can point to different directions along the orbit. The net torque on a single stellar orbit depends on the semi-major axis and eccentricity distribution of the disk (i.e. its surface density profile). The change of the eccentricity vector direction over one orbit and $t_\mathrm{M}$ can be in principle can be calculated using Eq. \eqref{eq: torque-h-e}. 

In the inner parts of the disk the resulting precession is prograde ($\vect{e}' \cdot \vect{v}_\mathrm{p}>0$) as most of the mass lies outside the orbits in an asymmetric configuration, and retrograde in the outermost parts of the disk as most disk mass is located within the orbits. The exact details where the prograde rotation turns into retrograde depend on the disk properties. For example, \cite{Madigan2018} reports a disk model with up to $10\%$ of the disk mass in the outer parts precessing retrograde while the majority of the disk stars precess in the prograde direction.

\subsection{Asymmetric eccentric disks and their (in)stability}

\begin{figure*}
\includegraphics[width=0.8\textwidth]{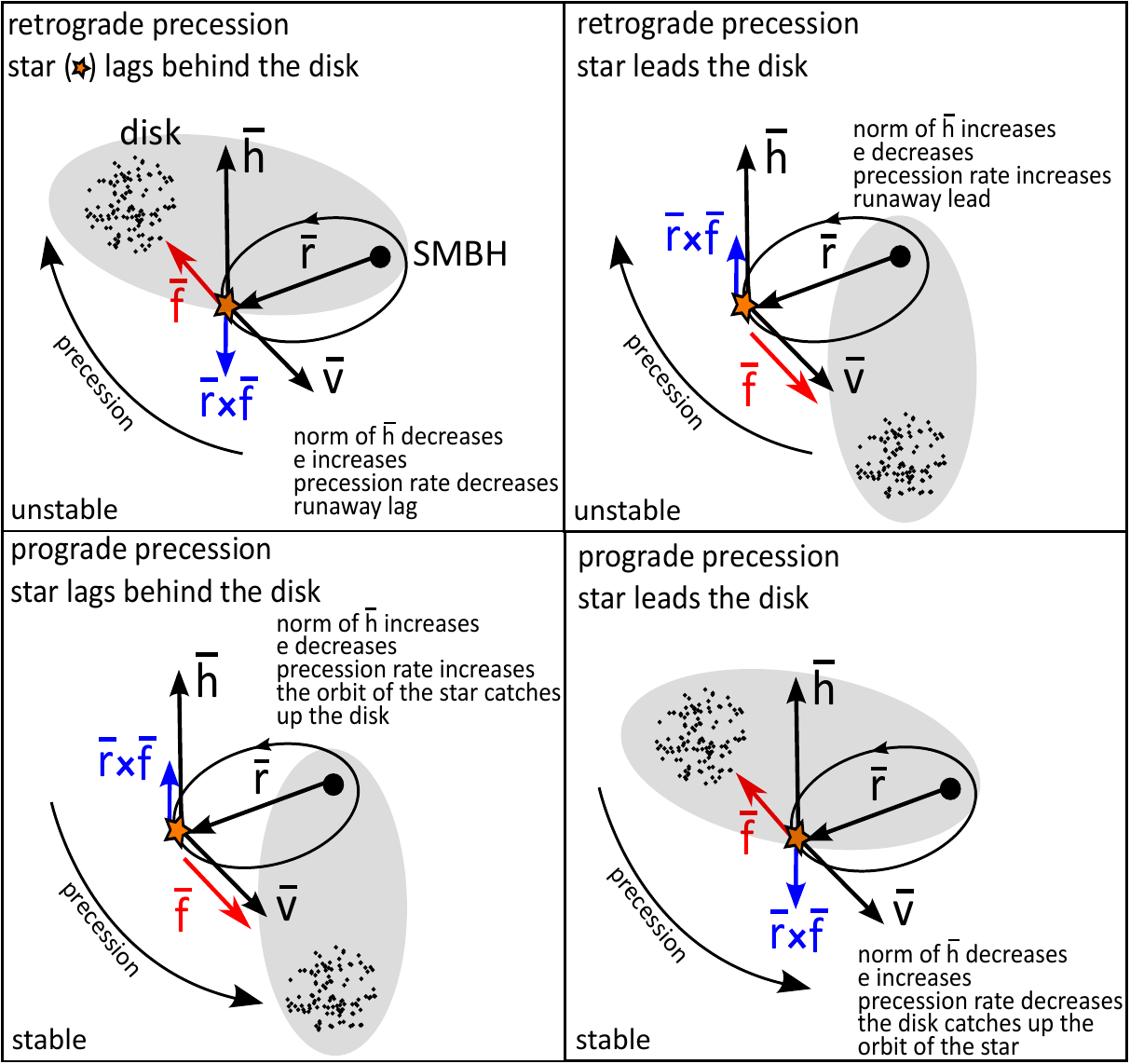}
\caption{A schematic illustration to understand the instability and and stability of asymmetric eccentric disks. Here all the precession processes have a Newtonian origin. The main body of the asymmetric eccentric disk is depicted here as a gray background with a point cloud perturbing a target star (yellow symbol), which is either leading the disk or lagging behind it. It is very important that the terms leading and lagging behind are defined with respect to the precession direction (curved arrow). The precession direction (with respect to the stars apocenter velocity) entirely defines whether the asymmetric eccentric disk is stable or unstable. If the precession is retrograde (top row) as in the case $M_\mathrm{cusp}\gg M_\mathrm{disk}$, the torques $\tau = \vect{r}\times\vect{f}$ from the disk will result in the star receding from its main body. Stars lagging behind the disk will increase their eccentricities while leading stars end up on more circular orbits. In the prograde case (bottom row) when $M_\mathrm{cusp}\ll M_\mathrm{disk}$ the situation is exactly the opposite. Leading stars will circularize and stars lagging behind the disk main body will reach higher eccentricities. In both prograde cases the star ends up being reabsorbed to the main body of the disk, so the disk is stable despite the eccentricity and precession rate oscillations.}
\label{fig: eccdisk-stability-instability}
\end{figure*}

\subsubsection{Asymmetric eccentric disks}
Circular and axisymmetric eccentric disks can be stable over long periods of time, but this is necessarily not the case for asymmetric eccentric disks for which the eccentricity vectors are apsidially aligned. In the simple case of a SMBH, a spherically symmetric stellar background cusp and an asymmetric eccentric disk (with a narrow eccentricity distribution), the stability of the asymmetric disk depends on how it precesses. Disks massive compared to the background stellar population ($M_\mathrm{disk}\gg M_\mathrm{cusp}$) precess prograde and are stable \citep{Tremaine1995,Jacobs2001,Madigan2018} while low-mass disks ($M_\mathrm{disk}\ll M_\mathrm{cusp}$) precess in the retrograde direction and rapidly disrupt \citep{Madigan2009,Subr2016}. 

\subsubsection{Stable massive disks}
We show a schematic illustration explaining the stability of massive ($M_\mathrm{disk}\gg M_\mathrm{cusp}$) asymmetric eccentric disks in the two bottom panels of Fig. \ref{fig: eccdisk-stability-instability}. The prograde precession of the disk orbits is caused by the mass of the disk itself. The prograde precession originates from the fact that in most points of the orbit of the star the perturbation from the disk to the star points away from the SMBH, causing a prograde torque. A disk star with an eccentricity higher than the mean eccentricity of the majority of disk stars will precess at a slower prograde rate than the disk stars, and thus begins to lag behind the bulk of the disk. Now the disk perturbs the lagging star and torques its orbit, increasing the magnitude $\norm{\vect{h}}$ of the angular momentum vector $\vect{h}=\vect{r}\times\vect{v}$ of the star and thus decreasing its eccentricity. Lower eccentricity leads to a faster prograde precession rate, and the star catches up the bulk of the disk and is reabsorbed into its main body.

The opposite process occurs for stars on initial orbits with lower eccentricity. They precess in the prograde direction faster than the bulk of the disk and thus begin to begin to lead the main disk. The torque from the main disk now works to decrease $\norm{\vect{h}}$ and increase $e_\star$, lowering the prograde precession rate. The main body of the disk can thus catch up the leading stars, again leading to the re-absorption of the star to the main disk. Thus, massive eccentric disks are stable against this type of eccentricity oscillations.

\subsubsection{Unstable low-mass disks}
The unstable case of the low-mass ($M_\mathrm{disk}\ll M_\mathrm{cusp}$) asymmetric eccentric disk is presented in the top panels of Fig. \ref{fig: eccdisk-stability-instability}. Now the massive background population causes the orbits the disk stars to precess in the retrograde direction. Compared to the prograde case, the sense of stars leading or lagging behind the disk is reversed as leading and lagging behind are defined with respect to the precession direction. Thus, the retrograde lagging behind case corresponds to the prograde leading case, and the retrograde leading case the prograde lagging behind case.

A star with a slightly higher eccentricity than the most disk stars will precess at a slower rate than the disk in the retrograde direction, and thus will begin to lag behind the main part of the disk. Now, the torque from the bulk of the disk causes $\norm{\vect{h}}$ to decrease, increasing the eccentricity of the star even further. This leads to an increasingly slow precession rate, and the star further recedes from the main part of the disk.

On the other hand, stars with lower eccentricity than in the bulk of the disk will precess at a faster retrograde rate and begin to lead the disk. The torque from the disk acts to increase $\norm{\vect{h}}$ and decrease $e_\mathrm{\star}$. This further increases the retrograde precession rate, again leading into a runway process. Thus, asymmetric eccentric disks in the regime ($M_\mathrm{disk}\ll M_\mathrm{cusp}$) are unstable against eccentricity oscillations. The initially narrow eccentricity distribution is broadened while the disk is disrupted. The stars lagging behind the disk will reach high eccentricities while the stars leading the disk become more circular. The entire process is termed the secular eccentric disk instability following \cite{Madigan2009}.

\subsection{Binary stars, the Hills mechanism and tidal disruption of stars}\label{section: 2-hills}

SMBHs can unbind binary stars resulting in the ejection of the one binary component and the capture of the another component around the SMBH \citep{Hills1988,Yu2003}. This is known as the Hills mechanism. Only binary stars close enough to the SMBH can become unbound, the approximate limit being the tidal radius $r_\mathrm{t}$ defined as
\begin{equation}
r_\mathrm{t} = \left( \frac{M_\bullet}{m_\mathrm{b}}\right)^{1/3} a_\mathrm{b},
\end{equation}
in which $m_\mathrm{b}$ and $a_\mathrm{b}$ are the mass and semi-major axis of the binary. Assuming a parabolic orbit for the binary center-of-mass, the semi-major axis $a_\mathrm{s}$ of the captured star around the SMBH is
\begin{equation}
a_\mathrm{s} \approx \frac{1}{k} \left( \frac{M_\bullet}{m_\mathrm{b}} \right)^{2/3} a_\mathrm{b}
\end{equation}
in which $k\sim0.3$--$1$ depends on the eccentricity and the phase of the binary \citep{Generozov2020}. The orbit of the captured star is very eccentric:
\begin{equation}
e_\mathrm{s} \approx 1 - \chi^\mathrm{k}\left( \frac{m_\mathrm{b}}{M_\bullet} \right)^{1/3}
\end{equation}
with $\chi^\mathrm{k}\sim1$--$2$. For example, a $m_\mathrm{b}=20 \;\mathrm{M_\odot}$ binary with a semi-major axis of $a_\mathrm{b}=1\;\mathrm{AU}$ disrupted by the Milky Way SMBH with $M_\bullet=4\times10^6 \;\mathrm{M_\odot}$ will result in a captured star with orbital elements of $a_\mathrm{s}\sim17$--$55$ mpc and $e_\mathrm{s}\sim0.97$--$0.98$.

The tidal forces of SMBHs affect single stars as well. A single star can be fully or partially disrupted at the separations smaller than the order-of-magnitude tidal (or Roche) radius $r_\mathrm{t}$ defined as
\begin{equation}
r_\mathrm{t} = \left( \frac{M_\bullet}{m_\mathrm{\star}}\right)^{1/3} R_\mathrm{\star}
\end{equation}
in which $R_\star$ is the radius of the star. Part of the disrupted stellar material can remain bound to the SMBH, and falls back into the SMBH in an accretion flow powering a luminous observable transient with a characteristic decay time-scale, a tidal disruption event \citep{Hills1975, Rees1988, Gezari2021}. The modeling of tidal disruption events (TDEs) is in general challenging and numerically expensive, requiring advanced (magneto)hydrodynamical solvers, general relativity and detailed models stellar structure. In addition, the parameter space of the problem is large (including the stellar models and orbits), and general fitting formulas from detailed simulations are rarely available.

\subsection{Relaxation processes}

While the Hills mechanism produces stars on very eccentric orbits around the SMBH, the cumulative eccentricity distribution of the S-stars in the Milky Way center is $F(e)=e^\beta$ with $\beta\sim2.6$ \citep{Gillessen2009}. Thus, an efficient mechanism would be required to relax the orbital eccentricities of the Hills mechanism origin S-cluster stars withing the age of the S-cluster, $\sim10$ Myr \citep{Habibi2017}.

The common non-resonant two-body relaxation changes both the energy and angular momentum of the stars (and hence $a_\star$ and $e_\star$) in a random-walk like manner. The two-body relaxation timescale $t_\mathrm{NR}$ of $N$ stars around a SMBH can be estimated as
\begin{equation}
t_\mathrm{NR} \sim \frac{1}{\beta^2 N}\frac{M_\bullet^2}{\tilde{m}^2} P_\star
\end{equation}
in which $\tilde{m}=\langle m^2\rangle/\langle m\rangle$ \citep{Binney2008}. Here $\beta^2=\ln{(M_\bullet/m_\star)}$ is a factor of the order of unity. The two-body relaxation time-scale $t_\mathrm{NR}$ for the S-stars is considerably longer than the ages of the stars ($\lesssim10$ Myr), so the effect of two-body relaxation on the S-stars is weak.

In near-Keplerian potentials the orbits of the stars are constrained by the symmetries of the potential, and torques on the stellar orbits are correlated and build up in a coherent manner \citep{Rauch1996}. While the energy (semi-major axis) is relatively unaffected by resonant relaxation, both the magnitude (scalar resonant relaxation, SRR) or the direction (vector resonant relaxation, VRR) of the angular momentum vector can change. Thus, scalar resonant relaxation can relax the eccentricities of the S-cluster stars while vector resonant relaxation can change the orientation of their orbital planes. 

The scalar resonant relaxation timescale $t_\mathrm{SRR}$ is approximately
\begin{equation}
t_\mathrm{SRR} \sim \frac{M_\bullet}{\tilde{m}} P_\star
\end{equation}
when mass precession (not the relativistic precession) is the dominant orbit precession mechanism. It has been argued that the scalar resonant relaxation timescale can be as low as $t_\mathrm{SRR}\sim10$ Myr, comparable to the age of the S-cluster, especially when the effect of compact remnants from the old spherical stellar population is included in the calculations (e.g. \citealt{Antonini2013,BarOr2018,Generozov2020}).

The timescale for vector resonant relaxation (VRR) in a spherical system around a SMBH \citep{Eilon2009,Kocsis2015,Fouvry2019} can be estimated as
\begin{equation}
t_\mathrm{VRR} \sim \frac{1}{\beta_\mathrm{v}}\frac{M_\bullet}{\left[M(r)\tilde{m}\right]^{1/2}} P_\star
\end{equation}
in which $M(r)$ is the mass enclosed by the orbit of the star, and $\beta_\mathrm{v}$ is again a constant of the order of unity. The $t_\mathrm{VRR}$ in the Milky Way center is less than a Myr the VRR being the fastest relaxation process in the S-cluster region \citep{Panamarev2022}. Estimates for the relaxation timescales in the Milky Way center are presented in Table \ref{table: precession_timescales}.


\section{Numerical methods}\label{section: finalnumber-3}

\subsection{N-body simulation code}

\subsubsection{Gravitational dynamics}

For the numerical simulations of this study we use the novel N-body simulation code \bifrost{} \citep{Rantala2021,Rantala2023}. \bifrost{} is an accurate GPU-accelerated direct-summation N-body code based on the hierarchical \citep{Rantala2021} fourth-order forward symplectic integrator technique \citep{Chin1997,Chin2005,Chin2007a,Dehnen2017a}. The main integrator of the code is supplemented with specialized secular and regularized few-body integrators for binary systems, close fly-bys, multiple systems and small clusters around supermassive black holes \citep{Rantala2020}.

The code includes optional post-Newtonian (PN) equations of motion for the simulation particles, the option for external background potential, simple prescriptions for stellar evolution, stellar mergers and tidal disruption events (TDEs), and supports arbitrary fraction of binary stars due to its effective MPI parallelization.
If PN equations of motion are enabled, the interactions between the stars and the supermassive black hole are post-Newtonian everywhere in the simulation domain, not just in the subsystem regions as it is commonly done in the literature. We include only the lowest-order PN1.0 term and no spin PN terms as their dynamical effect on the simulation results would be small in the setups we are interested in.

\begin{table}
\begin{tabular}{l l l}
\hline
\bifrost{} user-given parameter & symbol & value\\
\hline
forward integrator time-step factor & $\eta_\mathrm{ff}$, $\eta_\mathrm{fb}$, $\eta_\mathrm{\nabla}$ & $0.2$\\
subsystem neighbour radius, stars & $r_\mathrm{rgb,\star}$ & $0.25$ mpc\\
subsystem neighbour radius, SMBH & $r_\mathrm{rgb,\bullet}$ & $2.5$ mpc\\
constant time-step radius, SMBH & $r_\mathrm{\Delta t,\bullet}$ & $5\;r_\mathrm{rgb,\bullet}$\\
regularization GBS tolerance  & $\eta_\mathrm{GBS}$ & $10^{-8}$\\
regularization GBS end-time tolerance & $\eta_\mathrm{endtime}$ & $10^{-4}$\\
regularization highest PN order &  & PN1.0\\
forward integration highest PN order &  & PN1.0\\
secular binary integration threshold & $N_\mathrm{orb,sec}$  & $0$\\
secular binary highest PN order &  & Newtonian\\
\hline
\end{tabular}
\caption{The \textsc{bifrost} user-given parameters relevant for simulations of this study.}
\label{table: bifrost_options}
\end{table}

The user-given accuracy parameters of \bifrost{} in this study are displayed in Table \ref{table: bifrost_options}. In addition to the code options in \cite{Rantala2023}, we introduce a number of additional code functionalities and parameters. Massive particles (such as SMBHs) can now optionally have larger subsystem radii $r_\mathrm{rgb,\bullet}$ than the subsystem radii of stars $r_\mathrm{rgb,\star}$. We also introduce a time-step limiter for small particles (i.e. stars) in the vicinity of a massive particle such that all particles within $r_\mathrm{ngb,\bullet}<r<r_\mathrm{\Delta t, \bullet}$ acquire the minimum time-step of the particles within the region. Both these small updates increase the integration accuracy of stellar orbits around SMBHs.

\subsubsection{Simple stellar evolution}

The current version of the \bifrost{} code is not coupled to a binary stellar evolution module. In order to include a simple stellar evolution prescription into this study we use the \sevn{} rapid population synthesis code \citep{Spera2015,Spera2017,Spera2019,Mapelli2020,Iorio2023} to calculate the evolution of mass $m_\star(t)$ and radius $R_\star(t)$ of stars in our simulation. We use the metallicity $Z=Z_\odot$ throughout the study. The \sevn{} code itself is based on interpolating pre-tabulated sets of latest \parsec{} \citep{Bressan2012,Chen_Yang2015,Nguyen2022} stellar tracks. 

The main effect of stellar evolution in our study is the rapid mass loss and death of massive stars. First stars die at the age of $\sim2.5$ Myr. At the inferred age of the CW disk O-type stars in the Milky Way center, $t=6$ Myr, stars more massive than $\sim 29 \;\mathrm{M_\odot}$ have already ended their lives. At the end of our simulations, $t=10$ Myr, no stars more massive than $m_\star \sim 18 \;\mathrm{M_\odot}$ are alive anymore. Obviously, including or not including stellar evolution has a large effect on the number of O-type stars. Thus, incorporating even a very simple stellar evolution prescription into our simulations allows a better comparison with observations. Another motivation for including simple stellar evolution in the runs is the study of single stars in the S-cluster. Massive stars rapidly lose mass in a few Myr before the end of their lives, and entering the giant phase makes the stars extremely susceptible for tidal disruption by the SMBH due to the significantly increased stellar radius. The effect of binary stellar evolution on the results of this work will be addressed in a future study.

\subsubsection{Tidal disruption events}

We briefly describe here the updated the TDE prescription of \bifrost{} \citep{Rizzuto2022,Rantala2023}. First, we do not use the tidal capture and drag force routines used in the intermediate mass black hole growth study of \citep{Rizzuto2022} in this study. The stars within the tidal disruption radius from the SMBH are accreted by the SMBH and removed from the simulation. For this study we update the definition of the tidal disruption radius $\mathcal{R}_\mathrm{t}$ using the fitting formulas from detailed hydrodynamical tidal disruption simulations of \cite{Ryu2020}. The tidal disruption radius is defined as
\begin{equation}
    \mathcal{R}_\mathrm{t} = \Psi_\star(m_\star) \Psi_\bullet(M_\bullet) r_\mathrm{t} =   \Psi_\star(m_\star) \Psi_\bullet(M_\bullet) \left( \frac{M_\bullet}{m_\star}\right)^{1/3} R_\mathrm{\star},
\end{equation}
in which $R_\star$ is the radius of the star, $r_\mathrm{t}$ is the order-of-magnitude tidal disruption radius \citep{Kochanek1992} and the correction factors are given in \cite{Ryu2020}. The SMBH term is of the order of unity ($\Psi_\bullet\sim1.32$) for the Milky Way SMBH mass of $M_\bullet=4\times10^6\;\mathrm{M_\odot}$, and the star term $\Psi_\star(m_\star)\sim1/2$ for stars considerably more massive than $1\;\mathrm{M_\odot}$. Thus, massive stars such as B-type and O-type stars we are most interested in this study disrupt closer to the SMBH compared to the standard order-of-magnitude tidal disruption radius estimate. In all TDEs the star is fully accreted by the SMBH, i.e. we do not include a partial tidal disruption prescription in our code for this study. Possible caveats of the adopted simple tidal disruption model are elaborated in Section \ref{section: discussion-tde}.

\section{Simulation setup}\label{section: finalnumber-4}

\subsection{Motivation}\label{section: ic-motivation}

The dynamics and instability of asymmetric eccentric stellar disks of masses around $M_\mathrm{disk} \sim 10^4 \;\mathrm{M_\odot}$ have been widely studied in the literature (e.g. \citealt{Madigan2009,Subr2016}), as this disk mass is close the inferred mass of the CW and CCW disk stellar disks in the Milky Way center \citep{Bartko2010}. 

However, there are caveats in using simulations of $M_\mathrm{disk} \sim 10^4 \;\mathrm{M_\odot}$ stellar disks to study the structure and long-term evolution of the Milky Way center. The inferred age of the young, bright and massive stars ($6$ Myr) is enough that stars more massive than $\sim 29 \;\mathrm{M_\odot}$ have already reached the end of their life-times. Thus, a top-heavy $M_\mathrm{disk} \sim 10^4 \;\mathrm{M_\odot}$ stellar disk today was more massive at its formation. For extremely top-heavy power-law IMF slopes the mass loss of the stellar population in $6$ Myr is substantial. With $\alpha=0.25$ and $m_\mathrm{max}=120 \;\mathrm{M_\odot}$, a stellar population can lose $90\%$ of its mass in only $6$ Myr. For more moderately top heavy IMFs of $\alpha=1.3$ and $\alpha=1.7$ the mass loss in the same time period is $\sim70\%$ and $\sim50\%$. This encourages studying the dynamics of initially more massive eccentric disk systems in the context of the Milky Way center. However, the more massive initial disks do not necessarily fulfill the $M_\mathrm{disk} \ll M_\mathrm{cusp}$ criterion anymore for the secular eccentric disk instability \citep{Madigan2009}. As the eccentric disks of this mass scale do not fulfill the high-mass stability criterion $M_\mathrm{disk} \gg M_\mathrm{cusp}$ either, we expect that the disks in this intermediate mass range are unstable, but their disruption mechanism might differ from the model of \cite{Madigan2009}. Understanding the instability of these intermediate-mass asymmetric eccentric disks is the key motivation for this study.

\subsection{Asymmetric eccentric disk models}\label{section: disk-ic}
\subsubsection{Disk properties}

We construct asymmetric eccentric disk models of stars orbiting a SMBH embedded in a static spherical background cusp potential. In this study the mass of the SMBH $M_\bullet$ is set to the mass of the SgrA* SMBH, $M_\bullet = 4\times10^6\; \;\mathrm{M_\odot}$ (e.g. \citealt{Do2019,GRAVITYCollaboration2019,GRAVITYCollaboration2020,EHT2022_IV}) for Milky Way center comparison.

We sample the semi-major axes $a_\mathrm{\star}$ for the disk stars and binary star center-of-masses from a distribution of $f(a)$ in such a way that the surface density $\Sigma(R)$ of the disk is
\begin{equation}
    \Sigma(R) = \Sigma_\mathrm{0} \left(\frac{R}{R_\mathrm{0}}\right)^{-2}.
\end{equation}
The power-law slope of $-2$ is chosen to enable more straightforward comparison to previous simulation studies and Milky Way center observations \citep{Paumard2006, Madigan2009} The effect of choosing a different disk surface density profile is discussed later in Section \ref{section: discussion-surfacedensity}. The scale radius of the disk $R_\mathrm{0}$ and the surface density at the scale radius $\Sigma_\mathrm{0} = \Sigma(R_\mathrm{0})$ are fixed by setting the inner $R_\mathrm{in}$ and the outer $R_\mathrm{out}$ edge of the disk as well as its total mass $M_\mathrm{disk}$. We simulate in total three different disk masses, $1.0\times10^4 \;\mathrm{M_\odot}$ (sample L for low-mass and literature setup), $3.75\times10^4 \;\mathrm{M_\odot}$ (sample M for mid-mass) and $7.5\times10^4 \;\mathrm{M_\odot}$ (sample H for high-mass). We fix the inner and outer edge of the disk models to $R_\mathrm{in} = 0.05$ pc and $R_\mathrm{out} = 0.5$ pc. In each disk model all the stellar orbits share a common eccentricity $e_\mathrm{init}$. We model in total nine different disk eccentricities from $e_\mathrm{init}=0.1$ to $e_\mathrm{init}=0.9$ with constant increments of $0.1$. The initial inclinations of the orbits are small but non-zero, $\abs{i}<0.01$. The other orbital elements, the longitude of the ascending node $\Omega$, the argument of periapsis $\omega$ and the mean anomaly $\mathcal{M}$ have random values $\Omega,\omega,\mathcal{M}\in [0,2\pi]$.

The initial asymmetry of the disk models is enforced by aligning the eccentricity (Laplace-Runge-Lenz) vectors of Eq. \eqref{eq: laplacerungelenz} of the disk stars, i.e. the pericenters of the stellar orbits lie all in the same direction. This is illustrated in the right panel of Fig. \ref{fig: disk-illustration}.

\subsubsection{Stellar population properties}
The masses of the stars in the disk model are follow a standard power-law initial mass function
\begin{equation}
    \xi(m) = \derfrac{N}{m} \propto \left( \frac{m}{m_\mathrm{min}} \right)^\mathrm{-\alpha}
\end{equation}
from a minimum stellar mass of $m_\mathrm{min} = 0.08 \;\mathrm{M_\odot}$ to the maximum mass of $m_\mathrm{max} = 120 \;\mathrm{M_\odot}$. We study three different IMF slopes, the extremely top-heavy $\alpha=0.25$ and two moderately top heavy options $\alpha=1.3$ and $\alpha=1.7$. These choices are motivated by the observations of the Milky Way center which indicate a top-heavy IMF in the range from $\alpha=0.45\pm0.3$ \citep{Bartko2010} (central disks) to $\alpha=1.7\pm0.2$ (overall innermost $0.5$ pc). The high O/WR-B ratios observed in the CW and CCW disks (e.g. \citealt{vonFellenberg2022}) point towards even more extreme IMF in the innermost parts of the Milky Way center disk structures.

\subsubsection{Binary star population properties}

Star formation in the vicinity of SMBHs is poorly understood, and binary star formation even less so, and there is a very limited amount of observations available to constrain the properties of the binary population in the disks around SMBHs. In order to proceed, we assume that some very general binary population properties from star clusters and field stars remain valid in extreme environment of the galactic nuclei.

First, we assume that massive stars prefer having (massive) companion stars with mass ratios of $q=m_\mathrm{2}/m_\mathrm{1}>0.1$. The binary fraction as a function is chosen to reflect the observations in less extreme star formation environments \citep{Moe2017,Winters2019}. The binary fraction of B-type primary stars is $>90\%$ while practically every O-star in the initial conditions is in a binary. Low-mass stars are predominantly single stars.

Next, the eccentricity distribution $f(e_\mathrm{b})$ of binaries is close to the thermal eccentricity distribution, i.e.
\begin{equation}
    f(e_\mathrm{b}) = 2e_\mathrm{b},
\end{equation}
and that close binaries cannot have high eccentricities, $r_\mathrm{p} = a_\mathrm{b}*(1-e_\mathrm{b})>R_\mathrm{\star,1}+R_\mathrm{\star,2}$. Finally, the distribution of the initial semi-major axes is the log-uniform distribution i.e. the reciprocal distribution
\begin{equation}
    f(a_\mathrm{b}) = \frac{1}{a_\mathrm{b}} \frac{1}{ \log{(a_\mathrm{max})} - \log{(a_\mathrm{min})} }.
\end{equation}
The minimum $a_\mathrm{min}$ and maximum $a_\mathrm{max}$ semi-major axes are selected based on survivability of binary systems in the disk. Wide binaries are evaporated by the repeated perturbations from encountered stars, and very hard, especially eccentric binaries are destroyed by stellar mergers (e.g. \citealt{Generozov2020}). For this study we use two options, the standard population with $a_\mathrm{min} = 10^{-6}$ pc and $a_\mathrm{max} = 10^{-4}$ pc and and the hard population for which each binary shares a common semi-major axis of $a_\star = 3\times10^{-6}$ pc.

\subsection{Background potential}

We model the effect of the old stellar population of the nuclear star cluster of the Milky Way (e.g. \citealt{Schodel2014}) on the orbits of the stars by including an external potential $\phi_\mathrm{ext}$ in the simulation. The external potential corresponds to a spherical power-low background density distribution of
\begin{equation}
    \rho(r) = \rho_\mathrm{0} \left( \frac{r}{r_\mathrm{0}} \right)^{-\gamma}.
\end{equation}
We use the slope $\gamma=1.4$ throughout this study motivated by the observations of \cite{Genzel2003}.
The background scale radius $r_\mathrm{0}$ and density at the scale radius $\rho_\mathrm{0} = \rho(r_\mathrm{0})$ are chosen so that the enclosed mass $M(r)$ within $r_\mathrm{0} = 1$ pc is $M(r_\mathrm{0}) = 5\times10^5 \;\mathrm{M_\odot}$ \citep{Genzel2003,Schodel2007}.

The main effect of the spherical external potential is to cause the orbits of the disk stars to precess and thus contribute to the instability of the initially asymmetric eccentric disks. The fact that the spherical background cusp population is represented by a smooth external potential instead of live particles significantly lowers the computational costs, but also weakens the scalar and vector resonant relaxation effects in the simulations. The consequences of this are discussed later in this article.

\begin{table}
	\centering
	\caption{The main properties of the simulation samples L (low-mass, literature), M (mid-mass) and H (high-mass) containing nine simulations each.}
	\label{tab: table_simulations}
	\begin{tabular}{c c c c c c}
		\hline
		label & $M_\mathrm{disk}$ & $\alpha$ & $e_\mathrm{init}$ & stellar evolution & PN\\
		\hline
		L & $1.00\times10^4 M_\mathrm{\odot}$ & $1.30$ & $0.1$--$0.9$ & no & no\\
        M & $3.75\times10^4 M_\mathrm{\odot}$ & $0.25$ & $0.1$--$0.9$ & yes & no\\
        H & $7.50\times10^4 M_\mathrm{\odot}$ & $0.25$ & $0.1$--$0.9$ & yes & no\\
		\hline
	\end{tabular}
\end{table}

\begin{table*}
	\centering
	\caption{The simulations exploring the effect of the IMF, the binary population properties and PN and TDE prescriptions compared to the setup M from Table \ref{tab: table_simulations}.}
	\label{tab: table_simulations-2}
	\begin{tabular}{c c c c c c c}
		\hline
		Simulation set & $N_\mathrm{sim}$ & $M_\mathrm{disk}$ & $\alpha$ & $e_\mathrm{disk}$ & binary population & variants\\
		\hline
        M & 9 & $3.75\times10^4 M_\mathrm{\odot}$ & $0.25$ & $0.1$--$0.9$ & standard & --\\
        \hline
        M7b & 1 & $3.75\times10^4 M_\mathrm{\odot}$ & $0.25$ & $0.7$ & hard & TDE+PN\\
        $\mathrm{M}_\mathrm{1.3}$ & 3 & $3.75\times10^4 M_\mathrm{\odot}$ & $1.30$ & $0.7$ & hard & TDE+PN, noTDE+noPN, TDE+noPN\\
        $\mathrm{M}_\mathrm{1.7}$ & 2 & $3.75\times10^4 M_\mathrm{\odot}$ & $1.70$ & $0.7$ & hard & TDE+PN, noTDE+noPN\\
        H7b & 1 & $7.50\times10^4 M_\mathrm{\odot}$ & $0.25$ & $0.7$ & hard & TDE+PN\\
		\hline
	\end{tabular}
\end{table*}

\subsection{List of simulations}
The number of simulations performed for this study is in total 36. The simulations are divided into two sets. The first sample consists of the three sub-samples L, M and H (nine runs each) exploring the effect of initial disk mass $M_\mathrm{disk}$ and eccentricity $e_\mathrm{init}$ on the disk disruption mechanism and milliparsec star cluster formation around the SMBH. The main properties of these simulation runs are collected in Table \ref{tab: table_simulations}. 

The smaller simulation sample consists of seven runs further investigating the effect of the IMF, binary population properties, TDE prescription and PN equations of motion on the milliparsec stellar population around the SMBH, its formation and physical properties. The main parameters of the runs in this simulation sample are presented in Table \ref{tab: table_simulations-2}.

\section{Evolution of the eccentric disks}\label{section: finalnumber-5}

\subsection{The rapid instability of asymmetric eccentric disks}\label{section: disk-instability1}

\begin{figure*}
\includegraphics[width=0.9\textwidth]{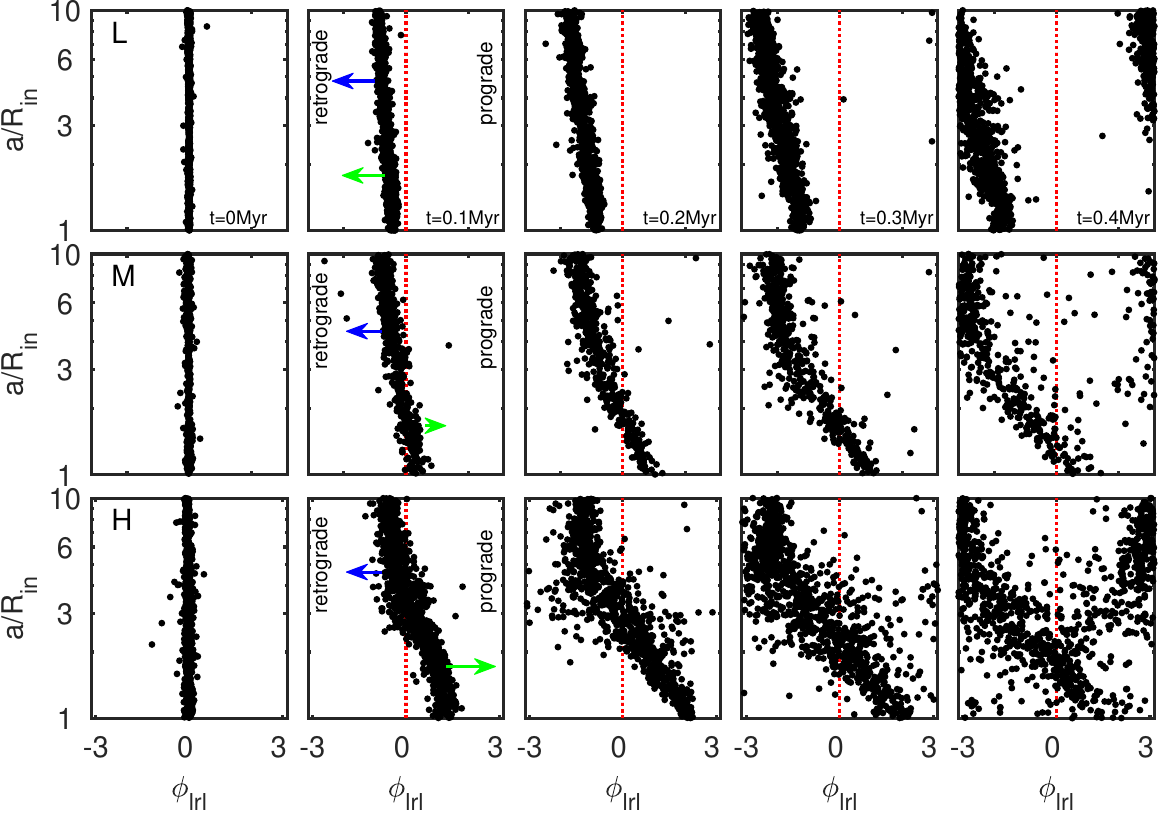}
\caption{The beginning of the disruption of the disk models L7 (top row), M7 (middle row) and H7 (bottom row) at five times from $t=0.0$ Myr to $t=0.4$ Myr. Each panels shows the eccentricity vector direction $\phi_\mathrm{lrl}$ of the disk particles at different semi-major axes from $a_\star$ from the SMBH. The low-mass disk model L7 precesses prograde at a rate almost independent of $a_\star$ while the standard secular eccentric disk instability of \citep{Madigan2009} proceeds as the distribution of $\phi_\mathrm{lrl}$ around its main value broadens. For the model M7 and especially the model H7 the beginning of the asymmetric disk disruption process is more rapid and qualitatively different than in the low-mass model. The reason for this is that in the models $M7$ and $H7$ the distribution of the disk to the orbit precession becomes non-negligible, breaking one of the assumptions of the eccentric disk instability model of \citep{Madigan2009}. While the outer parts of the disk precess retrograde, the inner parts of the disk either precess at a slow rate (M7) or precess prograde (H7). The initial alignment of the eccentricity vectors in setups M7 and H7 is rapidly lost as the different parts of the disk precess to opposite directions at different rates depending on $a_\star$. For the three models the mean values for $\phi_\mathrm{lrl}$ after $t=0.4$ are $\langle \phi_\mathrm{lrl}\rangle = -2.7 \pm 0.6$ (L7), $\langle \phi_\mathrm{lrl}\rangle = -2.6 \pm 1.4$ (M7) and $\langle \phi_\mathrm{lrl}\rangle = -3.1 \pm 1.6$ (H7).
}
\label{fig: phi-lrl-sim}
\end{figure*}

As expected from the fact that our disk models do not lie on the asymmetric eccentric disk stability regime $M_\mathrm{disk} \gg M_\mathrm{cusp}$, they are unstable and rapidly evolve into axisymmetric eccentric disks. That is, the directions of the disk star eccentricity (Laplace-Runge-Lenz) vectors lose their initial alignment and are rapidly randomized. The process lasts less than $1$ Myr for all the models in simulation setups L, M and H.

In order to quantify the disruption process, we study the distribution of eccentricities of the stars in the disk, and the distribution of directions of the eccentricity (Laplace-Runge-Lenz) vectors. We define the pericenter directions of the stellar orbits as
\begin{equation}
    \phi_\mathrm{lrl} = \arctan{\left( \frac{e_\mathrm{y}}{e_\mathrm{x}} \right)} - \phi_\mathrm{0}
\end{equation}
in which $e_\mathrm{x}$ and $e_\mathrm{y}$ are the planar components of the eccentricity vector $\vect{e}$. Here $\phi_\mathrm{0}$ is the reference direction of the initial eccentricity vectors. We show how the eccentricity vector directions $\phi_\mathrm{lrl}$ of stars with different semi-major axes $a_\mathrm{\star}$ evolve in Fig. \ref{fig: phi-lrl-sim}. The figure presents the initial evolution of the models L7, M7 and H7 during the first $0.4$ Myr of the simulations with intervals of $0.1$ Myr.

The lowest-mass model L7 evolves as similar setups in the literature. The asymmetric eccentric precesses in the retrograde direction with the precession rate being almost independent of the semi-major axes of the stars, as expected from Eq. \eqref{eq: timescale-massprecession} and the used background stellar profile power-law slope of $\gamma=1.4$. Stars with large $a_\star$ precess somewhat faster than stars with smaller $a_\star$. The secular eccentric disk instability as in \cite{Madigan2009} proceeds in the simulation as the distribution of the eccentricity vector directions $\phi_\mathrm{lrl}$ widens around its mean value as time goes on. After $0.4$ Myr of evolution the mean pericenter direction of the model L7 is $\langle \phi_\mathrm{lrl}\rangle = -2.7 \pm 0.6$ radians.

The models M7 and H7 behave qualitatively differently. While the outer parts of the disks above $r\sim3 R_\mathrm{in}$ precess in the retrograde direction just as the model L7, the inner parts of the eccentric disks either show little precession (M7) or actually precess in the prograde direction (H7). This is in contrast to the behavior of the setup L7 in which the stars precess retrograde at an almost the same rate, and the secular eccentric disk instability gradually widens the eccentricity vector direction distribution. In the model M7 $\langle \phi_\mathrm{lrl}\rangle = -2.6 \pm 1.4$ at $t=0.4$ Myr. The disk disruption especially in the model H7 is very different from the model L7, and is clearly not caused by the standard secular eccentric disk instability. While in L7 the pericenter direction distribution only gradually broadens, in H7 the initially narrow distribution rapidly flattens as the different parts of the disk are precessing at opposite directions. After $0.4$ Myr in the model H7 the mean eccentricity vector direction is $\langle \phi_\mathrm{lrl}\rangle = -3.1 \pm 1.6$.

As explained in Sections \ref{section: precession-theory-sphere} and \ref{section: precession-theory-disk}, the spherical background always causes retrograde precession, and the disk self-contribution to the precession can be prograde in the inner parts of the disk. Thus, we argue that the assumption of negligible disk self-precession of \cite{Madigan2009} i.e. ($M_\mathrm{disk} \ll M_\mathrm{cusp}$) is broken in the setups M and H. The disruption mechanism of the higher-mass disks differ from low-mass disks as the disks disrupt when its inner and outer parts precess at different directions. The prograde precession direction is not enough to stabilize the intermediate-mass disk models as opposed to the high-mass models \citep{Madigan2018}. We attribute this to the facts that a sizeable fraction of disk stars still precess retrograde, and that the prograde precession rate is different at different $a_\star$ from the SMBH.

\begin{figure*}
\includegraphics[width=0.85\textwidth]{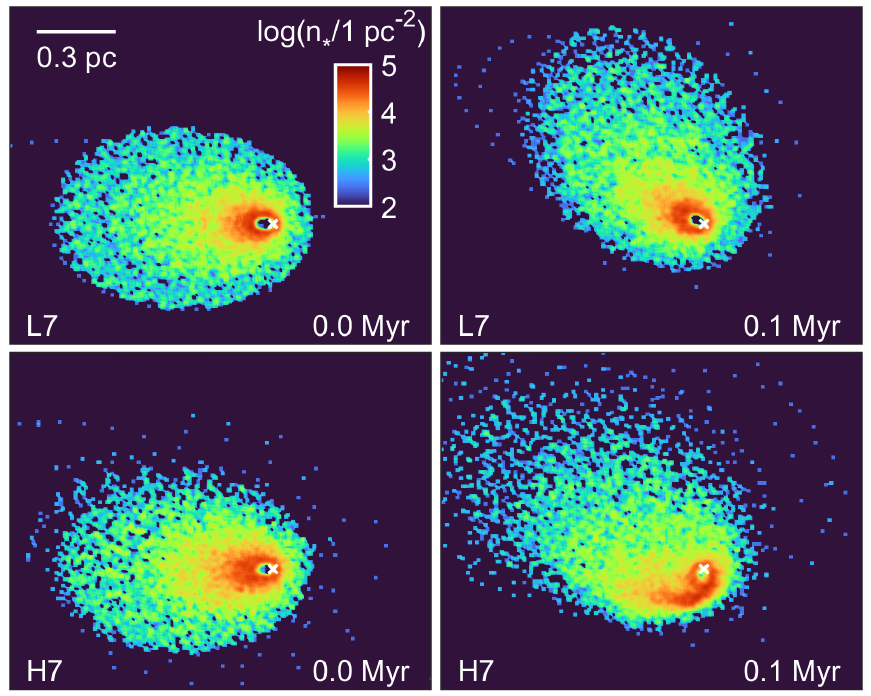}
\caption{The number surface density of stars in the models L7 (top row) and H7 (bottom row) at $t=0$ Myr and at $t=0.1$ Myr. Each panel is stacked from five \bifrost{} snapshots. The SMBH position is marked with a cross symbol in each panel. A short-lived right-handed spiral structure (below the SMBH in the bottom right panel) forms due to the outer parts of the disk precessing in the retrograde direction, the mid-disk not precessing and the inner parts of the disk precessing into prograde direction. The low-mass disk does not show such a feature.}
\label{fig: spiral}
\end{figure*}

Precessing elliptical orbits in thin, axisymmetric systems can give rise to spiral patterns, (e.g. \citealt{Kalnajs1973, Binney2008, Harsoula2021}) the most famous example being grand design spiral arms of disk galaxies. Spiral patterns indeed develop in the runs of our more massive setups $M$ and $H$, but they are short-lived as the disk disruption proceeds. One such spiral pattern from the run H7 is shown in Fig. \ref{fig: spiral} compared to the corresponding spiral-free disk in the low-mass model L7 at the same time. The spiral patterns are always one-sided and right-handed as the outer parts of the disks precess retrograde and the inner parts in the prograde direction.

\subsection{Eccentric disk instability in the intermediate-mass regime}\label{section: disk-instability2}

\begin{figure*}
\includegraphics[width=0.95\textwidth]{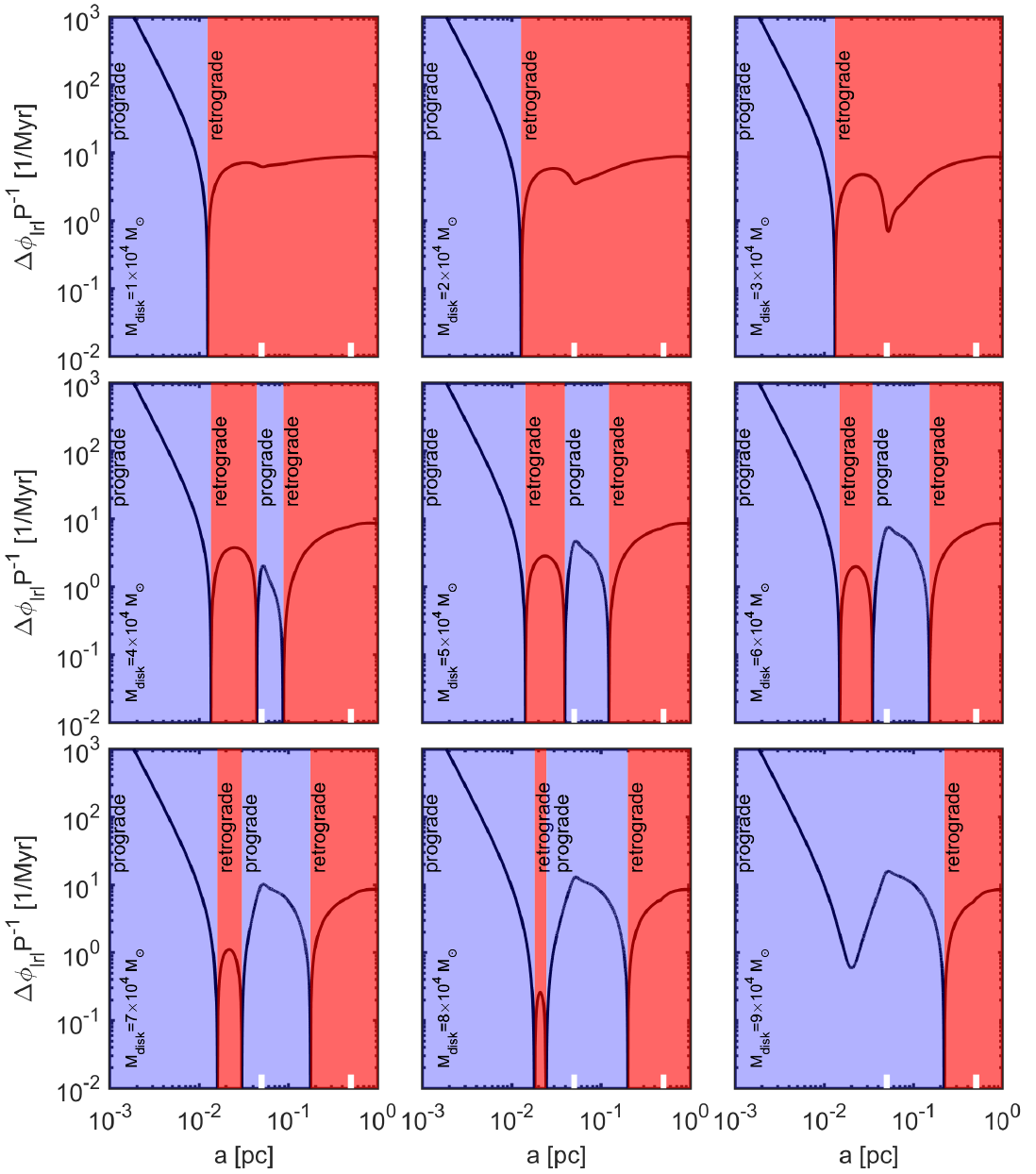}
\caption{The numerically estimated change of the eccentricity vector direction per orbital period $\Delta \phi_\mathrm{lrl}$ of a star as a function of its semi-major axis $a_\star$ around a SMBH in nine asymmetric eccentric disk models. The eccentricity of the disk is $e=0.7$ and the masses of the disks range from $10^4 \;\mathrm{M_\odot}$ to $9 \times 10^4 \;\mathrm{M_\odot}$. The mass of the SMBH and the properties of the stellar background cusp are the same in each panel. The two white tics at the bottom of each panel mark the inner and outer edges of the disks. In the panels of the top row ($M_\mathrm{disk}\leq3\times10^4\;\mathrm{M_\odot}$), near the SMBH ($a_\star \leq 10^{-2}$ pc) the precession is prograde (blue color) due to relativistic periapsis advance while further away from the SMBH the precession is retrograde (in red) due to mass precession from the spherical background. With higher disk masses (middle and bottom rows) the picture qualitatively changes as the disk mass itself becomes an important source of precession. First, a new region of weak prograde precession appears between $0.1$ pc $\lesssim a_\star \lesssim 0.2$ pc when $M_\mathrm{disk}=4\times10^4 \;\mathrm{M_\odot}$. This prograde region broadens and the magnitude of $\Delta \phi_\mathrm{lrl}$ further increases in higher-mass disks until the inner retrograde region completely vanishes when $M_\mathrm{disk}=9\times10^4 \;\mathrm{M_\odot}$. This simple model for initial disk precession directions explains well the features of the N-body simulations of Fig. \ref{fig: phi-lrl-sim}.}
\label{fig: disk-precession-070}
\end{figure*}

\begin{figure*}
\includegraphics[width=0.9\textwidth]{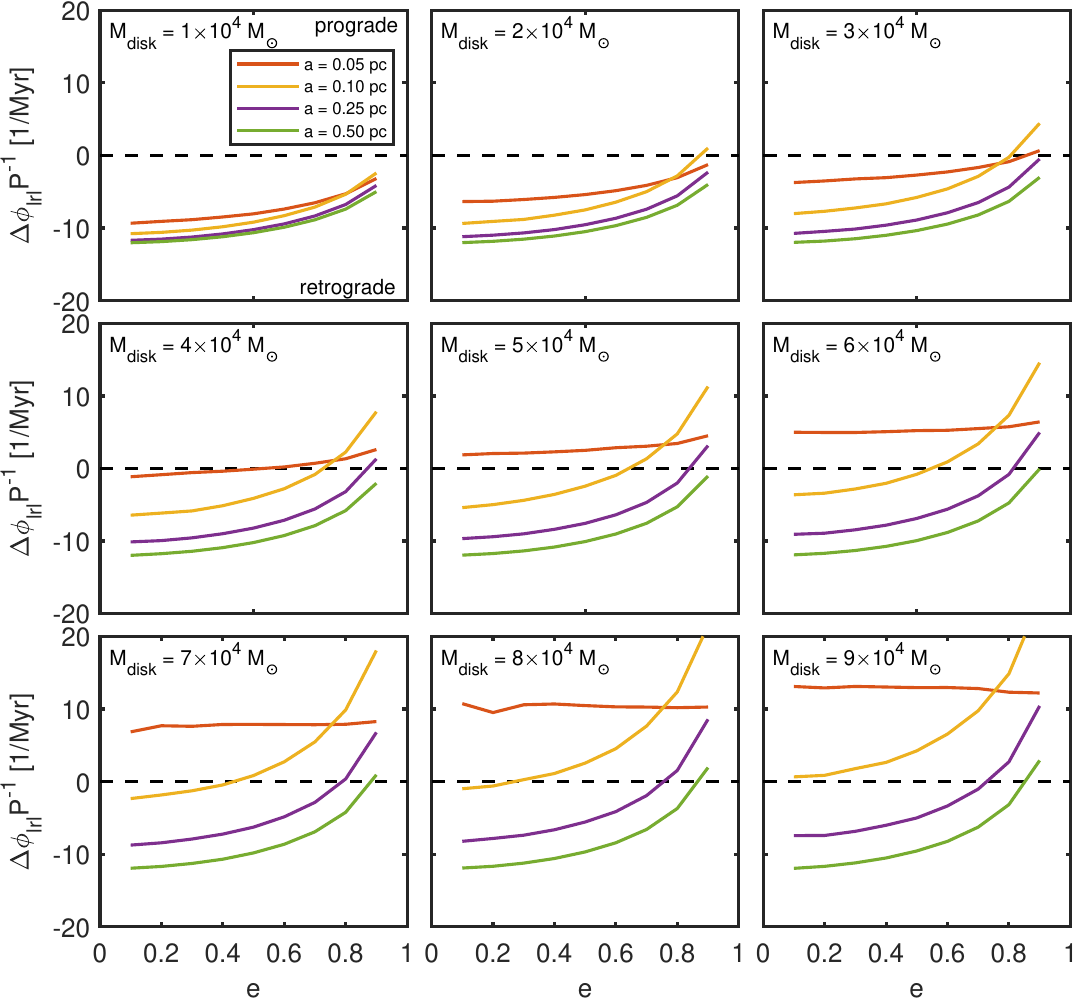}
\caption{The simple numerical precession showing the change of the direction of the eccentricity vector per orbit $\Delta\phi_\mathrm{lrl}/P$ for disk masses $10^4 \;\mathrm{M_\odot} \leq M_\mathrm{disk} \leq 9\times10^4 \;\mathrm{M_\odot}$ and initial eccentricities $0.1\leq e\leq 0.9$. The precession rate is measured at four different semi-major axes around the SMBH at $a_\star=0.05$ pc, $0.1$ pc, $0.25$ pc and $0.5$ pc. For $M_\mathrm{disk}=10^4 \;\mathrm{M_\odot}$ stars at all radii precess retrograde and the precession rate decreases by a factor of $\sim 3$ from low to high eccentricities. As Eq. \eqref{eq: timescale-massprecession} yields only a decrease by $\sim 40\%$ from the spherical background, the rest is attributed to the disk mass contribution itself. The inner edge of the disk ceases to precess retrograde around $M_\mathrm{disk} \sim 4\times10^4 \;\mathrm{M_\odot}$ at all disk eccentricities, after which it always precesses prograde with an increasing magnitude. At high disk masses, the precession rate at the inner edge of the disk is approximately constant. This is further elaborated in the main text. When orbiting outside the inner edge, the precession rate is always more prograde and less retrograde at high disk eccentricities at all disk masses. The outer parts of the disk almost always rotate in the retrograde direction, except with high eccentricities ($e_\mathrm{init}=0.9$) and high disk masses ($M_\mathrm{disk}\gtrsim 7\times10^4 \;\mathrm{M_\odot}$).
}
\label{fig: disk-precession-all-ecc}
\end{figure*}

In the previous section we demonstrated that the nature of the eccentric disk instability is different in the low-mass disk model L compared to the mid- and high-mass models M and H. This is because the self-contribution of the disk to the orbit precession becomes important in models M and H, breaking one of the assumptions of the secular eccentric disk instability of \cite{Madigan2009}.

We develop a simple semi-numerical torque model to explain the initial orbit precession directions and rates in our simulations. The model is mostly analytical containing a single numerical step. The instantaneous time derivatives of the reduced angular momentum vector $\vect{h}$ and the eccentricity vector are shown in Eq. \eqref{eq: torque-h-e}. Assuming that the non-Keplerian perturbations $\vect{f}$ are small, the total change of $\vect{e}$ per orbital period is obtained by integration the expressions over the unperturbed orbit. After a single orbital period the eccentricity vector is $\vect{e}(t_\mathrm{0}+P_\star) = \vect{e}(t_\mathrm{0}) + \Delta\vect{e}$ with 
\begin{equation}
\begin{split}
\Delta\vect{e} &= \int_\mathrm{0}^\mathrm{P_\star} \derfrac{\vect{e}}{t} \d{t}\\
&= \frac{1}{G M_\bullet} \int_\mathrm{0}^\mathrm{P_\star} \Bigl[ \vect{f}(t) \times \vect{h}  + \vect{v}(t) \times \Bigl(\vect{r}(t) \times \vect{f}(t)\Bigr)\Bigr] \d{t}.
\end{split}
\end{equation}
If the perturbation is planar then the reduced angular momentum vector $\vect{h}$ remains constant. In order to proceed, one should obtain the expression for the perturbation $\vect{f} = \vect{f}(\vect{r}(t)) = \vect{f}_\mathrm{cusp}(\vect{r}(t)) + \vect{f}_\mathrm{disk}(\vect{r}(t))$ at each point of the orbit, and integrate over a full orbital period to obtain $\Delta\vect{e}$. For the spherically symmetric background the non-Keplerian perturbation $\vect{f}_\mathrm{cusp}$ is simply
\begin{equation}
    \vect{f}_\mathrm{cusp} = \frac{G M_\mathrm{cusp}(r)\vect{r}}{\abs{\vect{r}}^3}
\end{equation}
and can be evaluated from the cumulative mass profile $M_\mathrm{cusp}(r)$ of the cusp in a straightforward manner. 

As opposed to the spherical case, for perturbing potentials without spherical symmetry calculating the perturbing acceleration $\vect{f}_\mathrm{disk}$ is challenging and we instead proceed numerically. This is the numerical step of the simple torque model. In the asymmetric eccentric disk case the change of the eccentricity vector of a star depends on the disk model and the semi-major axis of the star $a_\star$. We construct disk representations following the recipe described in Section \ref{section: disk-ic} consisting of $N_\mathrm{m}$ mass elements with the mass of each element being $m = M_\mathrm{disk}/N_\mathrm{m}$. Next, we discretize the unperturbed stellar orbit $\vect{r}(t), \vect{v}(t)$ into $N_\mathrm{p}$ discreet points $\vect{r}_\mathrm{k}$, $\vect{v}_\mathrm{k}$, equidistant in mean anomaly $\mathcal{M}$. The perturbing force $\vect{f}_\mathrm{k}=\vect{f}(\vect{r}_\mathrm{k})$ at each point $\vect{r}_\mathrm{k}$ is obtained by direct summing the Newtonian gravitational acceleration with a small Plummer softening of $\epsilon = 10^{-3}$ pc from each mass element $m$. Thus, the change of the eccentricity vector $\Delta\vect{e}$ over a single orbit is \begin{equation}
\Delta\vect{e} \approx \frac{1}{N_\mathrm{p}}\frac{P}{G M_\bullet}\sum_\mathrm{k=1}^\mathrm{N_\mathrm{p}} \Bigl[ \vect{f}_\mathrm{k} \times \vect{h}_\mathrm{k}  + \vect{v}_\mathrm{k} \times (\vect{r}_\mathrm{k} \times \vect{f}_\mathrm{k}) \Bigr].
\end{equation}
We also include the effect of the post-Newtonian PN1.0 perturbation from Eq. \eqref{eq: timescale-pnprecession} into our analysis. The orbits of stars within approximately $a_\star \lesssim 10^{-2}$ pc of the SMBH will always precess in the prograde direction, depending on the eccentricity of the star. In practise, we use $N_\mathrm{m} = 10^5$ mass elements per disk and $N_\mathrm{p} = 10^4$ sample points per stellar orbit. The final results are averaged by repeating the procedure $10$ times with different random realization of the disk in each sample. 

We present the precession direction analysis described above for a system consisting of a SMBH, spherical stellar background cusp and an asymmetric eccentric disk model in Fig. \ref{fig: disk-precession-070}. The mass of the SMBH and the properties of the cusp are identical to the ones in the simulations of these study. The disk masses range from $M_\mathrm{disk} = 10^4 \;\mathrm{M_\odot}$ to $M_\mathrm{disk} = 9\times10^4 \;\mathrm{M_\odot}$ with intervals of $10^4 \;\mathrm{M_\odot}$. The initial eccentricity of the disk is $e=0.7$. With low disk masses $M_\mathrm{disk}\leq3\times10^4$ particle orbits the precess retrograde, except in near the SMBH with semi-major axes $a_\star<10^{-2}$ pc where relativistic precession is important. With increasing disk mass, its contribution of the disk to orbit precession becomes increasingly large. The prograde precession rate decreases, especially in the vicinity of the disk inner edge, and becomes dependent on $a_\mathrm{\star}$. At disk mass of $M_\mathrm{disk}\sim4\times10^4$ the prograde picture qualitatively changes as a region of weak prograde precession appears around $0.05$ pc $\lesssim a_\star \lesssim 0.1$ pc. Now the disk has two prograde regions and two retrograde regions as the original outer retrograde region is split into two. The initially narrow prograde region gradually widens and the magnitude of precession increases, until at $M_\mathrm{disk}\sim9\times10^4 \;\mathrm{M_\odot}$ the inner retrograde region vanishes, and precession is prograde within $0.2$ pc of the SMBH.

The simple precession model presented explains the results of the N-body eccentric disk simulations of Fig. \ref{fig: phi-lrl-sim}. The low-mass literature setup L7 precesses retrograde at all radii, as expected, and the rate of precession is slightly larger at the outer parts of the disk. The outer parts of the model M precess retrograde just as in model L7, but the very inner edge of the disk slowly precesses in the prograde direction. This is exactly what is also seen in the panel $M_\mathrm{disk}=4\times10^4 \;\mathrm{M_\odot}$ of Fig. \ref{fig: disk-precession-070}: slow inner prograde precession and stronger outer retrograde precession. The behaviour of the highest-mass model H7 is also well explained by the simple precession models with masses $M_\mathrm{disk}\sim7\times10^4 \;\mathrm{M_\odot}$--$M_\mathrm{disk}=8\times10^4 \;\mathrm{M_\odot}$: prograde precession within $\sim 3 R_\mathrm{in} \sim 0.15$ pc of the SMBH and retrograde precession of equal magnitude outside this radius.

Whereas Fig. \ref{fig: disk-precession-070} shows the results of the simple precession model with initial eccentricity of $e=0.7$, Fig. \ref{fig: disk-precession-all-ecc} presents the results for nine initial eccentricities between $e=0.1$ and $e=0.9$ with intervals of $0.1$. The precession rates and directions of the stellar orbits are measured at four different semi-major axes from the SMBH, at $a_\star=0.05$ pc (inner edge of the disk), $a_\star=0.10$ pc, $a_\star=0.25$ pc and $a_\star=0.50$ pc (outer edge). 

With low disk mass $M_\mathrm{disk} = 10^4 \;\mathrm{M_\odot}$, the precession is retrograde at every radii, and the precession rate is larger with lower eccentricities, as expected from Eq. \eqref{eq: timescale-massprecession}. Going towards higher disk masses, the retrograde precession rates decrease and eventually turn prograde, beginning with disks of higher eccentricities $e\gtrsim0.7$ as the self-contribution of the disk to the precession becomes gradually more important. The inner edge of the disk precesses prograde when $M_\mathrm{disk}>4\times10^4 \;\mathrm{M_\odot}$ at all eccentricities while at $a_\star=0.1$ pc this occurs after $M_\mathrm{disk}>8\times10^4 \;\mathrm{M_\odot}$. At larger semi-major axes from the SMBH, at $a_\star=0.25$ pc and $0.50$ pc, the precession direction is almost always retrograde, except at high disk masses ($M_\mathrm{disk}\gtrsim7\times10^4 \;\mathrm{M_\odot}$) and high eccentricities ($e\gtrsim0.8$) where the precession rate is small or weakly prograde. 

When not orbiting at the inner edge of the disk, the precession rate always shows a dependence on the disk eccentricity. However, when the disk mass is high ($M_\mathrm{disk}\gtrsim7\times10^4 \;\mathrm{M_\odot}$), the precession rate at the inner edge of the disk is almost independent of the disk eccentricity. As the contribution of the spherical cusp is eccentricity-dependent (see e.g. the panel $M_\mathrm{disk}=10^4 \;\mathrm{M_\odot}$ in Fig. \ref{fig: disk-precession-all-ecc}), the disk contribution at the inner edge must be eccentricity-dependent as well to result in a constant precession rate. The prograde precession rate being higher at lower eccentricities can be understood in the following terms. When orbiting at the inner edge of the disk, there is no disk mass inside the pericenter distance from the SMBH. An orbit with low eccentricity also has a low apocenter and most of the disk mass is located outsize of the orbit, causing prograde torque (see Fig. \ref{fig: disk-torque}). With higher eccentricities and thus higher pericenters, there is increasingly more mass inside the orbit near the apocenter, making the precession less prograde by retrograde torques. We emphasize that the approximately constant precession rate at the inner edge of the disk is not a generic feature, but depends on the chosen masses and (surface) density profiles of the disk and the cusp, and the SMBH mass.

\subsection{The long-term evolution of the disk eccentricity distribution}

\begin{figure}
\includegraphics[width=\columnwidth]{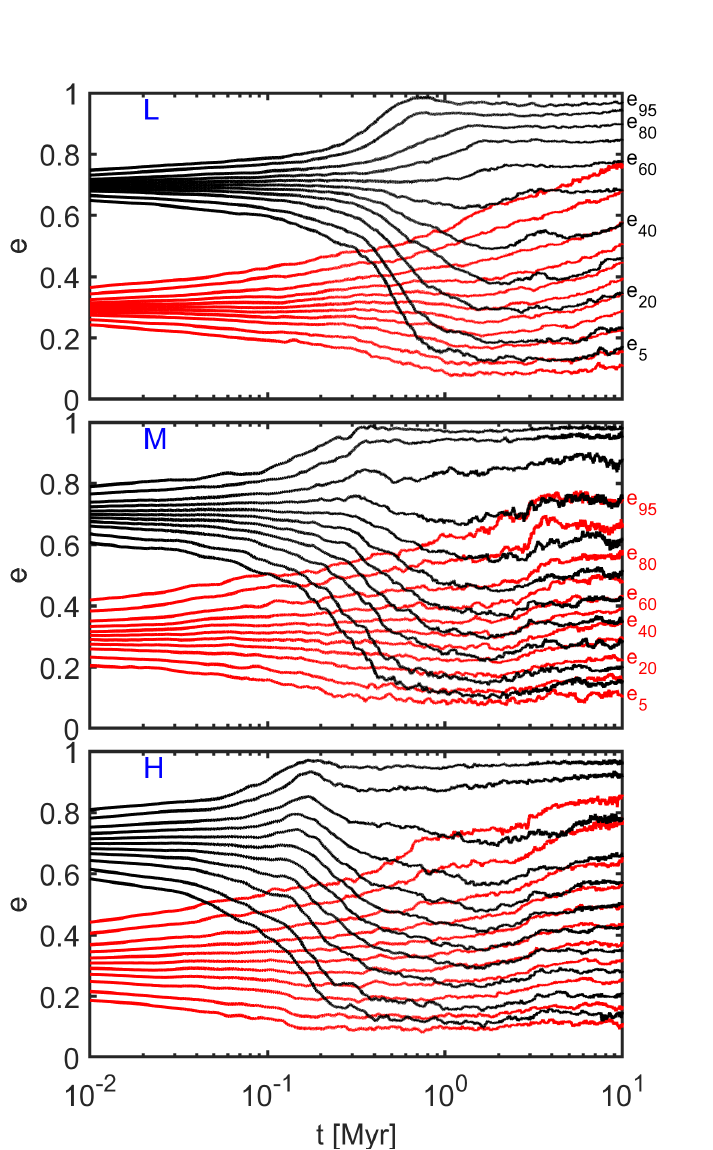}
\caption{The broadening of the initially narrow eccentricity distributions of the disks with initial eccentricities of $e_\mathrm{init}=0.3$ (red lines) and $e_\mathrm{init}=0.7$ (in black). The lines show the Lagrangian eccentricities (in the sense of the Lagrangian radii of a cumulative mass profile) $e_\mathrm{5}$, $e_\mathrm{20}$, $e_\mathrm{40}$, $e_\mathrm{60}$ $e_\mathrm{80}$ and $e_\mathrm{95}$. For example, $95\%$ of the particles have a lower eccentricity than $e_\mathrm{95}$. Binary stars are treated as their center-of-masses in the analysis. With the initial eccentricity of $0.7$, the disk stars can reach almost circular or radial orbits in less than $\sim 0.6$ Myr. Disk with initially lower eccentricity also has stars with moderately eccentric orbits, but only after several Myrs of evolution. The main difference between the models L, M and H is how rapidly the eccentricity distribution spreads. For the
model L7 the initial phase of rapid evolution is over after $\sim 0.6$ Myr, while the time-scale is $\sim0.4$ Myr and $\sim0.2$ Myr for the models M and H, respectively.}
\label{fig: lagrangian-ecc}
\end{figure}

Fig. \ref{fig: lagrangian-ecc} presents the evolution of the disk star eccentricities until the end of the simulations at $t=10$ Myr. Binary systems are treated as their center-of-masses. Stars with semi-major axes smaller than $a_\star=0.04$ pc or larger than $0.5$ pc are excluded from the analysis. Two initial eccentricities of $e_\mathrm{init}=0.3$ and $e_\mathrm{init}=0.7$ are shown from the three models L, M and H of different disk masses each. 

The initially narrow eccentricity distributions rapidly broaden in all the models. The broadening phase is rapid, just as the disruption of the alignment of the eccentricity vectors. This reflects the nature of the asymmetric eccentric disk instability in the disks with different masses just as seen in Fig. \ref{fig: phi-lrl-sim} for the spread of the eccentricity vector directions. Model L disrupts due to the secular eccentric disk instability while especially the model H disrupts more rapidly as the different regions to the disk precess to different directions. In the models L the rapid phase is essentially over in $\sim0.6$ Myr while in the models M and H the corresponding times are $\sim0.4$ Myr and $0.2$ Myr. With the higher initial eccentricity of $e_\mathrm{init}=0.7$, stars can reach very high eccentricities up to unity while for the initial eccentricity of $e_\mathrm{init}=0.3$, the majority of stars have eccentricities always less than $e_\star\sim0.8$.

\begin{figure*}
\includegraphics[width=0.8\textwidth]{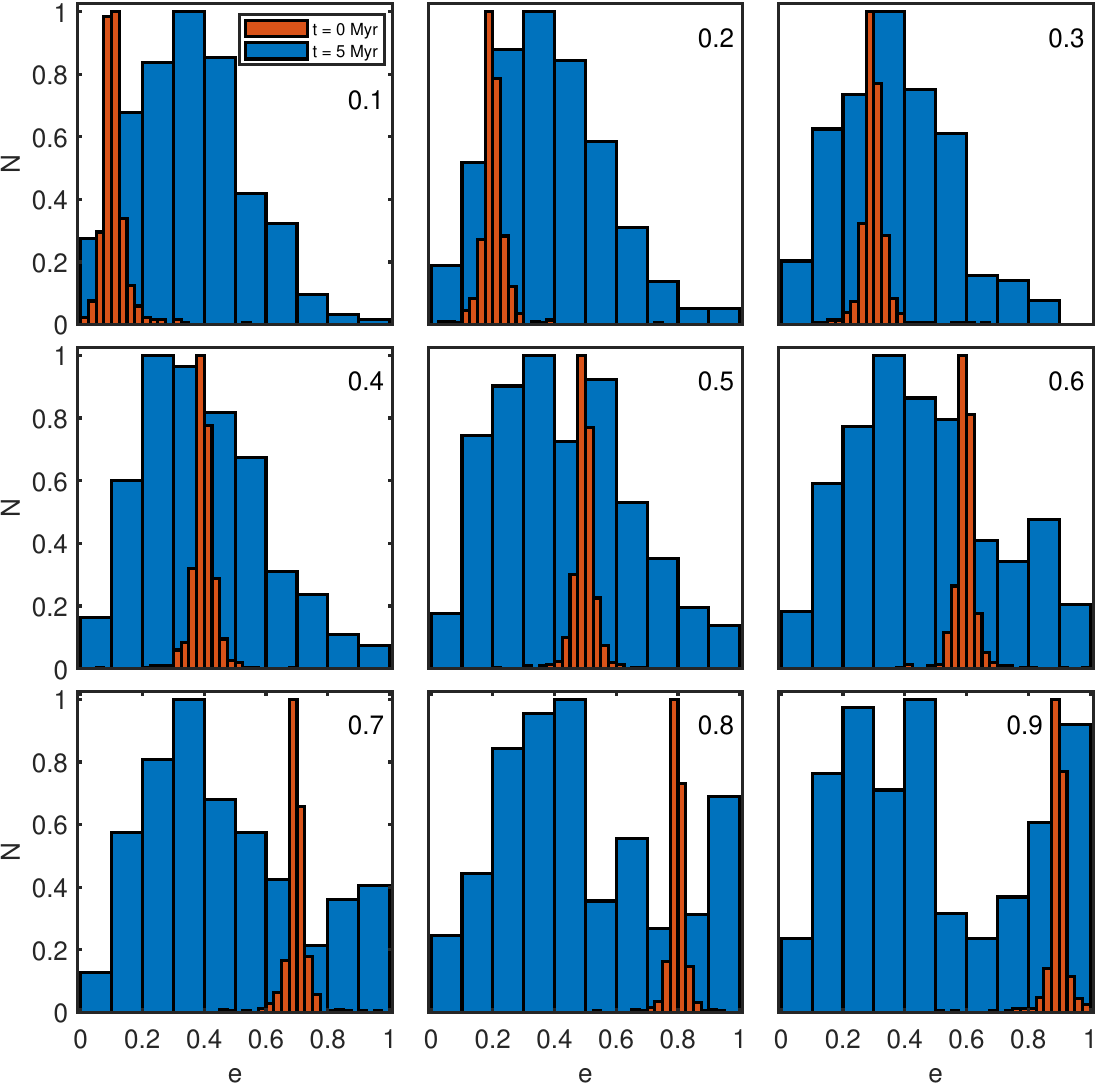}
\caption{The eccentricity distributions of the nine simulations of the disk model M at $0.04$ pc $<a_\star<0.5$ pc at two times, $t=0$ Myr and $t=5$ Myr. For the analysis the orbits of the binary stars around the SMBH are treated as the orbits of their center-of-masses. The originally narrow disk eccentricity distributions (in orange) considerably broaden in $5$ Myr, with the initial rapid evolution phase lasting only $\sim 0.5$ Myr. Interestingly, after $5$ Myr, the eccentricity distribution is broad for all initial eccentricities peaking between $0.2\lesssim e_\star \lesssim0.5$. While all setups produce stars on almost circular orbits, setups with initial eccentricities less than $\lesssim0.4$ do not have extremely eccentric stars beyond $e_\star>0.9$. For more eccentric initial setups with $e_\star\gtrsim0.7$, a secondary high-eccentricity peak begins to develop, becoming very prominent at initial eccentricity of $e_\mathrm{init}=0.9$.} 
\label{fig: ecc-histogram}
\end{figure*}

The histograms presenting the eccentricity distributions of the disks in model M, in total nine different initial eccentricities, are shown in Fig. \ref{fig: ecc-histogram}. For the analysis the binary systems in the disk are again treated as their center-of-masses. The eccentricity distributions are presented at two times, at $t=0$ Myr and later at $t=5$ Myr long after the unstable asymmetric eccentric disk has disrupted. At this point the eccentricity structure of the disks is in approximate equilibrium, as seen in Fig. \ref{fig: lagrangian-ecc} for the initial disk eccentricities of $e_\mathrm{init}=0.3$ and $e_\mathrm{init}=0.7$. After $t=5$ Myr of evolution all the nine models produce a broad disk eccentricity distribution from the initially narrow distributions. The broad distribution always peaks at $0.2\lesssim e \lesssim 0.5$ regardless of the initial disk eccentricity. All the models have stars on almost circular orbits with $e_\star<0.1$. However, only eccentric disks models with moderate to high initial eccentricities, $e_\mathrm{init}\gtrsim0.4$ produce stars with extremely high eccentricities of $e_\star>0.9$. A secondary peak of high-eccentricity stars develops in the models with initial eccentricity higher than $e_\mathrm{init}\gtrsim 0.7$ with the distribution being weakly bimodal in the model with initial eccentricity of $0.7$. In the initially very eccentric model with $e_\mathrm{init}=0.9$ the high-eccentricity peak of the disk eccentricity distribution becomes very prominent.

\subsection{The long-term evolution of the vertical disk structure}

\begin{figure}
\includegraphics[width=\columnwidth]{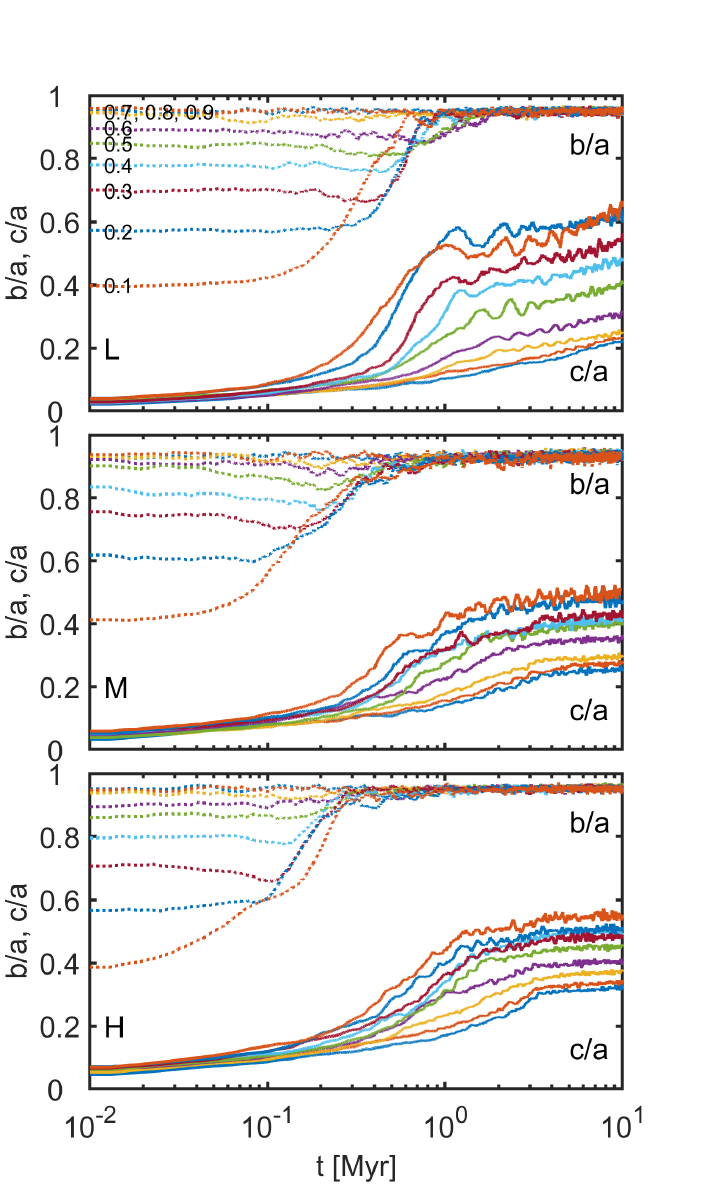}
\caption{The evolution of the shape of the disk models L, M, and H as characterized by the principal axis ratios $b/a$ and $c/a$ determined from the inertia tensors of the systems. The intermediate axis ratio $b/a$ rapidly reaches unity, as expected, as the initial alignment of the disk star eccentricity vectors is lost. The time-scale of $b/a$ reaching unity is very similar than the time-scale for the broadening of the initially narrow eccentricity distributions, from $\sim0.6$ Myr in model L to $\sim 0.2$ Myr in model H, shown in Fig. \ref{fig: lagrangian-ecc}. However, the vertical structure of the disks evolves more slowly than their azimuthal structure. We attribute this process to the dynamical heating of the disks in addition to the disk instability itself. The most eccentric initial models yield the most puffed-up, vertically extended thick disks. The final axis ratios $c/a$ after $10$ Myr of evolution range from $\sim 0.25$ to $\sim 0.5$ depending on the initial disk eccentricity while in the low-mass model L the highest $c/a$ can reach even somewhat higher values of $\sim0.6$.
}
\label{fig: axisratios}
\end{figure}

\begin{figure*}
\includegraphics[width=\textwidth]{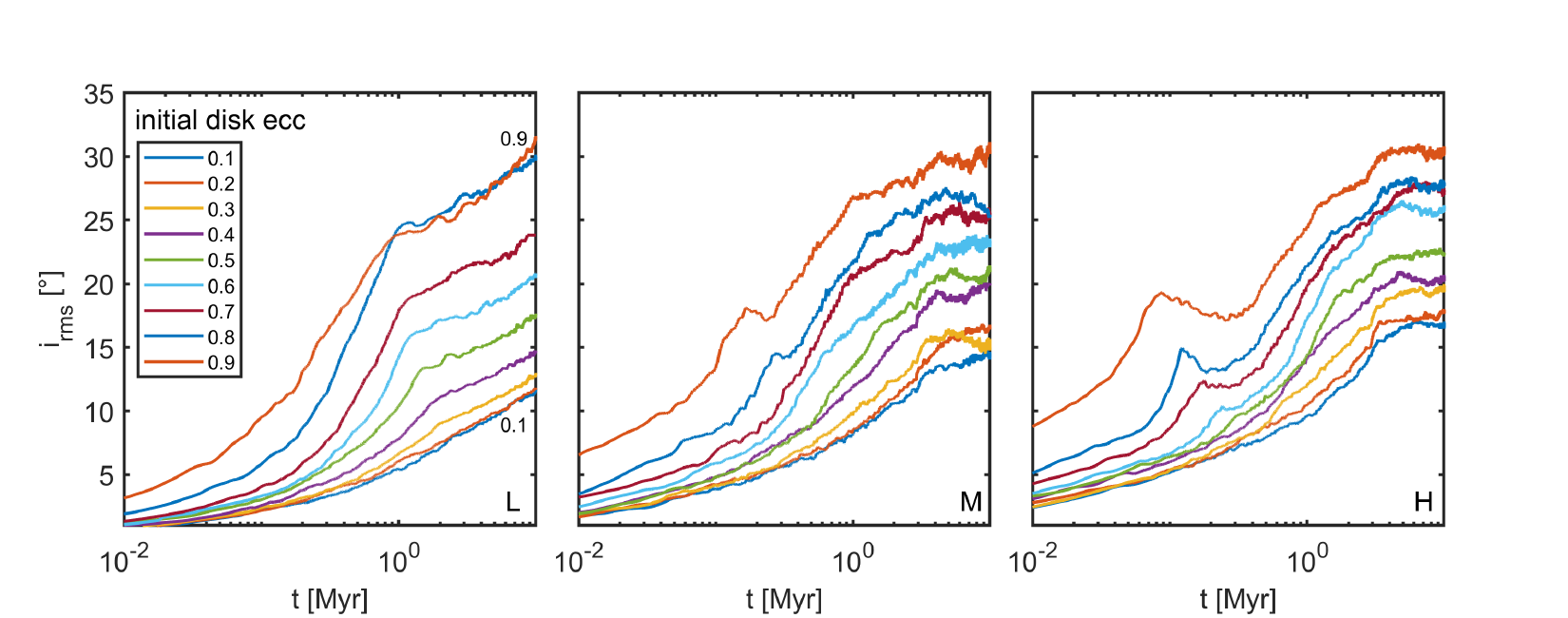}
\caption{The rms inclinations of the disk stars and binary center-of-masses within $0.04<a_\star<0.5$ pc in the simulated disk models. Already at $t=1$ Myr the rms inclinations range between $6\degree<i_\mathrm{rms}<25\degree$, $8\degree<i_\mathrm{rms}<27\degree$ and $10\degree<i_\mathrm{rms}<24\degree$ in the models L, M and H, respectively. As with the minor axis ratio $c/a$ in Fig. \ref{fig: axisratios}, the evolution of the rms inclination in the disk models slows down after $1$ Myr. The final rms inclinations after $10$ Myr of evolution are $12\degree<i_\mathrm{rms}<32\degree$, $14\degree<i_\mathrm{rms}<31\degree$ and $17\degree<i_\mathrm{rms}<31\degree$ for the three simulation sets. The early inclination feature around $t\sim0.1$ Myr in models M and H coincides with the appearance and dissolution of the spiral feature in the disk illustrated in Fig. \ref{fig: spiral}.}
\label{fig: inc-rms}
\end{figure*}

Next, we examine the evolution of the shape, and especially the vertical structure of the disk models. We first characterize the shape of the disk system by the ratios of its principal axis $a$, $b$ and $c$, with $a$ being the longest and $c$ the shortest axis. The axis ratios are calculated from the inertia tensor of the disk stars and binary center-of-masses within the 3D half-mass radii of the models ($r_\mathrm{h} \sim 0.2$ pc) as in \cite{Zemp2007}. Fig. \ref{fig: axisratios} presents the evolution of principal axis ratios $b/a$ and $c/a$ in each of the nine models of the simulation sets L, M and H. 

The evolution of the azimuthal direction, $b/a$, reflects the instability of the initially asymmetric eccentric disks. The intermediate axis ratio $b/a$ rapidly reaches unity in all the models on a time-scale corresponding to the broadening time-scale of the initially narrow eccentricity distributions of $0.6$ Myr, $0.4$ Myr and $0.2$ Myr for the models L, M and H, respectively. The minor axis ratio $c/a$ first increases in from an initial near-zero value to moderate values of $\sim 0.2$-$0.5$ within $0.2\lesssim t \lesssim 1$ Myr in the models. After $1$ Myr, the vertical evolution proceeds on a different, slower time-scale as. The minor axis ratio of the disk models keeps increasing in all the models, albeit more slowly than in the early disk disruption phase. The initially most eccentric disk models produce the most vertically extended disks with the final values of $c/a$ being $0.22 \lesssim c/a \lesssim 0.65$, $0.25 \lesssim c/a \lesssim 0.51$ and $0.33 \lesssim c/a \lesssim 0.55$ in the models L, M and H after $10$ Myr of evolution. The highest values of $c/a$ are obtained perhaps surprisingly in model L. There are two reasons for this. First, the in low-mass disk models of set L the vertical accelerations due to the disk gravity of the disk stars are smaller. Second, in the model L we do not include stellar evolution, so massive stars in the run do not die, so they can keep perturbing towards higher inclinations stars for a longer time.

Fig. \ref{fig: inc-rms} shows the root-mean-square (rms) inclinations $i_\mathrm{rms}$ of the disk stars for the simulation setups L, M and H. The rms inclinations of the disk models evolve as expected from the evolution principal axis ratios in Fig. \ref{fig: axisratios} with an initial rapid phase before $\sim 1$ Myr, and more gradual increase afterwards. The initial rapid phase approximately follows the theoretical expectation of the $t^{1/4}$ evolution of the rms inclination evolution \citep{Lissauer1993}. The final rms inclinations after $10$ Myr of evolution are $12\degree<i_\mathrm{rms}<32\degree$, $14\degree<i_\mathrm{rms}<31\degree$ and $17\degree<i_\mathrm{rms}<31\degree$ for the three models, with the initially most eccentric disk models resulting in higher rms inclinations of disk stars.

\subsection{Counter-rotating stars in the disks}

\begin{figure*}
\includegraphics[width=\textwidth]{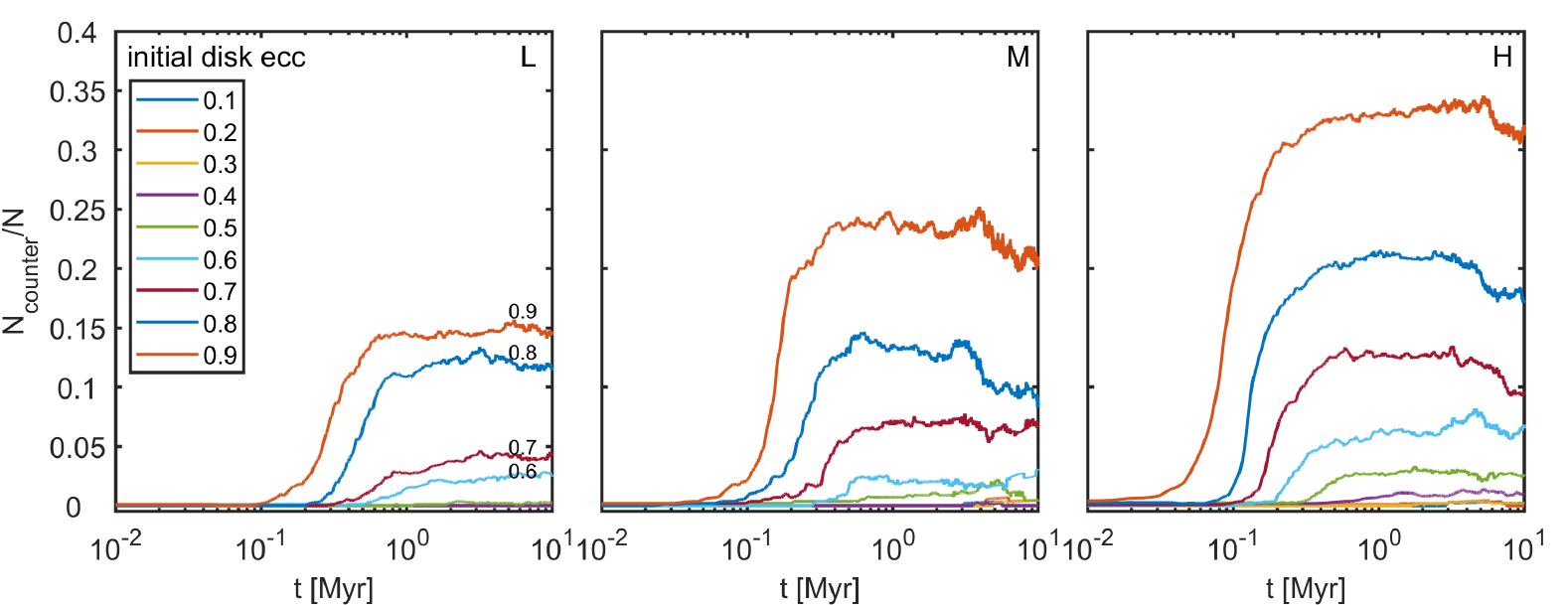}
\caption{The fraction of counter-rotating stars $f_\mathrm{c} = N_\mathrm{counter}/N$in the simulated disks. The models with initial eccentricity $e_\mathrm{init}\lesssim0.4$ have less than a percent of counter-rotating stars. Above this $e_\mathrm{init}$, the $f_\mathrm{c}$ rapidly increases to peak values of $f_\mathrm{c}\sim0.15$, $f_\mathrm{c}\sim0.25$ and $f_\mathrm{c}\sim0.34$ in the models L, M and H, respectively. Besides the initial eccentricity, the fraction of counter-rotating stars strongly depends on the initial disk mass. The build-up phase of $f_\mathrm{c}$ is rapid, consistent with the eccentric disk instability timescale in each model. After this, the evolution of $f_\mathrm{c}$ is slow.}
\label{fig: counter-rot}
\end{figure*}

Fig. \ref{fig: counter-rot} shows the number fraction of counter-rotating stars $f_\mathrm{c}=N_\mathrm{counter}/N$ in the disk as a function of time in the disk models L, M and H. Initially, all the stars are rotating in the same direction. The disk models with initial eccentricities less than $\sim 0.4$ end up having a very small ($\lesssim1\%$) fraction of counter-rotating stars. Disks initially more eccentric than this rapidly acquire a fraction of counter-rotating stars in the brief eccentric disk instability phase during the first Myr of the simulations, the time-scale being shorter for more massive disk models. This is consistent with the time-scales of the broadening of the initially narrow eccentricity distributions and the principal axis ratio $b/a$ in Fig. \ref{fig: lagrangian-ecc} and Fig. \ref{fig: axisratios}. After the initial rapid phase the $f_\mathrm{c}$ only mildly evolves.

The final $f_\mathrm{c}$ strongly depends on the initial disk mass and eccentricity. The maximum counter-rotation fractions are 
$f_\mathrm{c}\sim0.15$, $f_\mathrm{c}\sim0.25$ and $f_\mathrm{c}\sim0.34$ for the three disk models L, M and H, respectively. The stars than flip their rotation direction have preferentially high eccentricities when they do so. For example in the run H7, approximately $90\%$ of the stars have $e_\star>0.9$ at the moment of they turn counter-rotating. A majority of stars flipping their rotation direction are located in the inner half of the disk, $a_\star\lesssim0.3$ pc.

\subsection{Disk stellar content at later times}

For the purposes of this study we use a very simplified definition for O- and B-stars based on their initial masses: $2\;\mathrm{M_\odot} \leq m_\mathrm{B} \leq 20\;\mathrm{M_\odot}$ and $m_\mathrm{O} > 20\;\mathrm{M_\odot}$. Binary stars in the disk are counted as single objects as labelled using the type of the more massive component of the binary, i.e. OB-binaries are counted as O-stars. 

In the model L there are approximately $N_\mathrm{O} \sim 120$ O-stars in the disk at all times, however the number of B-stars is higher, $N_\mathrm{B} \sim 190$. As the model L specifically includes no stellar evolution and stars exceeding their life-times are not removed from the simulations, the number of O-stars is overestimated after $t\sim3$ Myr after which in reality O-stars would start dying.

\begin{figure}
\includegraphics[width=\columnwidth]{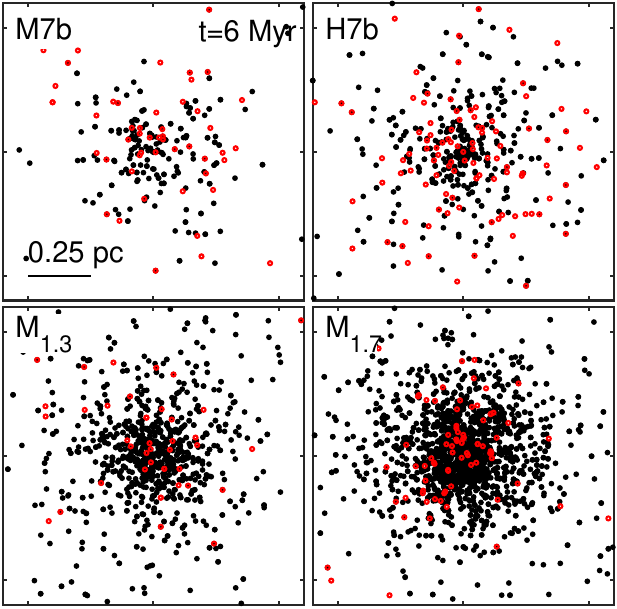}
\caption{The face-on projection of the disk showing the positions of B-type (black dots) and O-type (small red circles) stars after $t=6$ Myr of evolution.}
\label{fig: newruns-pr-xy}
\end{figure}

\begin{figure}
\includegraphics[width=\columnwidth]{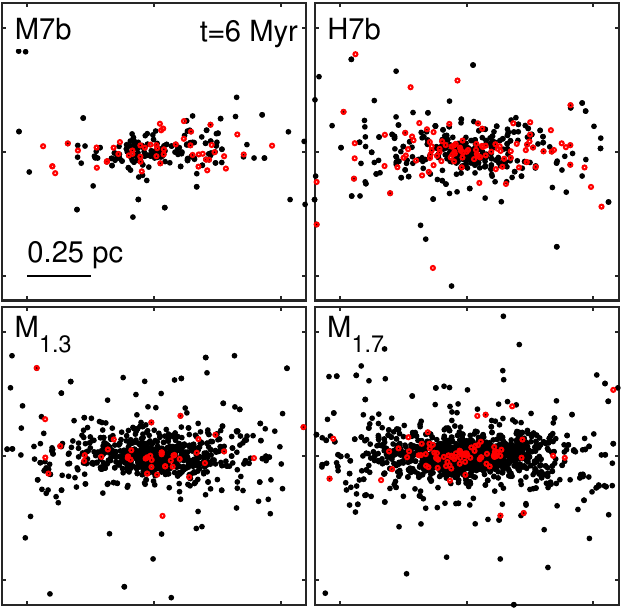}
\caption{Same as Fig. \ref{fig: newruns-pr-xy} but now in an edge-on projection. In the runs with an extremely top-heavy IMF (M7b, H7b in the top row) the B-type and O-type stars have similar vertical distributions. However, in the less top-heavy models ($\mathrm{M_{1.3}}$, $\mathrm{M_{1.7}}$, bottom row) the B-type stars form a somewhat thicker disk than the O-stars. The late-type stars (not shown in the panels) have an even more vertically extended distribution than the B-stars.}
\label{fig: newruns-pr-xz}
\end{figure}

The positions of the B-type stars and the O-type stars in the disks are shown in Fig. \ref{fig: newruns-pr-xy} and in Fig. \ref{fig: newruns-pr-xz} after $t=6$ Myr of evolution in face-on and edge-on projections, respectively. Just as by construction in simulation sets M and H (IMF slope $\alpha=0.25$), the models M7b and H7b have a comparable number of B-type and O-type stars in the disk at $t=6$ Myr. The numbers of the stars are $N_\mathrm{O}=102$, $N_\mathrm{B}=118$ (M7b), $N_\mathrm{O}=197$, $N_\mathrm{B}=209$ (H7b), $N_\mathrm{O}=102$, $N_\mathrm{B}=664$ ($\mathrm{M_{1.3}}$) and $N_\mathrm{O}=146$, $N_\mathrm{B}=1322$ ($\mathrm{M_{1.7}}$). The less extremely top-heavy setups have a considerably lower O-B star number ratio, as expected. The face-on position map of the disk stars in Fig. \ref{fig: newruns-pr-xy} show that the distribution of the B-type and O-type stars are comparable in each model and there is no obvious difference between the radial distributions of the stars of different masses. Thus, at $t=6$ Myr the radial distributions of the stars still reflect their original configuration at their formation time. 

The edge-on position map in Fig. \ref{fig: newruns-pr-xz} shows that while there is no difference in the thicknesses of the B-star and O-star disks in the models M7b and H7b, the B-star disks are somewhat more vertically extended than the disks of O-stars in the less top-heavy models. The late-type stars have even a more vertically extended distribution, especially in the disk setup $\mathrm{M_{1.7}}$. As all the disk models were originally very thin, we attribute the differences of the vertical disk heights of stars of different masses to the standard disk heating process. Interestingly, the O-star disk of the run $\mathrm{M_{1.7}}$ shows a weak warp structure. No warp structures are present in the other disks. 

Finally, the number of O-type and B-type stars in binaries is still large in the disks at $t=6$ Myr, $N_\mathrm{O+B}=87$ (M7b), $N_\mathrm{O+B}=178$ (H7b), $N_\mathrm{O+B}=567$ ($\mathrm{M_{1.3}}$) and $N_\mathrm{O+B}=798$ ($\mathrm{M_{1.7}}$). Initially, most of the B-type and O-type stars are located in binary systems. When comparing these numbers to the total numbers of B-type and O-type stars, we can see that the overall binary faction is still roughly $\sim0.5$ in the models at $t=6$ Myr. 

\subsection{Surviving binary stars in the disks at later times}

\begin{figure}
\includegraphics[width=\columnwidth]{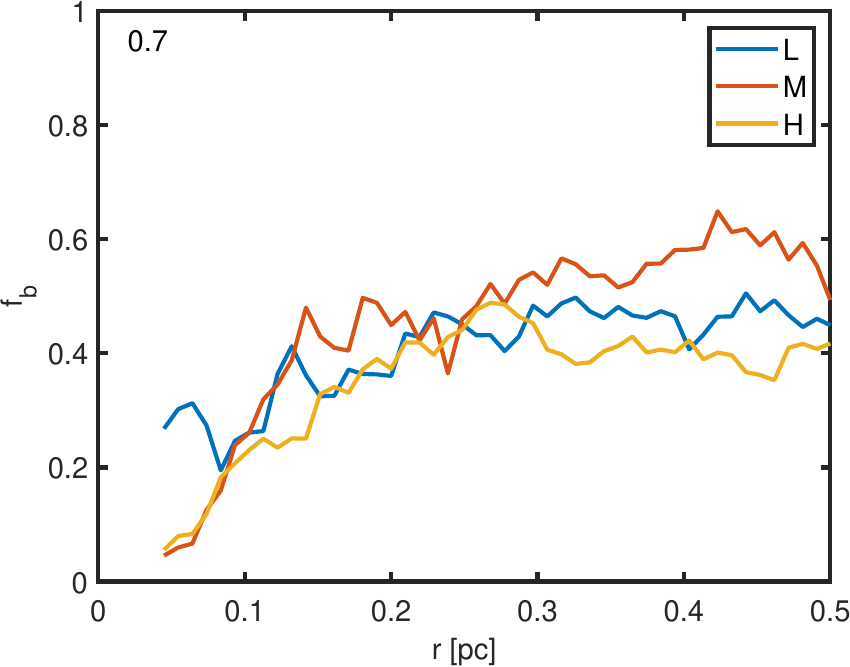}
\caption{The radial binary fraction $f_\mathrm{b} = N_\mathrm{b}/N$ of stars more massive than $2.1 \;\mathrm{M_\odot}$ in the disk models at $t=6$ Myr. In the outer parts of the disk, the binary fraction is $0.4 \lesssim f_\mathrm{b} \lesssim 0.6$. The binary faction begins to decrease below $r\sim0.15$ pc, reaching values of $0.05\lesssim f_\mathrm{b} \lesssim 0.25$ near the inner edge of the disk. The innermost binary star typically orbits the SMBH with a semi-major axis of $0.04$ pc $\lesssim a_\star \lesssim 0.065$ pc with a mild or moderate eccentricity of $0.2\lesssim e_\star \lesssim 0.5$.}
\label{fig: radial-fbin}
\end{figure}

By the end of the simulations, the disk models still retain binary stars. We present the radial dependence of the binary fraction of massive stars ($m_\star>2 \;\mathrm{M_\odot}$) at $t=6$ Myr in our simulations in Fig. \ref{fig: radial-fbin}. The radial separation of the binary star center-of-mass ranges from $0.04$ pc to $0.5$ pc in the analysis. In our models the binary fraction defined as $f_\mathrm{b} = N_\mathrm{b}/N_\mathrm{N}$ ranges from $0.4\lesssim f_\mathrm{b} \lesssim 0.6$ near the outer edge of the disk to $0.05\lesssim f_\mathrm{b} \lesssim 0.25$ at its inner edge. When comparing the orbital eccentricities of single stars and binary center-of-masses, we find that their distributions correspond to each other, except at high eccentricities which lack binary stars. This is well in line with the expectation that SMBHs drive nearby binaries to mergers, or unbind them. 

Even though the binary fraction decreases towards the inner edge of the disk, there is still a small number of binaries orbiting at these small separations from the SMBH. The innermost binary center-of-mass in our simulations has a semi-major axis of $0.04 \lesssim a_\star \lesssim 0.065$ pc with respect to the SMBH while having an eccentricity of $0.2 \lesssim e_\star \lesssim 0.5$, consistent with orbital elements of typical inner-disk single stars.

\section{High-velocity stars and tidal disruption events}\label{section: finalnumber-6}

\subsection{Escapers and high-velocity stars}

\begin{figure}
\includegraphics[width=\columnwidth]{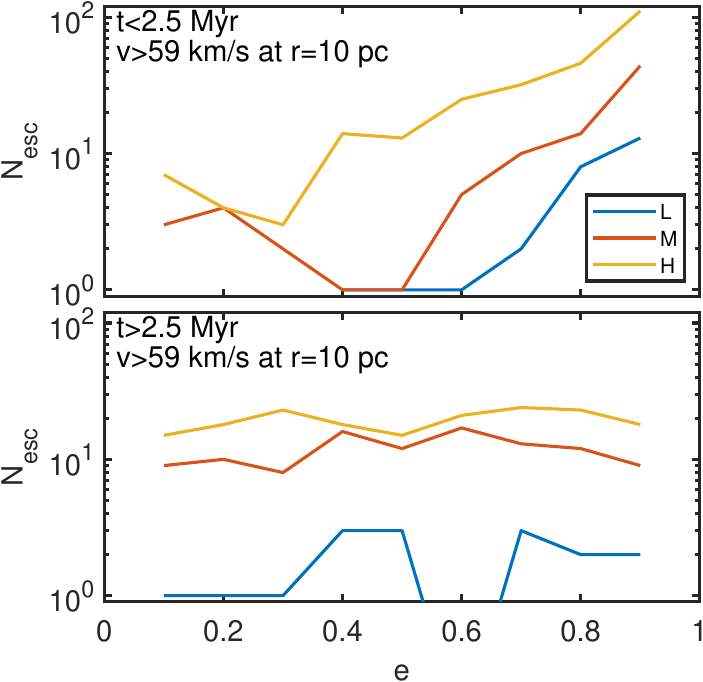}
\caption{The number of escapers i.e. unbound stars at a separation of $10$ of the SMBH in the simulations, corresponding to a minimum escape velocity of $\sim59$ km/s at this radius. The top panel shows the number of escaping stars in the early times of the simulations ($t<2.5$ Myr) before the most massive stars reach the end of their life-times (simulation sets M and H). Initially more massive and more eccentric disks have more escape events at early times, consistent with the Hills mechanism origin of the escaping stars. At later times ($t>2.5$ Myr, bottom panel) the main source of escaping stars are binary disruptions, mostly due to massive stars in binaries ending their lives.}
\label{fig: escapers}
\end{figure}

\begin{figure}
\includegraphics[width=\columnwidth]{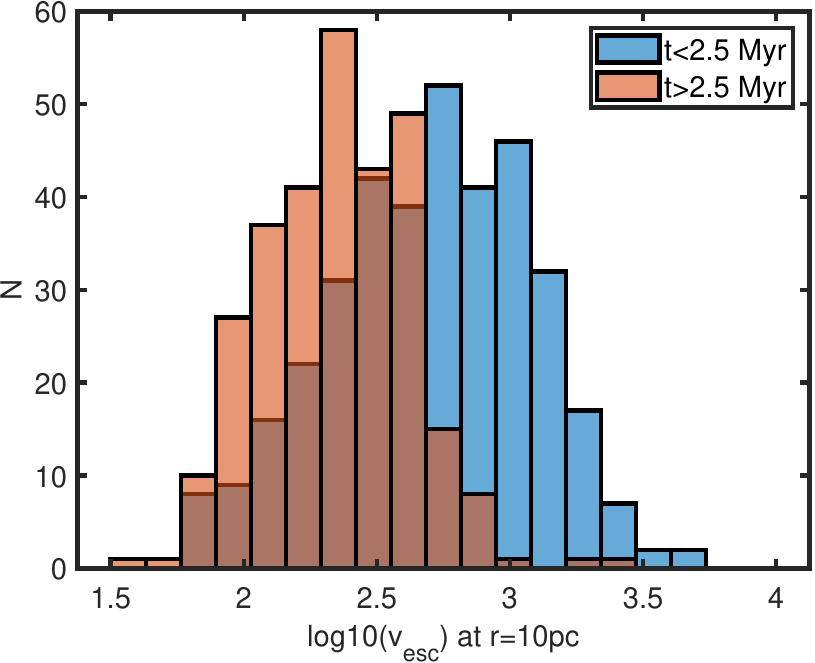}
\caption{The distributions of combined escape speeds of stars in all the simulations of this study. The escape events are divided into a Hills-dominated early ($t<2.5$ Myr, in blue) phase and a later ($t>2.5$ Myr, in orange) phase during which only few Hills ejections occur. While the two distributions are broad and overlap, only few late-time escapers reach the escape speed of $1000$ km/s. In total $81$ or $22\%$ of the early-time Hills escapers have escape speeds in the excess of $1000$ km/s. The mean logarithmic escape speeds of the two distributions are $\langle \log_\mathrm{10}{(v_\mathrm{esc})} 
\rangle = 2.70\pm0.38$ and $\langle \log_\mathrm{10}{(v_\mathrm{esc})} 
\rangle = 2.34\pm0.31$ corresponding to escape speeds of $217$ km/s and $500$ km/s.}
\label{fig: esc-speed-histogram}
\end{figure}

In the disk simulations the binary star center-of-masses can reach high eccentricities and thus low periapsis distances from the SMBH. The binary components can then become unbound to each other due to the Hills mechanism. The component unbound to the SMBH can reach high velocities, as explained in Section \ref{section: 2-hills}. Other mechanisms to accelerate stars into radial outward orbits with high speeds include disruption of binaries when one component reaches the end of its life-time, and complex and chaotic strong single-binary and binary-binary interactions. In our simulations, unbound stars are removed from the simulation at the separation of $10$ pc from the SMBH, corresponding to a minimum speed of $v_\mathrm{esc} \sim 59$ km/s at this radial separation. When an escape event occurs, we record the ID number, the mass, the escape direction and the velocity of the escaping star. In the simulations of this study the escapers are all single stars. Components of a previously disrupted binary may escape at very similar times, but the two stars are always well separated from each other at the escaper removal radius. 

We show the number of escapers $N_\mathrm{esc}$ as a function of the initial disk eccentricity in the models L, M and H in Fig. \ref{fig: escapers}. We divide the simulation time into two parts, the early period before $t<2.5$ Myr and the later period between $2.5$ Myr $<t<10$ Myr. The division time approximately corresponds to the life-time of most massive stars in our simulation, after which they are replaced with compact remnants (sets M and H). 

Before $2.5$ Myr, the primary reason for stars reaching the escape velocity is the binary break-up due to the Hills mechanism. We explicitly check the ID numbers of the escaping stars and stars on tightly bound orbits around the SMBH to confirm the Hills origin of the escaping stars in our simulations. Almost all the Hills binary break-ups occur at early times after the disruption of the asymmetric eccentric disks. In the low-mass setup L there are no escapers at early times in the simulations with the initial disk eccentricity less than $e_\mathrm{init}<0.7$, after which the number of escapers steadily increases with increasing initial eccentricity to $N_\mathrm{esc} = 13$ at $e_\mathrm{init}=0.9$. In the two more massive simulation setups M and H the trend is similar. Only a few escapers ($N_\mathrm{esc}<8$) originate from strong few-body interactions not involving the SMBH at low initial disk eccentricities, after which $N_\mathrm{esc}$ begins to increase. For the mid-mass set M this occurs around $e_\mathrm{init} \sim 0.5$-$0.6$ for the set M and for the most massive set H after $e_\mathrm{init} \sim 0.3$. The number of escapers increases rapidly, reaching $N_\mathrm{esc}=44$ and $N_\mathrm{esc}=111$ for the two sets.

After $2.5$ Myr the dominating mechanism of producing escaping stars in the disk simulations is the binary break-up due to death and removal of massive stars. In the low-mass simulation setup L, in which the removal is not performed, there are only few escaping stars after $t>2.5$ Myr, all due to strong interactions involving $N\geq3$ bodies. For the simulation setups M and H, there are $\sim 10$-$20$ escaping stars in each simulation, with no dependence on the initial disk eccentricity, as expected.

We present the distribution of speeds of the escaping stars at $r=10$ pc from the SMBH in Fig. \ref{fig: esc-speed-histogram}. As before, we divide the escape events into early ($t<2.5$ Myr) and late 
($t>2.5$ Myr) ones. As the number of escapers in many simulations is small, we study only the distribution of combined escape speeds from all escapes in our $27$ simulations. The both early and late escape speed distributions are well characterized by a log-normal distribution. The early-time mostly Hills-driven escape speed distribution has a mean logarithmic escape speed of $\langle \log_\mathrm{10}{(v_\mathrm{esc})} 
\rangle = 2.70\pm0.38$ corresponding to a speed of $\sim500$ km/s. The distribution is broad, and approximately $22\%$ of the early escapers have a high escape speed greater than $1000$ km/s. The two most extreme high-velocity stars move at the speed of $3500$ km/s and $5000$ km/s when exiting the simulation domain. After $2.5$ Myr, the escape speeds are on average lower, but the early and late distributions overlap. The log-normal fit yields a mean logarithmic escape speed of $\langle \log_\mathrm{10}{(v_\mathrm{esc})} 
\rangle = 2.34\pm0.31$. This corresponds to a speed of $217$ km/s. Only few late particles ($<4$) reach velocities higher than $1000$ km/s, and at two of these are late-time Hills ejections.

\begin{figure}
\includegraphics[width=\columnwidth]{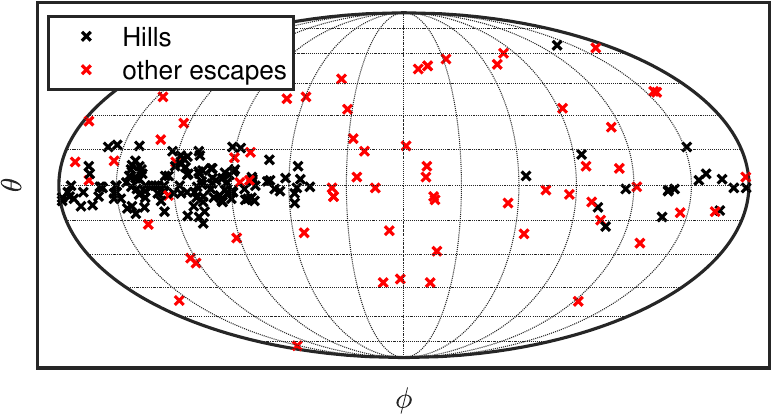}
\caption{The anisotropic distribution of early Hills escapers (in black) versus the more isotropic distribution of the late non-Hills escapers (in red). The data shown combines the simulations H7, H8 and H9. In the Mollweide projection, the initial apoapsis direction of the disk stars lies on the left of the image while the initially aligned periapsis direction is at the center. The stars ejected by the Hills mechanism preferentially escape near the disk mid-plane $\theta=0$ while the azimuthal direction $\phi$ shows a clear hemispherical asymmetry.}
\label{fig: esc-anisotropy}
\end{figure}

As pointed out by \cite{Subr2016}, the anisotropic direction distribution of the escaping stars is characteristic for the Hills mechanism with the binary supply from an asymmetric eccentric disk. We confirm this result in our simulations, presented in Fig. \ref{fig: esc-anisotropy}. Most of the stars ejected by the Hills mechanism early in the simulations escape to the directions close to the disk mid-plane $\theta=0$. At the azimuthal direction, the distribution shows a clear hemispherical asymmetry. The scatter of one standard deviation around the in the in-plane escape direction $\phi$ is $\sigma_\mathrm{\phi} = 57\degree$, $\sigma_\mathrm{\phi} = 45\degree$ and $\sigma_\mathrm{\phi} = 32\degree$ for the disk models H7, H8 and H9, respectively. 

The anisotropic escaper direction distribution reflects the nature of the supply of binary stars for the Hills mechanism. The direction of the escaping component of the disrupted binary is typically roughly opposite to the direction of the eccentricity vector of the just disrupted binary center-of-mass with respect to the SMBH. Thus, if the disk star eccentricity vectors are not yet completely misaligned even though the asymmetric eccentric disk is disrupting, the escaper direction distribution can still retain information of the orientation of the eccentric disk from the time the stars were ejected.

The late-time escapers not originating from the Hills mechanism show a more isotropic distribution, both having escapers all around the azimuthal angle $\phi$ and at high inclinations far above and below the disk plane $\theta = 0$.

\subsection{Accreted stars and TDEs}

\begin{figure}
\includegraphics[width=\columnwidth]{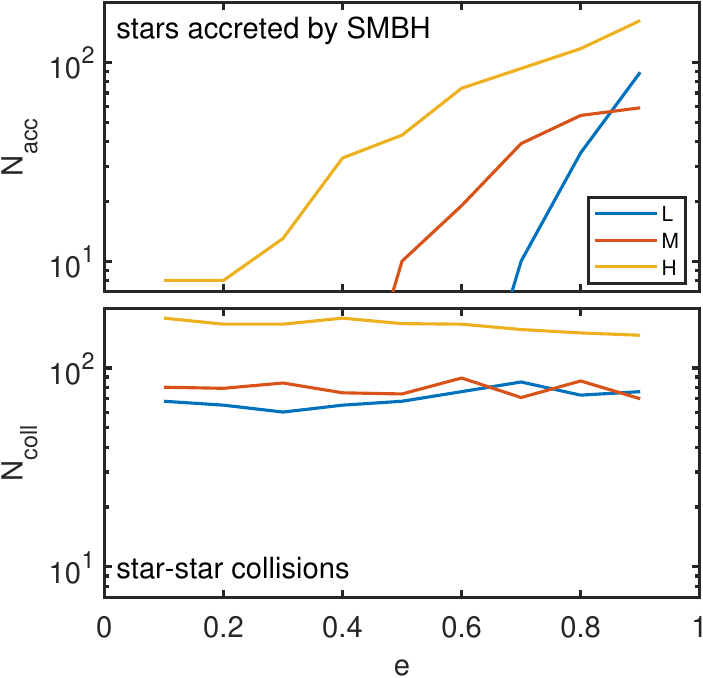}
\caption{Number of stars merging with the SMBH during the simulations (top panel) and with other stars (bottom panel) in the simulation setups L, M and H. While the number of stars accreted by the SMBH strongly depends on of the initial disk eccentricity, the number of star-star mergers has no such dependence.}
\label{fig: Ntde}
\end{figure}

We to study the total number of stars accreted by the SMBH and star-star mergers in the simulations of the samples L, M and H as a function of disk mass and initial disk eccentricity. These results are shown in Fig. \ref{fig: Ntde}. There are more than $\sim5$ star-SMBH mergers in the disk models with initial eccentricity higher than $0.7$ (L), $0.5$ (M) and $0.1$ (H). The number of total accreted stars strongly depends on the initial disk eccentricity. The vast majority of the accreted stars originate from almost parabolic orbits. The maximum number of total accreted stars is $N_\mathrm{acc} = 162$ in the simulation H9. Most accretion events occur early, immediately after the instability of the eccentric disks and the occurrence of stars on highly eccentric orbits. As most of the stars are accreted during the first Myr of the simulations, the merger rate in the simulation samples L, M and H roughly corresponds to a maximum TDE rate of $60$--$160$ $\mathrm{Myr}^{-1}$.

Star-star mergers occur in each of the simulation setups as presented in the bottom panel of Fig. \ref{fig: Ntde}. In contrary to the star-SMBH merger rate, the star-star merger rate does not show any dependence of the initial disk eccentricity. The actual number of star-star mergers ranges from $\sim60$ mergers in the models L to $\sim180$ in the setup H. 

\begin{figure}
\includegraphics[width=\columnwidth]{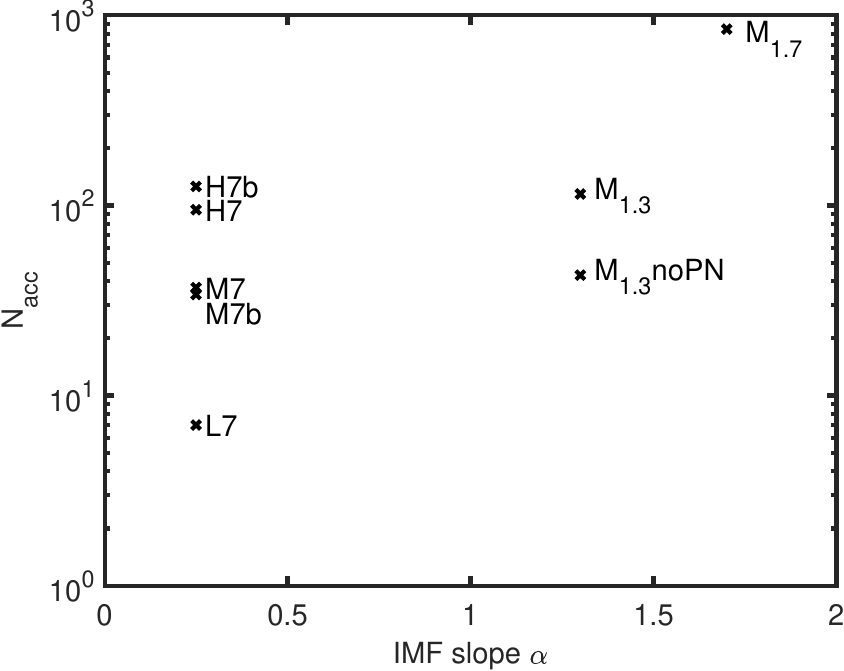}
\caption{The number of stars accreted by the SMBH ($N_\mathrm{acc}$) in selected simulation runs as a function of the IMF slope of the stellar population if the setup. First, the number of accreted stars increases with the increasing disk mass as the number of stars available for disruption increases, as already shown in Fig. \ref{fig: hills} and Fig. \ref{fig: Ntde}. Next, with a constant initial disk mass, the number of accreted stars with increasing (less top-heavy) IMF slope. The number of accreted stars ranges from less than $N_\mathrm{acc}<10$ events in the setup L7 to almost $N_\mathrm{acc}\sim10^3$ in the model $\mathrm{M_{1.7}}$. The binary population properties have a relatively small effect on $N_\mathrm{acc}$ when comparing runs M7, M7b, H7 and H7b. Runs without PN effects have less accreted stars as the mean orbital eccentricities are lower and pericenter distances higher compared to runs with PN effects included. This comparison can be only performed for the sample $\mathrm{M_{1.3}}$ as in the setups with other IMF slopes the runs without PN didn't include a TDE prescription either.
}
\label{fig: newruns-nacc-scluster}
\end{figure}

The number of accreted stars depend on the chosen merger prescription, the IMF and the assumed binary star population properties. We present the number of stars accreted by the SMBH ($N_\mathrm{acc}$) as a function of the IMF slope and the initial disk mass of the simulation models in Fig. \ref{fig: newruns-nacc-scluster}. As already demonstrated in Fig. \ref{fig: Ntde}, more massive eccentric disks result in a larger number of accreted stars, $N_\mathrm{acc}=7$ (L7), $N_\mathrm{acc}=37$ (M7) and $N_\mathrm{acc}=95$ (H7). 

Introducing the hard binary population has only a small to moderate effect (up to $\sim30\%$) on the number of accreted stars as the setup M7b has $N_\mathrm{acc}=34$ accretion events and H7b has $N_\mathrm{acc}=126$. Changing the IMF of the mid-mass disk model into a less extremely top-heavy one ($\mathrm{M_{1.3}}$) results in $N_\mathrm{acc}=115$ stars accreted by the SMBH, a number comparable to the high-mass setup H7b. Turning off the PN equations of motion in the model $\mathrm{M_{1.3}noPN}$ reduces the number of accreted stars by over a factor of $2.5$ as the stars end up on average on less eccentric orbits due to the missing suppression of the resonant relaxation effects. Finally, the least top-heavy model $\mathrm{M_{1.3}}$ undergoes $N_\mathrm{acc}=847$ star-SMBH mergers. The comparison of $N_\mathrm{acc}$ in the other models with and without PN than $\mathrm{M_{1.3}}$ is not possible as the runs without PN also had TDEs not enabled.

The vast majority of the stars are again accreted early. If the number of accreted stars within the first Myr is translated into a TDE rate, the models $\mathrm{M_{1.3}PNTDE}$ and $\mathrm{M_{1.7}PNTDE}$ have TDE rates of $R\sim90\;\mathrm{Myr}^{-1}$ and $R\sim770\;\mathrm{Myr}^{-1}$ within $1$ Myr of the simulation start. The model $H7$ has a TDE rate comparable to the setup $\mathrm{M_{1.3}PNTDE}$ while the other extremely top-heavy models have rates of $R\sim30\;\mathrm{Myr}^{-1}$ (M7) and $R<10\;\mathrm{Myr}^{-1}$ (L). Note that as before, the given TDE rates are not instantaneous but instead averaged over $1$ Myr, and that instantaneous TDE rates can be considerably higher. At later times the TDE rates are by several orders of magnitude lower. Overall, the first-Myr TDE rates considerably differ in the simulations with different disk masses and IMF slopes.

Finally, we estimate whether the TDEs in our simulations would be in principle observable. If the tidal disruption radius $\mathcal{R}_\mathrm{t}$ is smaller than the direct capture radius $r_\mathrm{cap} = 4 G M_\bullet/c^2$ (assuming a zero-spin SMBH), the star disrupts within the horizon and the event is not observable (e.g. \citealt{Coughlin2022}). By equating the order-of-magnitude tidal radius $r_\mathrm{t}$ and the capture radius we estimate that in the main sequence all our accreted stars disrupt outside the tidal radius. This is because the relatively low (Milky Way) mass of the SMBH in our simulations. For more massive SMBHs the tidal forces acting on a star would be weaker, and direct captures become more common. In the giant phase the disruption is further facilitated by the increased stellar radii, so direct captures are not expected in the latter phases of the simulations either.

\section{The milliparsec stellar population}\label{section: finalnumber-7}

\begin{figure}
\includegraphics[width=\columnwidth]{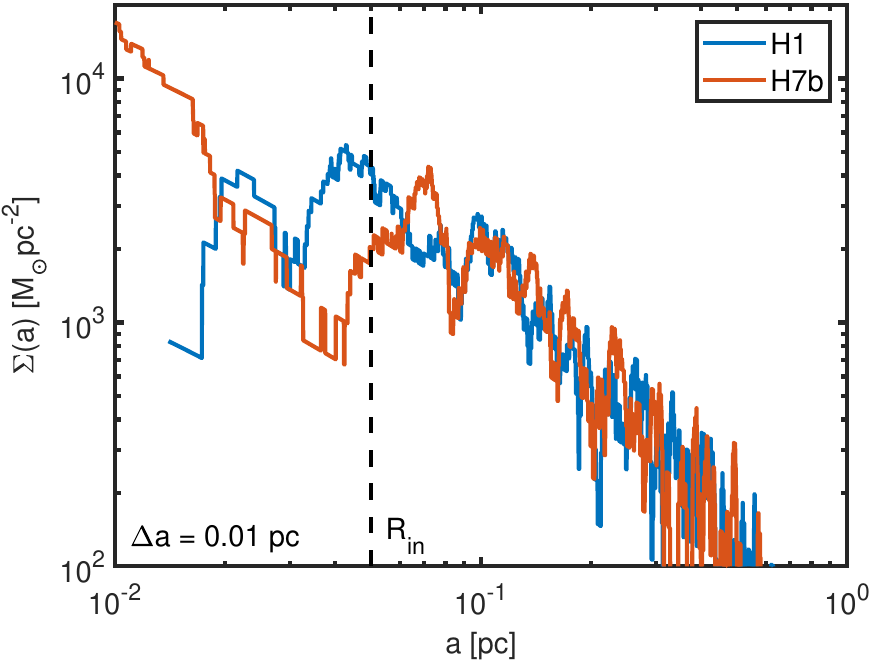}
\caption{The radial face-on surface mass density $\Sigma(a_\star)$ of the stars between the semi-major axis $a_\star=0.01$ pc and $a_\star=0.5$ pc in the runs H1 ($e_\mathrm{init}=0.1$, in blue) and H7b ($e_\mathrm{init}=0.7$, in orange) at $t=6$ Myr. The initial inner edge of the eccentric disk $R_\mathrm{in}=0.05$ pc is indicated with a vertical dashed line. The two surface density profiles are similar outside $R_\mathrm{in}$. Inside this radius, the surface density profile of the initially more circular model H1 first flattens and then begins to decrease within $a_\star\sim0.02$ pc. The initially more eccentric setup H7 has a local minimum in the surface density around $a_\star\sim0.04$ pc, inside which the surface density again increases. These milliparsec stars within this region originate from the Hills mechanism.}
\label{fig: radial-surfacedensity}
\end{figure}

\subsection{Radial surface density and eccentricity profiles}
During the simulation time a number of stars end up having a smaller semi-major axis around the SMBH than the original disk inner edge radius, $R_\mathrm{in} = 0.05$ pc. Hereafter, we term these stars the milliparsec stars. We present the radial surface density profile $\Sigma(a)$ of two of our models H1 ($e_\mathrm{init}=0.1$) and H7b ($e_\mathrm{init}=0.7$) in Fig. \ref{fig: radial-surfacedensity} after $t=6$ Myr of evolution. The radial bin size used is $\Delta a = 0.01$ pc. Outside $R_\mathrm{in}$, the two surface density profiles are similar. However, inside the initial disk inner edge the profiles differ. In the model H1, the surface density profile flattens inside $a_\star \sim 0.05$ pc, and finally drops inside $a_\star\sim0.02$ pc. The innermost stars in this model orbit at $a_\star = 0.014$ pc. However, the model H7b behaves qualitatively differently as the Hills mechanism has deposited stars at separations of few tens of milliparsecs from the SMBH. The model has a local minimum in the surface density profile near $a_\star\sim 0.04$ pc, inside which the surface density steadily increases as a power-law cusp.

The central concentration of stars in the run H7b within $a_\star \lesssim 0.04$ pc has a very steep three-dimensional density profile $\rho(a_\star) \propto a_\star^{-\alpha}$ with a slope of $\alpha\sim2.4$. This is roughly consistent with the results of \citep{Fragione2018}, who predict a power-law density profile $\rho(r) \propto r^{-\alpha}$ with $\alpha=9/4=2.25$ when Hills mechanism deposits stars around the SMBH. This central profile is steeper than the steady-state \cite{Bahcall1976} cusp which has a density profile slope of $7/4$.

\begin{figure}
\includegraphics[width=\columnwidth]{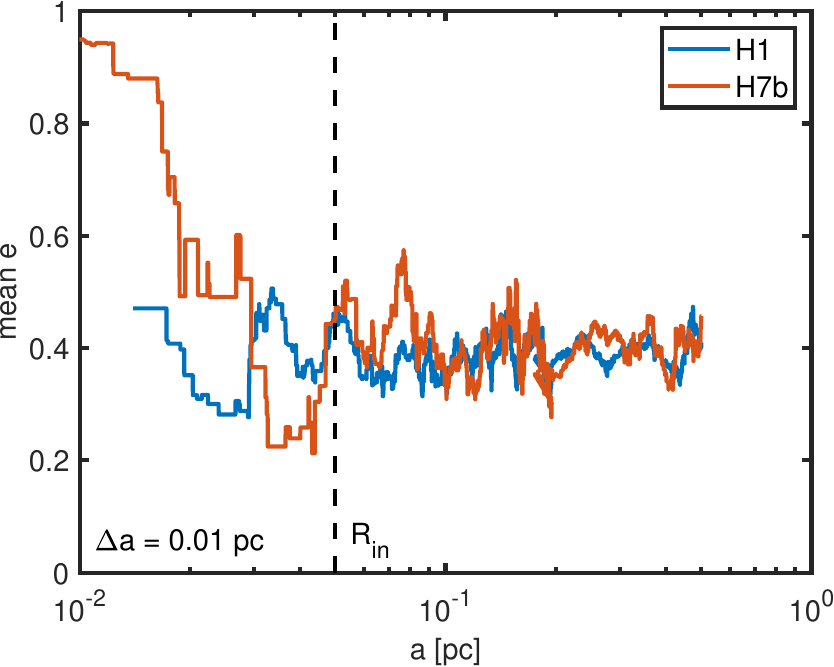}
\caption{The radial mean eccentricity profile of the simulations H1 ($e_\mathrm{init}=0.1$, in blue) and H7b ($e_\mathrm{init}=0.1$, in blue) at $t=6$ Myr averaged in bins with a width of $\Delta a=0.1$ pc. The initial inner edge of the disk is indicated with a vertical dashed line as in Fig. \ref{fig: radial-surfacedensity}. The initially more circular model H1 has a relatively constant mean eccentricity $\sim0.4$ at every part of the disk. In the model H7b the mean eccentricity has a similar value outside $R_\mathrm{in}$, but rapidly increases within $a_\star\lesssim0.04$ pc. This is again due to very eccentric stars deposited from binary systems into this region by the Hills mechanism.}
\label{fig: radial-ecc}
\end{figure}

Next, we study the radial mean eccentricity profile of the two models in the same $\Delta a = 0.01$ pc bins. The expectation is that in the models with a higher initial disk eccentricity the mean eccentricity of the milliparsec stars should be higher than in the initially more circular setups. This is because the disk migration origin milliparsec stars have similar mean eccentricity as the disk, and for the Hills mechanism origin milliparsec stars should be initially have $e_\star\sim0.98$. The mean eccentricity profiles are presented in Fig. \ref{fig: radial-ecc}. As expected, the model H1 has an almost constant mean eccentricity of $e_\star\sim0.4$ at all radii, including within $R_\mathrm{in}$. The mean eccentricity of the milliparsec stars in the model H1 is consistent with the main disk. However, in the model H7b the mean eccentricity strongly increases within $a_\star\lesssim0.03$--$0.04$ pc, reaching $e_\star\gtrsim0.9$ at the center. Again, this is consistent with the preferentially Hills mechanism origin of milliparsec stars in the model H7b. We note that the central mean eccentricity strongly depends whether PN effects are included in the simulations (H7b) or not (H7). This will be elaborated later in the study.

\subsection{The number of milliparsec stars}

\begin{figure*}
\includegraphics[width=\textwidth]{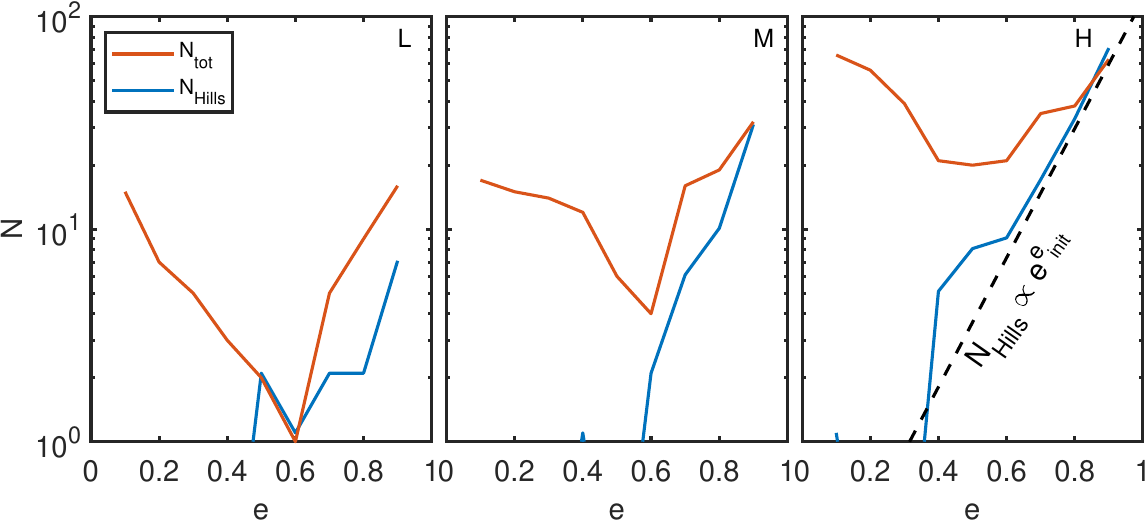}
\caption{The number of stars which resided at least $0.1$ Myr as a milliparsec star in the simulation setups L, M and H. The total number of stars is show in orange while the number of stars which originate from a Hills mechanism binary break-up is presented with the blue line. With initial disk eccentricities smaller than $e_\mathrm{init}\lesssim0.4$ there are no stars originating from the Hills mechanism. With higher initial eccentricities the number of Hills origin stars around the SMBH grows exponentially (dashed black line in the right plot) and is $N_\mathrm{Hills} \sim 70$ at maximum.}
\label{fig: hills}
\end{figure*}

In addition to producing high-velocity escaper stars, the Hills mechanism also deposits stars on tightly bound, very eccentric orbits around the SMBH. We trace the orbits of each star in the simulations from \bifrost{} snapshots (snapshot interval $\Delta t_\mathrm{snap} = 10^{-3}$ Myr) in order to find the stars which had semi-major axes smaller than $a_\mathrm{\star} < 0.04$ pc. The choice of the limiting semi-major axis of $a_\mathrm{\star} = 0.04$ pc is motivated by the local minimum of the surface density profile of the model with $e_\mathrm{init}=0.7$ in Fig. \ref{fig: radial-surfacedensity}. We also require that for the analysis the star has to reside within the milliparsec region for at least $t=100 \Delta t_\mathrm{snap}$. This procedure helps to filter out candidate stars which had an artificially small osculating $a_\star$ in the snapshot. We also trace the origin of the milliparsec to find out which fraction on the stars originated from a Hills mechanism binary break-up. For a star to be classified as a Hills origin milliparsec star we require that it was originally part of a binary system, and that its initial companion has escaped the simulated system with a high velocity ($v>v_\mathrm{esc}$).

The results of the milliparsec star tracing analysis are presented in Fig. \ref{fig: hills}. In the low-mass sample L, the total number of milliparsec never exceeds $N_\mathrm{tot} = 16$ stars. There are no stars of Hills origin in the models with a low initial eccentricity of $e_\mathrm{init}\lesssim0.5$, and the maximum number of the Hills mechanism origin milliparsec stars in the sample L is $N_\mathrm{Hills} = 7$ in the initially very eccentric models. It is also worthwhile to note that models with a low initial eccentricity produce more milliparsec stars of disk inner edge origin than more eccentric models.

The more massive simulation setups M and H have more milliparsec stars, including those with Hills mechanism origin, compared to the low-mass setup L, as expected. Disk models with initial eccentricities lower than $e_\mathrm{init}\lesssim0.4$ have no Hills origin milliparsec stars while the maximum number of disk inner edge origin stars is $N_\mathrm{disk} = 17$ and $N_\mathrm{disk} = 66$ in the models M9 and H9, respectively. With higher initial disk eccentricities the number of Hills mechanism origin milliparsec stars grows rapidly, and is in fact exponential as their numbers follow well the relation $N_\mathrm{Hills} \propto \exp{(e_\mathrm{init})}$. The maximum number of $N_\mathrm{Hills}$ reaches $N_\mathrm{Hills}=31$ and $N_\mathrm{Hills}=71$ in the setups M9 and H9.

\begin{figure}
\includegraphics[width=\columnwidth]{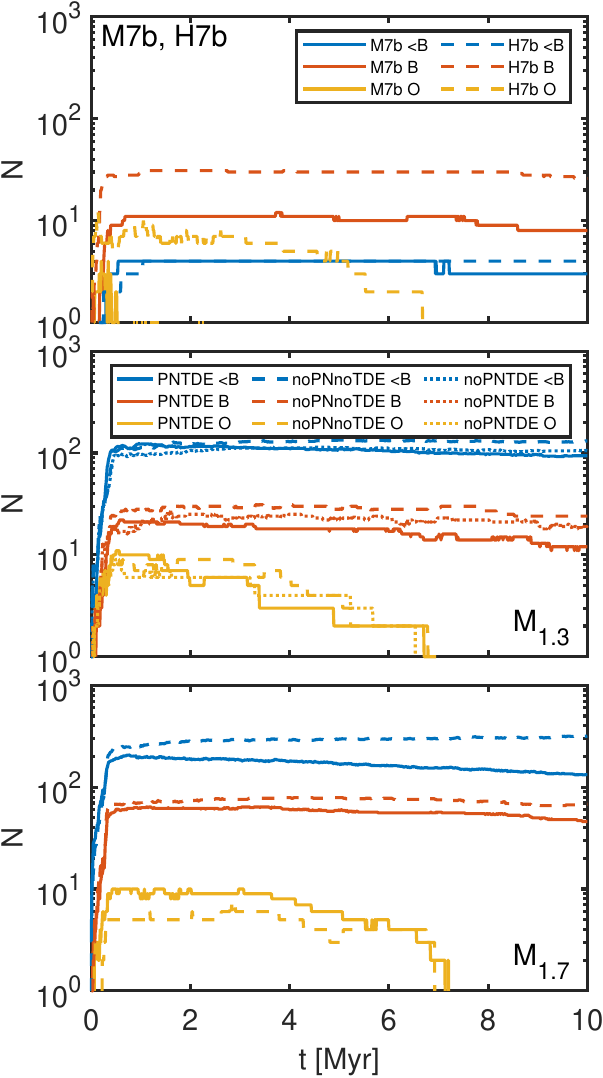}
\caption{The number of simulated O-type (O, in yellow), B-type (B, in orange) and late-type (<B, in blue) milliparsec stars as a function of time. The top panel shows the results of the setups M7b (solid lines) and H7b (dashed lines) The number of O-stars never exceeds $N_\mathrm{O}=12$ and reaches zero at $t=6.5$ Myr. In the setup M7b all the O-stars are early disrupted by the SMBH. The number of B-stars is close to constant $N_\mathrm{B}\sim10$ (M7b) and $N_\mathrm{B}\sim30$ (H7b) while the number of late-type stars is small, $N_\mathrm{<B}\lesssim4$. The middle panel shows the runs $\mathrm{M_\mathrm{1.3}}$ with a less extreme IMF ($\alpha=1.3$). The variant without both PN terms and TDEs (dashed line) has somewhat more stars in the milliparsec population than the runs which include TDEs (dotted line) and both PN and TDEs (solid line). The number of O-stars behaves as in the run H7b while the number of B-type stars is $N_\mathrm{B}\sim10$--$20$ in the run variant $\mathrm{M_\mathrm{1.3}PNTDE}$. The number of late-type stars is now larger than the number of B-stars, $N_\mathrm{<B}\sim100$. The bottom panel shows the results of the setup $\mathrm{M_\mathrm{1.7}}$ which are qualitatively similar as in setup $\mathrm{M_\mathrm{1.3}}$. The main difference of the two setups is the somewhat larger number of B-type and late-type stars in the setup $\mathrm{M_\mathrm{1.7}}$ due to its less top-heavy IMF.
}
\label{fig: newruns-N-scluster}
\end{figure}

Next, we study the effect of the IMF, PN equations of motion and including a TDE prescription on the number of milliparsec stars. The number of milliparsec stars in the simulation setups M7b, H7b, $\mathrm{M_{1.3}}$ and $\mathrm{M_{1.7}}$ is presented in Fig. \ref{fig: newruns-N-scluster}. As opposed to Fig. \ref{fig: hills}, here we only show the total number of milliparsec (regardless of their Hills or disk edge origins) but now as a function of time. We categorize the stars based on their mass into types <B ($m_\mathrm{\star}<2 \;\mathrm{M_\odot}$), B ($2 \;\mathrm{M_\odot}<m_\star<20 \;\mathrm{M_\odot}$) and O ($m_\star>20 \;\mathrm{M_\odot}$) as before.

The milliparsec population always has considerably more B-type stars than O-type stars. The number of of O-stars is always $N_\mathrm{O} \lesssim 12$ and decreasing rapidly with time as O-stars die due to single stellar evolution or get tidally disrupted by the SMBH. The last O-type stars in the milliparsec population are dead by $t\sim 6$ Myr -- $7$ Myr. We specifically highlight the simulation setup M7b in which all the few initial O-stars ($N_\mathrm{O}=4$) were disrupted by the SMBH before $t=0.5$ Myr, and the number of B-type milliparsec stars was $N_\mathrm{B} \sim 10$ until the end of the simulation at $t=10$ Myr. A larger simulation sample beyond a single realization of the each initial model would be required to assess how common the early O-star disruption scenario actually is.

The number of B-type stars in the milliparsec population ranges from $N_\mathrm{B}\sim10$ in the setup M7b to $N_\mathrm{B}\sim65$--$80$ in the setups $\mathrm{M_\mathrm{1.3}PNTDE}$ and $\mathrm{M_\mathrm{1.3}noPNnoTDE}$. In the runs with stellar accretion onto the SMBH enabled (label TDE in Fig. \ref{fig: newruns-N-scluster}), the number of stars of each type is somewhat smaller than without the TDE prescription. The PN equations of motion (label PN) result in a slightly smaller overall number of milliparsec stars. This is because PN precession suppresses relaxation effects and thus the Hills origin milliparsec stars retain their high eccentricities (and hence low pericenter distances) and are more susceptible to tidal disruption. This is further discussed in the Section \ref{section: results-ecc-pn}.

A unambiguous difference between the simulation setups with an extremely top-heavy IMF (M7b, H7b) and setups with only a moderately top-heavy IMF ($\mathrm{M_\mathrm{1.3}}$, $\mathrm{M_\mathrm{1.7}}$) is the number of late-type stars in the milliparsec population. While the extremely top-heavy models less than $N_\mathrm{<B}\lesssim5$ late-type stars, the moderately top-heavy models have $N_\mathrm{<B}\sim130$ and $N_\mathrm{<B}\sim300$, respectively.

\subsection{Milliparsec population eccentricity distribution: the importance of the PN equations of motion}\label{section: results-ecc-pn}

\begin{figure}
\includegraphics[width=\columnwidth]{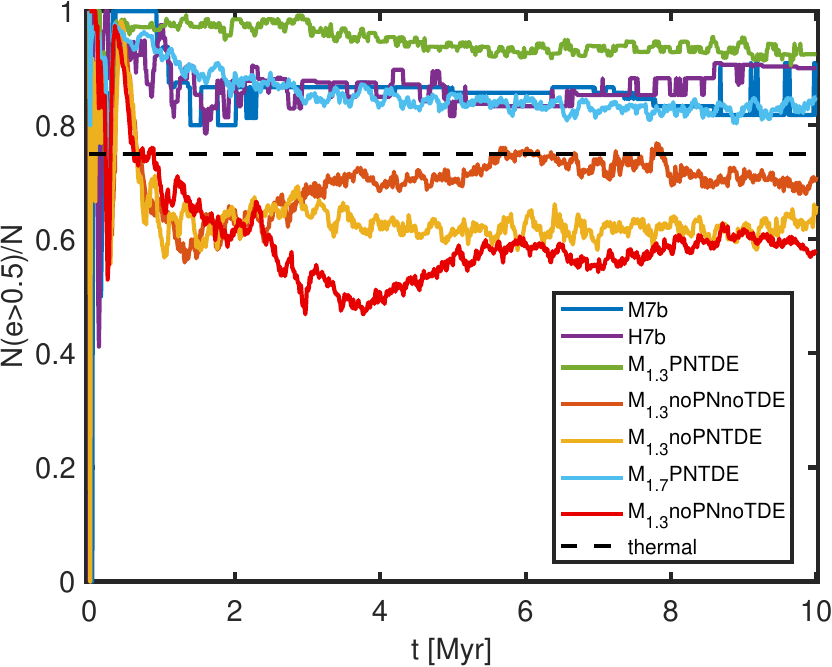}
\caption{The fraction of milliparsec stars $f_\mathrm{0.5} = N(e_\star>0.5)/N$ with eccentricities higher than $e_\star>0.5$ as a function of time in the various simulation setups. The thermal eccentricity distribution $f(e)=2e$ (dashed black line) has a fraction of $f_\mathrm{0.5}=3/4$. Regardless of the disk mass, the IMF and the TDE prescription, the simulation setups with PN equations of motion enabled (solid green, blue and purple lines) have a higher fraction of high-eccentricity stars than the thermal distribution. On the other hand, in the runs without PN (solid orange, yellow and red lines), the fraction of high-eccentricity stars is smaller than for the thermal distribution. This is due to the initially very eccentric orbits of the Hills origin milliparsec stars and the suppression of relaxation effects due to the PN precession, as discussed in the text.
}
\label{fig: newruns-ecc-scluster}
\end{figure}

We characterize the evolution of the eccentricity distribution of the models M7b, H7b, $\mathrm{M_{1.3}}$ and $\mathrm{M_{1.7}}$ in Fig. \ref{fig: newruns-ecc-scluster}. As the number of milliparsec stars is initially ($t\lesssim0.2$ Myr) small and the distribution does not always necessarily follow an exponential distribution $F(e) \propto e^{\beta}$, we instead use a non-parametric proxy for the eccentricity distribution. Namely, we investigate the fraction of eccentric milliparsec stars which we define as $f_\mathrm{0.5} = N(e_\star>0.5)/N$, in which $N$ is the total number of stars. For the common thermal eccentricity distribution $f_\mathrm{0.5} = 3/4$. 

The fraction of high-eccentricity stars $f_\mathrm{0.5}$ is high at the time of the milliparsec population formation ($t\sim0.5$ Myr), reflecting the predominant Hills origin of milliparsec stars in the simulations. At later times whether the PN equations of motion are enabled in the simulations determines how the fraction of high-eccentricity stars evolves. Initial disk mass, number of milliparsec stars, TDE prescription and the IMF have a smaller effect than the PN. Without PN, the fraction of high-eccentricity stars ranges becomes smaller than the thermal fraction of $3/4$ already at $t\sim1$ Myr, and ranges afterwards from $f_\mathrm{0.5} = 0.46$ to $f_\mathrm{0.5} = 0.75$. While the semi-major axis and hence the orbital energy of the milliparsec stars change very slowly, the norm of the angular momentum and thus eccentricity evolve rapidly. This is characteristic for scalar resonant relaxation (e.g. \citealt{Merritt2011}). With PN, the fraction of high-eccentricity stars always remains high, $f_\mathrm{0.5}>0.8$. This is because PN precession effects efficiently suppresses SRR effects. 

\begin{figure}
\includegraphics[width=\columnwidth]{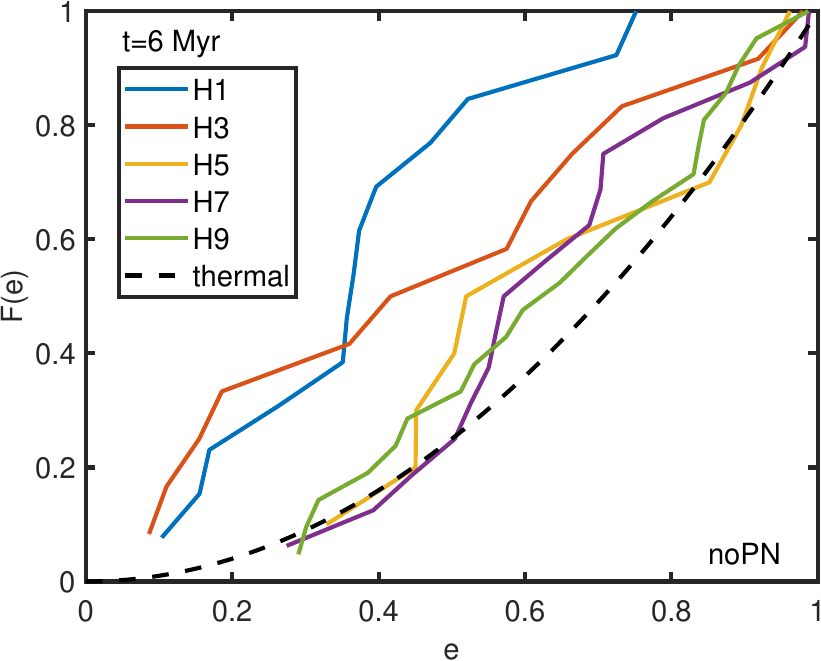}
\caption{The cumulative eccentricity distributions $F(e)$ of the milliparsec stars in the simulation sample H at $t=6$ Myr. Even though most of the stars originate from the Hills mechanism in runs H7 and H9, the milliparsec stars thermalize in $t=6$ Myr. Initially more circular setups H1 and H3 have a clearly sub-thermal distribution, and the most circular model H1 does not have stars with $e_\star>0.75$. These runs do not include PN equations of motion and relaxation effects are overestimated as discussed in the main text. Simulations including PN effects are presented in a later figure.}
\label{fig: ecc-thermal-Hsample-noPN}
\end{figure}

\begin{figure}
\includegraphics[width=\columnwidth]{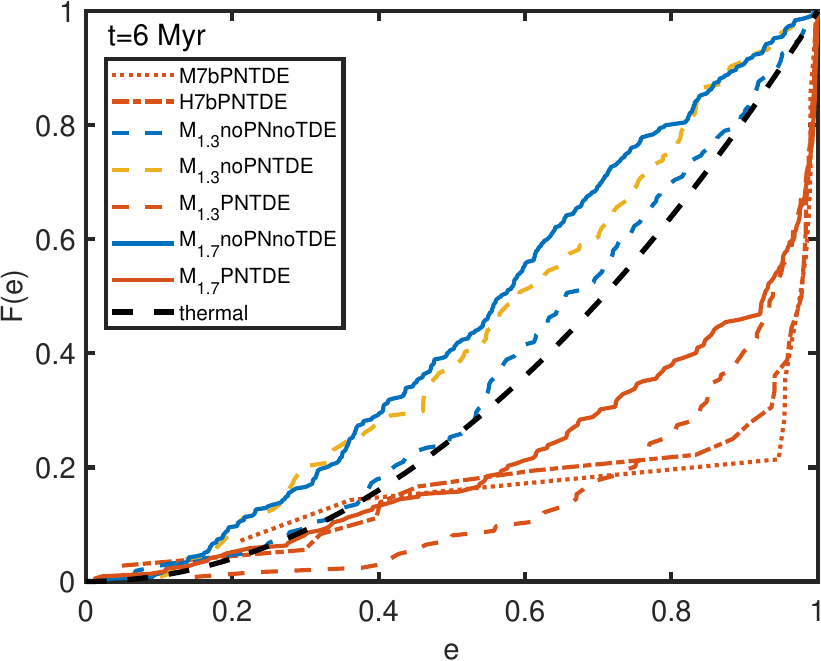}
\caption{The cumulative eccentricity distribution of $F(e)$ of the milliparsec stars at $t=6$ Myr. The thermal distribution $F(e)=e^2$ is indicated using the dashed black line. With the PN effects enabled (orange lines), the VRR is suppressed and the eccentricity distributions remain superthermal. The eccentricity distributions are the most concentrated to very high eccentricities ($e_\star>0.9$) with the setups including extremely top-heavy IMFs (M7b, H7b). With less extremely top-heavy IMFs ($\mathrm{M_\mathrm{1.3}PNTDE}$ and $\mathrm{M_\mathrm{1.7}PNTDE}$), the eccentricity distribution becomes gradually less and less superthermal. Without PN (blue and yellow lines), relaxation proceeds and the resulting eccentricity distributions are subthermal. Whether TDEs are included or not appear to have only a small effect on the overall results.}
\label{fig: newruns-Fe-scluster}
\end{figure}

We present the cumulative eccentricity distribution $F(e)$ of the milliparsec stars at $t=6$ Myr of the runs H1--H9, M7b, H7b and the in total five variants of the setups $\mathrm{M_{1.3}}$ (three variants) and $\mathrm{M_{1.7}}$ (two variants) in Fig. \ref{fig: ecc-thermal-Hsample-noPN} and in Fig. \ref{fig: newruns-Fe-scluster}. In the set of nine runs H1--H9 without PN, the cumulative eccentricity distribution is always more circular that the $F(e) = e^2$ thermal distribution. The models with a large number of initially very eccentric milliparsec stars H7 and H9 have relaxed to have a mildly subthermal eccentricity distribution. The initially more circular models H1 and H3 are strongly subthermal, and especially the model H1 includes no milliparsec stars with eccentricities higher than $e_\star\gtrsim0.75$. In Fig. \ref{fig: newruns-Fe-scluster}, in the runs with PN equations of motion enabled, the eccentricity distribution remains strongly superthermal, i.e. there is little evolution after the formation of the milliparsec population as the scalar resonant relaxation effects are suppressed. With PN, simulation setups with more top-heavy IMFs result in more superthermal eccentricity distributions at $t=6$ Myr. Without PN, the eccentricity distributions of the milliparsec stars are close to thermal ($\mathrm{M_{1.3}noPNnoTDE}$) or subthermal. The less top-heavy IMFs result in more subthermal eccentricity distributions. The effect of including or not including the TDE prescription appears to have only a small effect on the eccentricity distributions.

To summarize, the inclusion of the PN equations of motion in the simulations has a profound effect on the evolution of the eccentricity distribution on the milliparsec stars. With PN, the Hills mechanism origin milliparsec stars can retain their high eccentricity as the vector resonant relaxation effects are properly suppressed. Not including PN equations of motion in the simulations can result in close to thermal eccentricity distributions, such as in the simulation run $\mathrm{M_{1.3}noPNnoTDE}$. We emphasize that PN effects at least of the order of PN1.0 should always be included in the simulations of stars around SMBHs at milliparsec separations to properly capture the suppression of the resonant relaxation effects.

\subsection{The (an)isotropy of the milliparsec stellar population: the importance of PN and background modeling}\label{section: results-inc-pn}

\begin{figure*}
\includegraphics[width=0.8\textwidth]{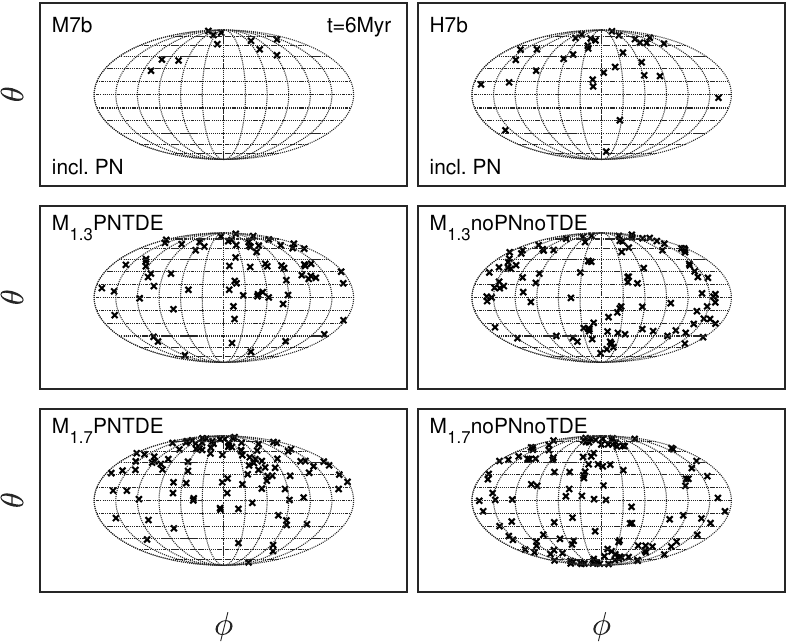}
\caption{The angular momentum vector directions of the milliparsec stars at $t=6$ Myr. The original angular momentum vector directions lie at the top of each panel. The extremely top-heavy models ($\alpha=0.25$) in the top row (especially M7b) yield disky, anisotropic milliparsec populations. Models without PN equations of motion are more isotropic than their PN counterparts because of the missing suppression of the relaxation effects. There milliparsec populations become increasingly is isotropic with less top-heavy IMFs (increasing $\alpha$) of the setups. As further discussed in the text, modeling the spherical background as live particles instead of a smooth external potential would enhance relaxation effects, rendering the milliparsec stars more isotropic. 
}
\label{fig: newruns-inc-scluster}
\end{figure*}

We show the distribution of the angular momentum vector directions of the milliparsec stars at $t=6$ Myr in Fig. \ref{fig: newruns-inc-scluster}. The two extremely top-heavy models ($\alpha=0.25$) yield either disky (M7b) or lopsided (M7b) angular momentum vector direction distributions. The milliparsec stars become more isotropic with less extremely top-heavy IMF, i.e. with increasing $\alpha$. Comparing the simulations without PN equations of motion to their PN counterparts, we see that without PN, the milliparsec star populations are more isotropic. In general, relativistic precession effects tend to suppress relaxation and oscillation effects, such as the von Zeipel -Livod-Kozai (ZLK) oscillations \citep{vonZeipel1910,Lidov1962,Kozai1962,Ito2019} which can exchange orbital eccentricity for inclination for hierarchical systems. 

In this study, the spherical background cusp potential is represented as a smooth external potential instead of live particles in order to limit the computational costs. Including a live population of particles representing the old stellar population around the SMBH would enhance the relaxation effects, especially VRR \citep{Panamarev2022}, and make the milliparsec stellar populations more isotropic than in the runs with a smooth external background. It has been shown in idealized simulation setups that relaxation effects from the background stellar population can rapidly render the stars at milliparsec sepatations from the SMBH isotropic, even when PN effects are properly modeled \citep{Hamers2014}.

\section{Discussion}\label{section: finalnumber-8}

\subsection{Comparison to observations -- the Milky Way center stellar disks}

\subsubsection{Background and motivation}

Even though stellar disks within the gravitational influence radii of SMBHs are presumed to be common in the galactic nuclei in the local Universe, resolved stellar disk dynamics at the level of single stars can only be observed from the nucleus of the Milky Way galaxy. The second-closest galactic nucleus, the center of M31, contains a lopsided eccentric disk structure, but it is very massive compared to the surrounding spherical stellar cusp ($M_\mathrm{disk}\gg M_\mathrm{cusp}$). Thus, comparison of the M31 eccentric disk dynamics to the eccentric disk simulations of this study for which $M_\mathrm{disk}\lesssim M_\mathrm{cusp}$ is not very meaningful. Hence, the Milky Way center is the only available comparison for our simulations. 

Recent observations indicate that the central parsec contains two almost perpendicular eccentric disk structures (CW and CCW disks, e.g. \citealt{Bartko2009}) and possibly two other outer filamentary structures \citep{vonFellenberg2022}. Obviously, our simulation results starting from initial conditions with a single asymmetric eccentric disk are far too simplistic to explain these complex disk structures of young, massive stars. However, we can still compare a number of observable properties of the individual disks to our simulation results. We focus on the mainly on the CW disk and to a lesser extent to the CCW disk and the outer filamentary structures. 

The main disk structural parameters varied in our asymmetric eccentric initial conditions are the initial disk mass, the initial disk eccentricity and the IMF of the stellar population. Our full simulation sample probes the initial disk mass range from $M_\mathrm{disk}=10^4\;\mathrm{M_\odot}$ to $7.5\times M_\mathrm{disk}=10^4\;\mathrm{M_\odot}$. Due to stellar evolution, extremely top-heavy disk models can lose up to $50\%$--$90\%$ of their mass in $t=6$ Myr, as discussed in Section \ref{section: ic-motivation}. Thus, the initially massive disk models have masses corresponding to the present-day CW and CCW disk masses at later times. The initial eccentricities cover the almost entire parameter space from almost circular to very eccentric disks, i.e. $0.1\leq e_\mathrm{init}\leq0.9$. The IMF determines the number of stars in the disk, as well as number ratios of stars of different masses and spectral types. We tested both extremely top-heavy (power-law slope of $\alpha=0.25$) and less top-heavy IMFs ($\alpha=1.3$, $\alpha=1.7$). For most of the comparisons we choose the simulation state at $t=6$ Myr which is consistent with the age estimates of the young, massive stars in the Milky Way center. Unless otherwise stated, the observed disk structural parameters are from \cite{vonFellenberg2022}.

\subsubsection{Disk star eccentricity distribution}\label{section: discussion-disk-eccentricity}

We have shown in Fig. \ref{fig: ecc-histogram} that the late-time eccentricity distribution of the disk stars is to a large extent almost independent of the initial eccentricity of the disks. This is despite of the fact that the initial eccentricity distributions are very narrow and distinct from each other. Initial disk setups with eccentricities of $0.1\leq e_\mathrm{init} \leq 0.6$ result in qualitatively very similar eccentricity distributions with a broad peak around $0.2\lesssim e_\star \lesssim 0.5$. In the setups with a higher initial disk eccentricity than $e_\mathrm{init}\gtrsim0.6$, a secondary peak of high-eccentricity stars appears. This feature becomes more prominent with increasing initial disk eccentricity. The simulated eccentric disks at $t=6$ Myr are consistent with the observed eccentricity distributions of the CW and CCW disks and the outer filaments. The CW and CCW disks have a median eccentricity of $0.4$--$0.5$ while the outer filaments have a somewhat higher median eccentricity, $0.7$. The CCW disk also contains high-eccentricity stars, but not beyond $\gtrsim0.9$. However, the highest initial eccentricity of $e_\mathrm{init}=0.9$ produces a large number of extremely eccentric stars ($e_\star>0.9$) which appear to be missing from the Milky Way center disks, and it is thus disfavoured. Based on these results, it is difficult to infer the initial disk eccentricity $6$ Myr ago with certainty from the CW disk eccentricity distribution observed today.

\subsubsection{The missing disk warps}\label{section: warp}

The CW disk appears to be significantly warped in both angles on the sky \citep{Bartko2009,Genzel2010}. The angle between the angular momentum vectors at the inner edge of the CW disk compared to its outer edge is $\sim60\degree$. 

Our simulated disks do not show any strong warp features. The disks of simulated O-type stars show warp-like features in Fig. \ref{fig: newruns-pr-xz} in the run $\mathrm{M_{1.7}}$ and possibly in H7b, but both of these features are weak. Hence, we conclude that the simulation setups of this study do not reproduce the kind of warped disk features observed in the Milky Way center CW disk.

There are two plausible explanations for the missing disk warps in our simulations, both connected to the chosen initial conditions. First, our initial setups only include a single stellar disk embedded in a smooth spherical external background. If the disk stars were all test particles, their orbital planes would be conserved, also preventing any warp structure formation. In addition, the Milky Way center has at least two sub-parsec scale disks, the almost perpendicular and coeval CW and the CCW disks. It has been shown using N-body simulations that the interaction of such two disks can lead to significant warping of the disks \citep{Löckmann2009}.

Second, \cite{Kocsis2011} have shown using analytical arguments and Monte Carlo simulations that an initially circular, razor-thin disk embedded in a spherical background stellar population can rapidly develop a significant warp due to vector resonant relaxation. Spiral structure and warp formation in disks is also possible in N-body simulations, e.g. \cite{Perets2018} with self-consistent field methods. This brings back the question of the modeling of the spherical background in our simulations. In this study, we modeled the background using a smooth static external potential instead of a live cusp of stellar particles for computational cost reasons. Thus, our simulations lack the VRR effects on the disk stars originating from the spherical background. We already discussed the need of including a live background population in the context of the simulated milliparsec star eccentricity distribution in Section \ref{section: results-ecc-pn} and angular momentum vector direction distributions in Section \ref{section: results-inc-pn}. The missing disk warp issue adds a third motivation to pursue these significantly computationally more expensive simulations in the future.

\subsubsection{Disk stellar population and the IMF}

The observed numbers of different type of massive and evolved stars can be used to constrain the IMF and age of the Milky Way center young stellar population. Both the ratio of O/WR stars to B stars ($f_\mathrm{OWR-B} = N_\mathrm{O/WR}/N_\mathrm{B}$) and the ratio of O and B stars to WR stars $f_\mathrm{OB-WR} = N_\mathrm{OB}/N_\mathrm{WR}$ are sensitive to the IMF and the age of the stellar population (e.g. \citealt{Lu2013}).

In our simulations, the disk stellar type content at the observed age of the CW disk is largely determined by the IMF and the assumed solar metallicity. It should be noted that the definition of a B-type or an O-type star in this study is extremely simplified and only based on the initial mass of the star. The initial conditions of the simulation sets M and H with an extremely top-heavy IMF ($\alpha=0.25$) are constructed to have an equal number of type B and O-stars after $t\sim6$ Myr of evolution. Consequently, all our other models with a less top-heavy IMF ($\alpha=1.3$ in $\mathrm{M_{1.3}}$ and $\alpha=1.3$ in $\mathrm{M_{1.7}}$) will have an excess of B-stars compared to the number of O-stars which is not observed in the Milky Way center \citep{Perets2010}. While the models M and H have more O-type stars than B-type stars in the disks at $t\lesssim6$ Myr and thus qualitatively match the observed O/WR-B ratio in the disk, a more profound question is whether these models can simultaneously match the other observational constraints from the Milky Way center.

Finally, a crucial assumption in setting up our initial conditions was that the IMF is constant everywhere in the disk from the inner edge of $R_\mathrm{in}=0.05$ pc to the outer edge of $R_\mathrm{out}=0.5$ pc. Recent observations show that in the Milky Way center, the inner disk structures have a higher O-B star ratio than the outer filaments \citep{vonFellenberg2022} while the IMF slope of all the stars within the central $\sim0.5$ pc is certainly higher than $\alpha=0.25$ \citep{Lu2013}. The possibility of an IMF gradient is also consistent with a number of predictions that the IMF could have been more top-heavy closer to the SMBH than in the outer parts of the star-forming gas disk \citep{Hobbs2009}.

Including an IMF gradient in our disk setup will open new possibilities to better satisfy the various observational constraints from the Milky Way center. Hence, the IMF gradient in the initial disk is an important step towards better-motivated numerical simulations of the Milky Way center disks in the future. The key question here is whether a simulation with an initial IMF gradient simultaneously reproduce the O-B number ratios in the disks and more other observational constraints than a simulation with a fixed IMF slope.

\subsubsection{Binary stars in the disks}
Little is known about the binarity of stars in the Milky Way center. Still, a few stars have been undoubtedly identified as binary stars, most of which are eclipsing binaries with O/WR components within the central $0.2$ pc (\citealt{Ott1999,Martins2006,Rafelski2007,Pfuhl2011}). No binary stars have been detected in the S-cluster region (e.g. \citealt{Chu2018}). There are indications of decreasing binary fraction below $r\lesssim0.02$ pc from the central SMBH compared to the field binary fractions of massive stars \citep{Chu2023}, which is consistent with the simple theoretical expectation that SMBHs unbind binary stars or make them merge into single stars. 

Our simulations in this study indicate that the binary fraction of massive stars the disks is decreased from further away, at $r\lesssim 0.1$--$0.2$ pc. Further away from the SMBH, the binary fraction remains unchanged and is $f_\mathrm{b}\sim0.5$ at the disk age of $t=6$ Myr. However, these results strongly depend on the assumptions of the binary population properties in the disk initial conditions. Still, the results show a possibility that the binary fraction of the massive stars in the CV and CCV disks and outer filaments in the Milky Way center at the present day could be moderate or high.

\subsection{Comparison to observations -- the S-stars in the Milky Way center}

\subsubsection{O-type stars and the S-cluster}
The innermost massive O/WR stars in the Milky Way center are located outside the S-cluster, orbiting at the sharp inner edge of the present-day CW disk. In our simulations, the milliparsec stars in this region originate both from the Hills mechanism (initially high eccentricity) and from the inner edge of the disk (initially low to moderate eccentricity). Both mechanism bring O-type stars in the young, central milliparsec population predominantly before $t=1$ Myr. A number of solutions to the missing O/WR stars in the S-cluster have been proposed. First, the S-cluster could be old enough that the O-stars are already dead. However, this does not explain why the CW disk prominently still contains O-type stars if the S-cluster and the CW disk are indeed coeval. A refined solution is that the O-stars in the S-cluster have already entered the giant phase of their evolution, and have been tidally disrupted by the SMBH. As another related and more complex solution, \cite{Chen2014} proposed a formation of a so-called rapidly evolving region (RER) near the inner edge of the disk which can drive massive stars migrate close to the SMBH due to ZLK-like resonances and are tidally disrupted. Finally, it has been noted that the Hills mechanism deposits more massive (and thus more O-type) captured companions at larger separations from the SMBH \citep{Generozov2020}.

Most of the O-stars in our simulations that end up milliparsec population via the Hills mechanism have tidally disrupted either in the main sequence or in the giant phase and, or are massive enough and simply have died of old age by $t=6$--$7$ Myr. This is illustrated in Fig. \ref{fig: newruns-N-scluster}. One of our simulations, M7b, has all its milliparsec O-type stars tidally disrupted already early in its evolution. A simulation sample far larger than included in this study would be needed to assess how common this early disruption scenario actually is.

\subsubsection{The number of S-stars and the O/WR-B ratio of the disks}
The S-cluster is observed to contain approximately $30$--$40$ B-type stars (e.g. \citealt{Gillessen2017}) at the inferred age of $t=6$ Myr of the disks \citep{Paumard2006}. Our simulation that produce milliparsec stars typically have from a few stars up to $N\sim80$ milliparsec stars. The number of milliparsec stars in our simulations depends on multiple parameters. In the following the notation $P\uparrow$ means that the increasing the value of the parameter $P$ or enabling physical process $P$ increases the number of milliparsec stars. The notation $P\downarrow$ indicates the opposite. The most important parameters affecting the number of milliparsec stars are the initial disk eccentricity ($e_\mathrm{init}\uparrow$, exponentially), initial disk mass ($M_\mathrm{disk}\uparrow$), IMF ($\alpha\uparrow$) and binary star population properties (e.g. $a_\mathrm{b}\downarrow$), the inclusion of the PN equations of motion (PN $\downarrow$) and the TDE prescription (TDE $\downarrow$). Thus, we conclude that our models can produce milliparsec stellar populations with a comparable number of stars as in the S-cluster in the Milky Way center. 

We emphasize the most important question about the number of milliparsec stars in our simulations is not whether a model produces an exact match to the observed number of S-stars. Rather we argue that a more relevant question is whether a model can produce a correct number of S-stars while simultaneously also matching the numbers of O and B-type stars and their ratio at $t=6$ Myr. The same applies for other observational constraints from the Milky Way center disk structures. The work in the literature has not extensively focused in this important point while studying the Hills channel of S-cluster formation via direct numerical simulations \citep{Perets2010}. It seems that there is no easy, evident solution for the problem.

The initial conditions of the simulation sets M and H with an extremely top-heavy IMF ($\alpha=0.25$) are constructed to have approximately an equal number of type B and O-stars ($100$--$200$) in the disk after $t=6$ Myr of evolution. The specific models M7b and H7b yield $N_\mathrm{B}\sim10$ and $N_\mathrm{B}\sim30$ B-type stars in the milliparsec region, and as demonstrated above their numbers can be adjusted up or down by changing the various parameters in new runs. We argue that these runs are our best match in both the number of S-stars and the numbers of B and O stars and the O-B ratio in the eccentric disk. 

\subsubsection{IMF gradients in the initial disk?}
It is probable that the IMF slope $\alpha$ in the central parsec of the Milky Way is not as top-heavy as $\alpha=0.25$ used in our most top-heavy models. As the models have an equal number of O-stars and B-stars approximately at $t=6$ Myr, a population consisting predominantly of O/WR stars at that time as observed in the CW disk \citep{vonFellenberg2022} would require a close-to-flat IMF. The most top-heavy IMF in the literature for the central disks is $\alpha=0.45\pm0.3$ \citep{Bartko2010}. Less top-heavy IMF slopes have been proposed, e.g. $\alpha=0.85$ \citep{Paumard2006} and $\alpha=1.7\pm0.2$ \citep{Lu2013}. We find no apparent radial mass segregation in the eccentric disks during the simulation runs i.e. more massive O-stars do not migrate towards the inner edge of the disk as shown in Fig. \ref{fig: newruns-pr-xy}. This suggests that observed O/WR-B ratio in the CV disk might be primordial and not caused by stellar migration. An IMF gradient suggested by hydrodynamical star formation simulations (e.g. \citealt{Hobbs2009}) is an attractive solution to the O/WR-B ration and the global IMF slope issue. Near the inner edge of the disk the IMF is almost flat ($\alpha\approx0$ with a cutoff $m_\mathrm{max}$) and the stars there are predominantly O-type stars. Moving outwards from the inner the disk $\alpha$ increases, resulting in a less extreme top-heavy global IMF in the central parsec. Performing simulations using disk initial conditions including an IMF gradient is an interesting prospect and we will perform such simulations in future work.

\subsubsection{Late-type stars in the S-cluster}
Main sequence late-type stars (in the sense of later type than B) i.e. stars below the mass of a few $\mathrm{M_\odot}$ cannot be yet resolved as individual stars in observations of the Galactic center. Our simulations show that the number of late-type stars is strongly dependent on the IMF of the initial disk. As demonstrated in Fig. \ref{fig: newruns-N-scluster}, there are less late-type stars than B-type stars if the IMF is extremely top-heavy ($\alpha=0.25$) by a factor of $\sim3$--$5$. On the contrary, in the simulations with a less extremely top-heavy IMF ($\alpha=1.3$ and $\alpha=1.7$) there are more late-type stars than B-stars in the milliparsec population. The ratio of the number of late-type stars and the B stars $N_\mathrm{<B}/N_\mathrm{B}$ decreases from $N_\mathrm{<B}/N_\mathrm{B}\sim5$ to $N_\mathrm{<B}/N_\mathrm{B}\sim3$ when the IMF slope increases from $\alpha=1.3$ to $\alpha=1.7$. Thus, the number of late-type stars in the S-cluster, if possible to directly observe in the future, can provide useful information from the IMF of the Milky Way center disks.

\subsubsection{S-stars and the original disk eccentricity}
We noted in Section \ref{section: discussion-disk-eccentricity} that based on the observed present-day CW and CCW disk eccentricity distributions it is difficult to infer the original eccentricity distribution of the disks at the time of their births $6$ Myr ago. This is because almost all of our initial disk setups yield eccentricity distributions comparable to the observed distribution after $6$ Myr of evolution despite the very large differences of the initial models. Only the most eccentric models with $e_\mathrm{init} = 0.9$ are disfavoured as they result in a very large number of stars with $e_\star>0.9$, which is not observed in the disks today. On the other hand, in order to Hills mechanism to operate in the simulations, binary star center-of-masses on highly eccentric orbits are required. In our simulations, Hills origin milliparsec population stars begin to be produced when the initial eccentricity of the disk exceeds $e_\mathrm{init}\gtrsim0.5$--$0.6$. Based on these facts, an approximate simulation-based constraint can be placed on the initial disk eccentricity, $0.6\lesssim e_\mathrm{init}\lesssim0.9$. It should be noted that this simulation constraint is relies on the assumption that Hills mechanism is the source of a majority of young stars in the S-cluster.

\subsubsection{The eccentricity and angular momentum vector direction distributions: the need for a live background stellar cusp}
The fact that the eccentricity distribution of our simulated milliparsec stars is too superthermal and their angular momentum direction distribution anisotropic too compared to observations of the S-cluster is not necessarily an outstanding problem. This is because of the missing live stellar background population in this study and the use of a smooth static spherical external potential instead. It has been established in the literature in simplified models that the spherical background can render the inclination (angular momentum direction) distribution isotropic \citep{Subr2016} and relax the initially very eccentric stellar orbits close to the observed eccentricity distribution \citep{Hamers2014}. As already discussed in the Section \ref{section: warp}, the vector resonant relaxation from the background could also naturally alleviate the missing disk warp issue in our simulations. Including a live spherical background population is thus an utmost priority for future simulation studies.

\subsection{Comparison to previous simulation work}

\subsubsection{Asymmetric eccentric disk instability mechanism}

Numerical eccentric disk studies in the mass range $M_\mathrm{disk} \lesssim M_\mathrm{cusp}$ have often focused in studying systems resembling one of the disky structures and the central S-star cluster in the Milky Way center. However, only relatively few self-consistent simulations including the evolution and the disruption of the eccentric disks and following the formation of the S-cluster via the Hills mechanism have been reported in the literature. 

Representative simulation studies such as \cite{Madigan2009} or more recently \cite{Generozov2020} and \cite{Generozov2021} start from a single eccentric disk with a mass comparable to the present-day Milky Way center CW disk ($M_\mathrm{disk}\sim10^4\;\mathrm{M_\odot}$). After the eccentric disk instability the most eccentric orbits are identified and populated with binary stars. The binary star encounters with the SMBH are then integrated in isolation using a few-body code. Importantly the chosen initial disk mass is the inferred present-day Milky Way center CW disk mass, even though the disk mass at its formation $4$ Myr -- $8$ Myr ago was higher due to subsequent stellar mass loss and deaths of massive stars. The mass loss in $6$ Myr can be substantial, especially if the IMF of the original stellar population was extremely top-heavy. We have shown in Section \ref{section: disk-instability1} and Section \ref{section: disk-instability2} that if the initial mass of the asymmetric eccentric disk is higher than few times $\sim10^4\;\mathrm{M_\odot}$, its exact disruption mechanism differs from the standard secular eccentric disk instability of \cite{Madigan2009}. Most importantly, the assumption of the negligible contribution of the disk on the precession rates of the stars breaks down. The disk self-contribution has important non-trivial consequences. The asymmetric eccentric disk is still unstable in this intermediate-mass regime $M_\mathrm{disk} \sim M_\mathrm{cusp}$ within a certain radius, as opposed to the stable case $M_\mathrm{disk}\gg M_\mathrm{cusp}$ as observed in the M31 center. The nature of the instability in the intermediate-mass disk regime is a precession-direction instability: the asymmetric eccentric disk is very rapidly ($t\lesssim0.6$ Myr) disrupted as different parts of the disk precess at different directions. While the outer parts of the asymmetric eccentric disk precess to the retrograde direction just as in Madigan's secular eccentric disk instability, the inner parts of the disk precess in the prograde direction due to torques from the disk itself. A curious consequence of the precession direction instability is the appearance of a short-lived right-handed one-sided spiral pattern in the disrupting disk in Fig. \ref{fig: spiral} due to precessing elliptical orbits of the disk stars.

\subsubsection{TDE rates in eccentric disk setups}
We note that TDE rates from our simulations can appear somewhat lower than in the eccentric instability studies in the literature, namely \cite{Madigan2018}, who report a very high TDE rate of $0.3$--$3\; \mathbf{yr}^{-1}$ per galaxy. In our simulation models, the TDE rates range roughly from $10$ Myr$^{-1}$ up to $1000$ Myr$^{-1}$ depending on the initial setup.

There are several reasons for the difference. Briefly, \cite{Madigan2018} assumes a significantly more massive stellar disk ($M_\mathrm{disk} = 10^6\;M_{\odot}$) than we have adopted for this study, and the given TDE rates are averaged over somewhat smaller time windows ($0.03$ Myr--$0.3$ Myr) than in our study. In more detail, first, their TDE rate estimate does not originate directly from a simulation, but is based on a number of assumptions, namely the M31-inspired SMBH-disk mass ratio yielding $M_\mathrm{disk} = 10^6\;M_{\odot}$. Next, based on numerical simulations, it is estimated that $\sim10\%$ of the stars (roughly $N_\star \sim 10^5$) could disrupt during $10^2$--$10^3$ orbital periods. As the orbital period at the initial inner edge of the disk is $P_\star \sim 300$ yr, the estimate of $0.3$--$3\; \mathbf{yr}^{-1}$ per galaxy is reached.

In our simulations, the ratio of disrupted stars to the total number of stars in the simulation typically ranges from $\sim5\%$--$10\%$, a value only slightly lower than adopted by \cite{Madigan2018}, due to our more detailed TDE procedure. Thus, the differences of the TDE rates can be fully attributed to the different assumed IMFs and chosen initial disk masses, as both less top-heavy IMFs and more massive disks provide more stars to be disrupted by the SMBH.

\subsubsection{The importance of PN effects}

The work closest resembling the approach of our study is \cite{Subr2016}. The authors start with a disk setup with a top-heavy IMF and a high binary fraction, but include no external background potential, stellar evolution or PN equations of motion. The asymmetric eccentric disk unstable due to axisymmetric planar von Zeipel-Lidov-Kozai effect and produces eccentric binary center-of-mass orbits and results in a formation of a star cluster within the inner edge of the disk. However, we note that in simulations studying the Milky Way center S-cluster formation at least a smooth background potential should be included as one otherwise ignores important mass precession effects around the SMBH. \cite{Subr2016} find that a thermal S-cluster eccentricity distribution can be reached within a reasonable time, and the observed isotropic angular momentum vector distribution can be reproduced as well if a small spherical particle population ($N=500$ stars with $m_\star = 1 \;\mathrm{M_\odot}$) is included. 

However, we caution that these results may be overly optimistic because of the pure Newtonian nature of the simulations. We have shown in Sections \ref{section: results-ecc-pn} and \ref{section: results-inc-pn} that the milliparsec star eccentricity and angular momentum direction distributions at $t=6$ Myr in our models crucially depend on whether PN equations of motion are included in the simulation or not. Without PN, the relaxation effects are artificially strong, rendering the simulated central star clusters on average less eccentric and more isotropic than they should in reality be. We emphasize that PN accurate equations of motion should always be used in simulation codes when studying the stellar dynamics in the vicinity of SMBHs.

\subsubsection{The importance of spherical background modeling}

It has been questioned in the literature whether the observed CW disk thickness supports the idea of an initially dynamically cold CW disk. \cite{Cuadra2008} argued against the cold-disk origin of Milky Way center stellar disks because of too low rms inclinations and the lack of counter-rotating stars in their simulations. The main difference between the simulations of this study and \cite{Cuadra2008} is that their study uses equal-mass stars in the disks, and does not include external potential. We note that the maximum rms inclinations in our studies are approximately two times higher, and our simulation setup includes runs with moderate counter-rotation fractions up to $\sim30\%$. We attribute this difference to the eccentric disk instability, which is partially driven by the spherical background which the runs of \cite{Cuadra2008} do not include.

The simulations of this study use a smooth spherical external background potential instead of a live stellar background population, and no N-body studies using a setup similar to ours and including the live background have been reported in the literature. As the background is smooth, relaxation effects of the disk and the milliparsec stars caused by the background are neglected. A number of issues in our simulation results can be related to the missing relaxation from the background. First, our disks do not show any significant warping. It has been shown using Monte Carlo simulations that vector resonant relaxation induced by the background stellar population can cause strong warping of the disk \citep{Kocsis2011}. The second issue is the S-cluster relaxation problem: which physical relaxation processes can thermalize and isotropize the S-cluster in its age? \cite{Hamers2014} argues that including an observationally motivated (e.g. \citealt{Hopman2006}) background live stellar cusp is enough to relax the eccentricity distribution of the to the observed superthermal distribution ($F(e)=e^{2.6}$, \citealt{Gillessen2009}), even when PN terms are included in the equations of motion of the stars. Their setup involves integrating $N=19$ S-stars with $a_\mathrm{\star}\lesssim0.032$ pc on initially highly eccentric orbits $0.93\leq e_\star \leq 0.99$ as test particles in a background of $N=4.8\times10^3$ stellar remnants with a density profile slope of $\gamma=2$. Each of the remnant has a mass of $m_\star = 10\;\mathrm{M_\odot}$. Already at $t=6$ Myr, a superthermal distribution with $F(e)=e^{2.6}$ is reached, consistent with observations of the orbits of the S-cluster stars.

Testing the realistic live background relaxation scenario would be extremely desirable for our eccentric disk and milliparsec population models as well in order to test whether the additional relaxation effects are strong enough to make our milliparsec star eccentricity distributions less superthermal and more isotropic. Unfortunately, including a stellar component extending well beyond the outer edge of the eccentric disk would significantly increase the particle number and thus computational costs. Assuming the smooth external density profile used in this study extending to $4R_\mathrm{out} = 2$ pc and a background particle (stellar remnant) mass of $10\;\mathrm{M_\odot}$ the background particle number is $N_\mathrm{bkg} \sim1.5\times10^5$, two-three orders of magnitude more than in the pure disk simulations of the current study. While simulating a dense star cluster of a few times $10^5$--$10^6$ particles for $10$ Myr is nowadays not a problem per se \citep{Wang2016,Wang2020b}, the fact that the star cluster is fully embedded in the influence radius of the SMBH in which the particle velocities and accelerations are high makes the problem far more computationally challenging than the traditional N-body star cluster simulations (e.g. \citealt{Aarseth2003}). Even though the simulations of million-body systems around SMBHs are computationally expensive, they are not prohibitively so and have recently become feasible. We highlight the simulations direct N-body simulations of \cite{Panamarev2019} as an example of such studies. Motivated by this, we will perform eccentric disk simulations with a live nuclear star cluster background in a future study.

\subsection{Uncertainties in parameters and numerical models}

\subsubsection{Top-heavy IMF cutoff and very massive stars}

The IMF in the Milky Way center disks is top-heavy (e.g. \citealt{Paumard2006,Bartko2010,Lu2013}), and at the inferred age of $6$ Myr \citep{Paumard2006} stars with initial masses above $\sim 30 \;\mathrm{M_\odot}$ have already died. If the IMF is extremely top-heavy ($\alpha\lesssim1.0$), in addition to the power-law slope of the IMF, another important question is whether the IMF has a high-mass cut-off at $m_\mathrm{max}$. For these extemely top-heavy IMFs the cutoff determines the number of very massive stars (VMSs). This study adopted a cut-off of $m_\mathrm{max} = 120 \;\mathrm{M_\odot}$ but this choice is of course somewhat arbitrary. For the most top-heavy IMF we studied, $\alpha=0.25$, the fraction of very massive stars ($>100\;\mathrm{M_\odot}$) is $\sim13\%$) for $m_\mathrm{max}=120 \;\mathrm{M_\odot}$. If the cutoff mass is increased to $m_\mathrm{max}=250 \;\mathrm{M_\odot}$, the fraction of stars with $>100\;\mathrm{M_\odot}$ is $\sim50\%$ and $>200\;\mathrm{M_\odot}$ is $\sim16\%$. With an even larger cutoff mass of $m_\mathrm{max} = 500 \;\mathrm{M_\odot}$ the fractions of stars over $100 \;\mathrm{M_\odot}$, $200 \;\mathrm{M_\odot}$ and $400 \;\mathrm{M_\odot}$ are $70\%$, $50\%$ and $16\%$, respectively. Thus, the number of VMSs is extremely dependent on the IMF cutoff mass for extremely top-heavy IMFs. From the N-body code point of view, higher stellar masses pose no problem in the mass range in which stellar evolution tracks from fast population synthesis codes are available. In the literature single stellar evolution tracks up to $m_\mathrm{max} = 500 \;\mathrm{M_\odot}$ exist (e.g. \citealt{Szecsi2022}).

Adopting a high value for the cut-off mass $m_\mathrm{max}$ for an extremely top-heavy IMF would result in VMSs both in the disk and the milliparsec region early in their evolution. The maximum residence time of such massive stars near the SMBH is only a few Myr due to their short life-times, smaller than the typical estimates of the current age of the S-cluster stars. The very massive stars naturally act as massive perturbers in the S-cluster, enhancing ZLK oscillations and resonant relaxation effects, and are conveniently already dead by the present day. The effect of a massive perturber (typically in intermediate-mass black hole) on the relaxation of the S-cluster has been studied in the literature (e.g. \citealt{Generozov2020}). The remnants of such stars (e.g. \citealt{GRAVITYCollaboration2023}) are an attractive candidate for the detection of gravitational waves from the galactic center.

\subsubsection{Prescriptions for tidal disruption events}\label{section: discussion-tde}

The massive stars in the S-cluster can prematurely die if they are tidally disrupted by the SMBH, ending their perturbing effect on the S-stars. TDEs and the dynamics of S-stars are inherently coupled. This is because after full tidal disruption, the disrupted star does not contribute to resonant relaxation effects or ZLK oscillations of the S-star orbital elements. As TDEs occur in reality, also simulations of S-cluster stars should always have a recipe for tidal disruption of stars. The chosen TDE prescription and its details affect the long-term evolution of the simulated S-clusters. We show in the middle panel of Fig. \ref{fig: newruns-N-scluster} that whether a TDE prescription is included in the simulation can affect the number of S-stars by up to $20\%$--$30\%$. 

In this study, we adopted a somewhat simple prescription based on the tidal radius argument \citep{Kochanek1992} with a correction factor based on detailed hydrodynamical simulations of TDEs \citep{Ryu2020}. We note that the tidal disruption simulations of \cite{Ryu2020} assume a main-sequence stellar structure. It has been shown that the tidal disruption radius strongly depends on the stellar age due to the evolved structure of the star, and stars close to the end of their main sequence lifetime may not be disrupted even at pericenter separations of $r_\mathrm{p} \sim r_\mathrm{t}/3$ \citep{Golightly2019}. In addition, \cite{Coughlin2022} suggest that the tidal disruption radius of likely $<r_\mathrm{t}/5$ for massive and evolved stars. Furthermore, in our simple model the stars can only fully disrupt, e.g. there are no partial nor repeated TDEs. Based on these considerations, our simulations most probably overestimate the TDE rate while somewhat underestimating the number of S-star like milliparsec stars. However, we emphasize that most of the tidal disruption events in our simulations occur during or just after the eccentric disk instability when $t<1.5$ Myr. At this point even the most massive stars have not had time to evolve away from the main sequence, and the assumption of the main sequence tidal disruption is likely still valid.

\subsubsection{The initial surface density profile of the asymmetric eccentric disk}\label{section: discussion-surfacedensity}

For this work we chose a power-law eccentric disk surface density profile with a slope of $-2$ in order to facilitate comparison with previous simulation studies (e.g. \citealt{Madigan2009}) and Milky Way center observations (e.g. \citealt{Paumard2006}), and to restrict the already fairly large parameter space of the simulation models. In the following we briefly discuss the implications of choosing a different initial asymmetric eccentric disk surface density profile.

Consider a single star in the asymmetric eccentric disk with a certain semi-major axis $a$. If the star lags or leads the main bulk of the disk, the star is pulled towards the main disk. This perturbing acceleration can be divided into a radial and a tangential component. The radial component affects the precession rate of the orbit while the tangential component dictates how rapidly the star recedes from the main disk, or is re-absorbed back into it, as illustrated in Fig. \ref{fig: eccdisk-stability-instability}.

First, if the disk self-contribution to the precession rate of the stars is negligible (as in the case of low-mass disks), the different disk surface density profile does not affect the precession rate of the star. However, it has an effect on the time-scale of the secular eccentric disk instability as it affects the tangential component of the perturbing acceleration from the bulk of the disk. 
Second, if the disk itself is massive enough to affect the precession rate of the star (the case of intermediate-mass disks), steeper disk surface density profiles lead to increased retrograde precession, as more disk mass is located within the orbit of the star with a fixed semi-major axis $a$. The opposite is true for shallower disk surface density profiles. Thus, we expect that for intermediate-mass asymmetric eccentric disks with a steeper surface density profile than the slope of $-2$ the instability of the disk follows more the characteristics of the secular instability of \cite{Madigan2009} than the more complex precession direction instability presented in this work. The opposite is again true for eccentric disks with a shallower surface density profiles: with a given stellar orbit there is less disk mass within the orbit, and prograde orbit precession is more prevalent. This facilitates the onset of the more complex precession direction instability.

We note that the considerations presented here are only approximate and simplified, and additional detailed N-body simulations would be required to assess the effect of the varying eccentric disk surface density profile slope on the results of this study. As the parameter space of the presented study is already fairly large, we do not perform such simulations for the purposes of this work.

\section{Summary and conclusions}\label{section: finalnumber-9}

We study the 10 Myr evolution of asymmetric eccentric disks with top-heavy single and binary stellar populations around SMBHs in galactic nuclei using direct N-body simulations. First we would like to better understand the eccentric disk instability in the regime in which the self-contribution of the disk to the orbit precession cannot be ignored and study the properties of the resulting disks. Second, we want to self-consistently study the formation of milliparsec populations of stars within the inner edge of the disk around the SMBH. Previous N-body studies have focused on more simplified initial conditions or individual aspects of the formation process (e.g. \citealt{Madigan2009,Hamers2014,Subr2016,Generozov2020}). Finally, we compare our simulated disks and milliparsec stars to observations of the Milky Way central parsec with the conclusion that no model studied simultaneously matches the various observational kinematic and stellar population constraints from the S-cluster and the central stellar disks.

We run a simulation sample of $36$ asymmetric eccentric disk models with varying initial disk masses, the initial disk eccentricities, the stellar IMFs and the binary star population properties. The initial disk masses range from $M_\mathrm{disk}=1.0\times10^4 \;\mathrm{M_\odot}$ to $M_\mathrm{disk}=7.5\times10^4 \;\mathrm{M_\odot}$. In the literature, most simulation studies focused on the low mass regime to match the present day observed Milky Way center CW disk mass with no stellar evolution. However, the disk is $t\sim6$ Myr old, and being top-heavy may already have lost a substantial amount of mass due to stellar mass loss and massive stars dying. Thus, the original initial masses of the Milky Way center disks most probably lie in the upper mass range. We simulate a wide range of initial disk eccentricities from $e_\mathrm{init}=0.1$ to $e_\mathrm{init}=0.9$ and three different power-law IMF slopes, the extremely top-heavy $\alpha=0.25$, and the moderately top heavy IMFs with $\alpha=1.3$ and $\alpha=1.7$. We also perform simulations with and without post-Newtonian equations of motion, and with and without a tidal disruption event prescription. The fixed parameters include the SMBH mass and the power-law static external potential representing the spherical background stellar cusp, both matched to  Milky Way measurements.

The lowest-mass asymmetric eccentric disks undergo the secular eccentric disk instability described in \cite{Madigan2009} and produce a thick, axisymmetric disk in less than $1.5$ Myr. Our more massive disk models are unstable as well, but the physical mechanism for the asymmetric eccentric disk disruption is different from \cite{Madigan2009}. For disk masses higher than a few times $10^4\;\mathrm{M_\odot}$ the contribution of the disk itself to the precession of disk star orbits becomes non-negligible, breaking one of the assumptions of the disk instability model of \cite{Madigan2009}. The negligible contribution is usually stated as $M_\mathrm{disk} \ll M_\mathrm{cusp}$. Based on our results, the disk contribution to the orbit precession already becomes important when $M_\mathrm{disk}/M_\mathrm{cusp}\sim0.07$. The reasons for this surprisingly low disk-cusp ratio are that the mass distribution of the disk is planar as opposed to the more massive spherical background, and that the disk is asymmetric. While in the standard secular eccentric disk instability the (retrograde) orbit precession is only caused the spherical background cusp, in the more massive setups the self-contribution of the disk to the orbit precession becomes important. As the eccentric disks are asymmetric, the net torque over an orbit can result in prograde precession in the disk. The prograde precession proceeds from the inner towards the outer parts of the disk with increasing $M_\mathrm{disk}$. Thus, the alignment of the eccentricity vectors of the disk stars is rapidly lost, in less than $1$ Myr,  as the inner and outer parts of the disk precess in different directions. We term this form of the asymmetric eccentric disk instability the precession direction instability. The elliptical orbits precessing at different rates and directions also generate a short-lived one-sided right-handed spiral pattern in the disrupting disk. This structure torques the eccentricities of the disk stars, and the initially narrow eccentricity distributions broaden on shorter time-scales than in the standard eccentric disk instability.

We construct an analytical model for estimating the precession rates and directions of the stars in the disks as a function of $M_\mathrm{disk}$, $e_\mathrm{init}$, the semi-major axis of the star $a_\mathrm{\star}$ and the density profile of the background cusp. The only numerical component in the model is the calculation of the acceleration of the star by the other disk stars at different points of the orbit. The simple model agrees well with the simulation results. The model can be used to predict the precession rates and directions of disk stars and thus the nature of the eccentric disk instability of arbitrary disk setups without running simulations.

The late-time eccentricity distribution of the disk stars always peaks between $0.2 \lesssim e_\star \lesssim 0.5$, regardless of the initial eccentricity of the disk model. Overall, the disk eccentricity distributions are very similar. Only when $e_\mathrm{init}\gtrsim0.7$, a secondary high-eccentricity peak begins to appear in the distribution. This peak reaches the height of the moderate-eccentricity peak at $e_\mathrm{init}\gtrsim0.9$. If the initial disk eccentricity is very high, the eccentricity distribution is bimodal with a secondary peak at $e_\star \sim 0.9$. The vertical height of the disk rapidly increases from its initial thin configuration, and keeps increasing at a slower rate after the eccentric disk instability is over. The rms disk star inclinations reach values typically between $10\degree\lesssim i_\mathrm{rms}\lesssim 30\degree$ in the models. Isolated asymmetric eccentric stellar disks can attain a sizeable fraction of counter-rotating stars, but only if their initial eccentricity is $e_\mathrm{init}>0.4$. Initially more eccentric and more massive disk models end up having up to $30\%$ of stars on counter-rotating orbits. The majority of stars which flip their rotation direction have very high eccentricities ($e_\star>0.9$) at the moment of the flip, consistent with the results of \cite{Madigan2018}.

Simulation models with initial eccentricity higher than $e_\mathrm{init}\gtrsim0.5$--$0.6$ produce up to $\sim100$ escapers and high(hyper)-velocity stars. 
The escaper and high-velocity star direction distribution from the Hills mechanism binary break-ups in our simulations is strongly anisotropic, consistent with the simulation results of \cite{Subr2016}. While the escape direction distribution of non-Hills escapers is close to isotropic, stars ejected by the Hills mechanism predominantly escaper close to the disk mid-plane $\theta=0$. In the azimuthal direction $\phi$ the escaper direction distribution is a moderately narrow cone reflecting direction of the eccentricity vectors of the stars near the inner edge of the disrupting disk. The standard deviation of the in-plane escaper direction distribution gets narrower with the increasing initial disk eccentricity and can be as narrow as $\sigma_\mathrm{\phi}\sim30\degree$. Around $\sim20\%$ of the escaping stars have speeds exceeding $1000$ km/s at $r=10$ pc from the SMBH.

If a tidal disruption event prescription is included in the runs, a large number of stars can be disrupted by the SMBH, most during the first Myr of the simulation. Typically, the number of disrupted stars is at least comparable to the number of stars in the simulated milliparsec cluster. The TDE rate in our simulations during the first Myr strongly depends on the IMF, and varies from $R<10\;\mathrm{Myr}^{-1}$ to $R<770\;\mathrm{Myr}^{-1}$. At later times the TDE rate drops by several orders of magnitude.

Populations of stars closely orbiting the SMBH form in our simulations inside the original inner edge of the disk. We term these simulated stars the milliparsec stars. For initial disk eccentricities $e_\mathrm{init}\lesssim0.5$ the milliparsec population consists predominantly of stars originating from the inner edge of the disk. With higher $e_\mathrm{init}$ there are enough binary stars on a high-eccentricity low-pericenter orbits around the SMBH so that the Hills mechanism can operate. The number of Hills mechanism origin milliparsec stars depends on the initial disk mass, initial disk eccentricity (exponential dependence) the IMF and binary star population properties. The maximum number of B-type milliparsec stars is $\sim80$ in our simulations. Most our simulated milliparsec populations contain a few O-type stars until $t=6$--$7$ Myr. Compared to the observed Milky Way center S-cluster with $\sim30$--$40$ members \citep{Gillessen2017} we conclude that our simulation models can produce central clusters with a comparable number of B-stars in the S-cluster. However, an IMF slope of $\alpha=0.25$ (or lower) is required to simultaneously match the observed number of O/WR stars and the high O/WR-B ratio in the CW disk. The global IMF slope of the young stars within the central parsec of the Milky Way is not that low, for example \cite{Lu2013} provide an estimate of $\alpha=1.7\pm0.2$. When comparing the eccentricity and angular momentum distributions of our simulated milliparsec stars to the S-cluster observations (e.g. \citealt{Gillessen2009,Gillessen2017,Burkert2023}), we find that our clusters are too superthermal in eccentricity and too anisotropic. This fact points to a missing source of relaxation in our simulations. The most obvious candidate for the missing relaxation effect is the resonant relaxation from a realistic  stellar cusp background population. In this study as we model the background with a static smooth external potential. 

We emphasize the importance of including PN equations of motion in the simulations of stars around SMBHs, which is not always the case in the literature. Without PN effects, the resonant relaxation effects are not properly suppressed and relaxation is too strong compared to reality. In this unrealistic case, the simulated milliparsec can have a close to thermal eccentricity distribution, just as in the simulations of \cite{Subr2016} who did not include PN terms in their equations of motion. With the proper PN treatment, we find that the resulting milliparsec stellar population is strongly superthermal compared to the S-cluster observations.

We note that simulation setups with a single, asymmetric eccentric disk embedded in a spherical background as our models are too simplistic to reproduce all the properties of the Milky Way center with all its complexity, such as the S-cluster, the warped CW disk, the almost perpendicular CCW disk and the outer filamentary structures. While our simulations can reproduce a number of individual properties of Milky Way center, such as the number of S-stars and the eccentricity distribution of the central disks, it is challenging to simultaneously fulfill the various observed kinematic and stellar population constraints in a single simulation. While the disk binary Hills mechanism origin of the S-stars remains plausible, more refined setups are needed to better understand the Milky Way center using numerical simulations.

We identify three key improvements for future models. First, resonant relaxation effects in our simulations might be underestimated as we used a fixed spherically symmetric external potential instead of including a live stellar background in the study. Due to the too weak scalar and vector resonant relaxation, the simulated S-clusters might remain superthermal in eccentricity and anisotropic in angular momentum vector direction distribution at $t=6$ Myr. In has been shown in the literature that the scalar resonant relaxation originating from the background stars can indeed relax the initially very eccentric S-cluster star orbits within the age of the stars \citep{Antonini2013,Hamers2014,Generozov2020}. In addition, it has been argued that vector resonant relaxation from the background stars can induce strong disk warps \citep{Kocsis2011}, which are observed in the Milky Way center CW disk but not appear in our simulation. Thus, including a live stellar background seems a promising approach to solve multiple different issues at once. At present, representing the full background population with individual stars is computationally too challenging.

Second, including an IMF gradient in our initial disk setups. Matching the observed number of O/WR stars and the O/WR-B star ratio in the CW disk in our simulations with a global IMF would require an almost flat IMF slope ($\alpha<0.25$). This substantially differs from the overall observed IMF slope in the central parsec, which is less top-heavy with $\alpha=1.7\pm0.2$ (e.g. \citealt{Lu2013}). An IMF gradient would allow having the high O/WR-B ratio in the inner parts of the disks while globally having a less extreme total IMF. Such a variation of the IMF within the disks has been suggested by both recent observations \citep{vonFellenberg2022} and star formation simulations \citep{Hobbs2009}.

Finally, a single and binary stellar evolution code properly coupled to our simulation code would allow a more straightforward comparison of observed Milky Way center stars of various spectral types and evolutionary stages to simulation results. We plan to couple the very recent \sevn{} binary stellar evolution package \citep{Iorio2023} to our \bifrost{} code in a future study.

\section*{Data availability statement}
The data relevant to this article will be shared on reasonable request to the corresponding author.

\section*{Acknowledgements}
The authors thank Taeho Ryu for tidal disruption event prescriptions. The numerical simulations were performed using facilities hosted by the Max Planck Computing and Data Facility (MPCDF), in Garching, Germany. TN acknowledges support from the Deutsche Forschungsgemeinschaft (DFG, German Research Foundation) under Germany's Excellence Strategy - EXC-2094 - 390783311 from the DFG Cluster of Excellence "ORIGINS".


\bibliographystyle{mnras}
\interlinepenalty=10000
\bibliography{references}




\bsp	
\label{lastpage}
\end{document}